\documentclass[acmlarge]{acmart}

\usepackage{caption}
\usepackage{subcaption}
\usepackage{color}
\usepackage{xcolor} 
\usepackage{textcomp}
\usepackage[dvipsnames]{xcolor}
\usepackage{graphicx}

\newcommand{\textchange}[1]{\textcolor{black}{#1}}

\AtBeginDocument{%
  }

\setcopyright{cc}
\setcctype{by}
\acmJournal{IMWUT}
\acmYear{2026} \acmVolume{10} \acmNumber{3} \acmArticle{151}
\acmMonth{9} \acmDOI{10.1145/3832000}

\begin{document}


\title{Evaluating Glanceable Multi-Device Family Health Tracking with Smartwatches and Home Displays}



\author{Lucas M. Silva}
\authornote{Corresponding author.}
\correspondingauthor
\orcid{0000-0001-9795-9071}
\affiliation{
  \institution{University of Iowa}
   \country{USA}
}
\email{lucas-silva@uiowa.edu}

\author{Evropi Stefanidi}
\orcid{0000-0001-5547-6426}
\affiliation{%
  \institution{TU Wien}
  \country{Austria}
}
\email{evropi.stefanidi@tuwien.ac.at}

\author{Aehong Min}
\orcid{0000-0002-3790-2126}
\affiliation{%
  \institution{University of California Irvine}
  \country{USA}
}
\email{aehongm@uci.edu}

\author{Franceli L. Cibrian}
\orcid{0000-0002-7084-6904}
\affiliation{%
  \institution{Chapman University}
  \country{USA}
}
\email{cibrian@chapman.edu}

\author{Jesus A. Beltran}
\orcid{0000-0003-3533-3983}
\affiliation{%
  \institution{California State University Los Angeles}
  \country{USA}
}
\email{abeltr99@calstatela.edu}

\author{Cassie Zeiler}
\orcid{0009-0002-5651-6674}
\affiliation{%
  \institution{University of California Irvine}
  \country{USA}
}
\email{czeiler@hs.uci.edu}

\author{Sabrina E. B. Schuck}
\orcid{0000-0002-9709-2241}
\affiliation{%
  \institution{University of California Irvine}
  \country{USA}
}
\email{sabrina@uci.edu}

\author{Kimberley D. Lakes}
\orcid{0000-0003-3698-274X}
\affiliation{%
  \institution{University of California Riverside}
  \country{USA}
}
\email{klakes@medsch.ucr.edu}

\author{Gillian R. Hayes}
\orcid{0000-0003-0966-8739}
\affiliation{%
  \institution{University of California Irvine}
  \country{USA}
}
\email{gillianrh@ics.uci.edu}

\author{Daniel A. Epstein}
\orcid{0000-0002-2657-6345}
\affiliation{%
  \institution{University of California Irvine}
  \country{USA}
}
\email{epstein@ics.uci.edu}

\renewcommand{\shortauthors}{Silva et al.}

\begin{abstract}
While ubiquitous computing research has explored diverse devices for personal health tracking, we know less about multi-device designs for family informatics, where health management is inherently collaborative. To understand how families adopt and perceive ubiquitous access to shared health data across contexts, we evaluated smartwatch-only, home display-only, and combined designs for tracking moods and goals, domains central to family health behavior regulation. 44 people across 12 families alternated between these designs over nine weeks. Log analysis revealed that mood tracking and goal reporting were significantly more frequent with the home display present compared to smartwatch-only use, despite an overall decline in \textchange{mood tracking} over time. Tracking peaked in afternoons, dropped on weekends, and occurred 2.6$\times$ more at home, with children tracking more consistently than adults across all designs. From interview analysis, we learned how family data glanceability on smartwatches supported opportunistic tracking and awareness while apart, whereas displays reminded families to self-track and collaborate during home routines including members that avoided wearables (\textit{e.g.}, non-participants). Multi-device redundancy accommodated diversity in routines, mobility patterns, and device preferences among members in the same family. We discuss opportunities for multi-device family informatics that accommodates different preferences through inclusive, glanceable, and adaptable ubiquitous data sharing.
\end{abstract}

\ccsdesc[500]{Human-centered computing~Ubiquitous and mobile computing}
\ccsdesc[300]{Human-centered computing~Ubiquitous and mobile devices}
\ccsdesc[500]{Human-centered computing~Empirical studies in collaborative and social computing}
\keywords{Family Informatics, Smartwatch, Home Display, Mood Tracking, Goal Tracking, Health Behavior Regulation}


\maketitle

\section{INTRODUCTION}

Ubiquitous computing research has investigated how systems with different devices facilitate manual self-tracking of behaviors and tasks that impact wellness \cite{Luo2020,ModEatMobileHCI,Kim2017,Kim2021, Epstein2020}. Beyond traditional smartphone apps, prior work has explored voice assistants and smart speakers for exercise \cite{Luo2020} or diet \cite{ModEatMobileHCI,silva2025Foodytalk}; ambient displays for health recovery \cite{Jones2021a}, mood \cite{jang2023}, or everyday tasks and habits \cite{Bressa2022}; and smartwatches for goal tracking \cite{silva2023unpacking} and activity labeling \cite{kim222MyMove}. Such research on multiple devices aim to make self-tracking more accessible and easier across contexts and health domains \cite{ModEatMobileHCI, Yuan2022,Luo2021,Luo2020,Seiderer2017,kim222MyMove}, valuing diverse modes of interaction to diminish the notorious challenge of self-tracking burden and abandonment \cite{Choe2017, Cordeiro2015a, Epstein2016b, Lazar2015}. Because personal health management \textchange{involves and is bolstered by} social support and collaboration \cite{Figueiredo2021,murnane2018social,saksono2024SocioCog}, multi-device approaches to health \textchange{management with tracking may be particularly valuable for caregiving and family contexts, where health} is interconnected among multiple people navigating different daily routines and spaces.

Tracking within \textchange{caregiving} settings (\textchange{\textit{e.g.}, family informatics \cite{pina2017personal}; socially-enabled health technologies \cite{saksono2024SocioCog}}) brings opportunities for collaborative health behavior management \cite{pina2017personal,saksono2024SocioCog,parkCollabTrackingReview}, but also brings social and technical challenges related to device and data integration for shared uses \cite{pina2020dreamcatcher,oygur2020raising}. \textchange{Such informatics systems for caregiving} can benefit overall family health by distributing data collection burden, promoting awareness of issues, enabling caregiving of members in need, and more \cite{pina2017personal, silva2024Codesign, Stefanidi2022,saksono2024SocioCog, Moon2025Fluidtrack, Grimes2009}. \textchange{Systems that enable social interaction features (\textit{e.g.}, health data sharing or messaging \cite{Dennis2024,snapi,Li2020}) might also improve people's ability to sustain personal health behaviors \cite{saksono2024SocioCog}}. However, prior work has typically centered \textchange{such systems on a single device for specific caregiving contexts, such as} phones for parental tracking of children's health (\textit{e.g.}, \cite{epsteinBabyTemptrack,Wang2017quantbaby,silva2023unpacking,oygur2020raising}), \textchange{computer dashboards for monitoring adult or elder member's chronic conditions (\textit{e.g.}, \cite{Yamashita2018,yamashita2017}),} or shared home displays for family use during a specific location and time (\textit{e.g.}, bedtime or morning routine \cite{pina2020dreamcatcher, Sonne2016}). While these single-device approaches have shown much promise, they may also be limiting opportunities for continuous, multi-tracking that could span individual contexts when members are alone and shared moments at home. \textchange{Yet}, we have a limited understanding of how multi-device systems might support or fall short in supporting families in tracking and reflecting on health behaviors pervasively across their diverse daily contexts while maintaining opportunities for collaborative sensemaking.

To investigate how \textchange{\textit{\textbf{multi-device systems}} can support continuous \textit{\textbf{mutual caregiving}} through} health tracking across contexts, we conducted a nine-week deployment with 12 families (20 adults, 24 children) using our FamilyBloom system \cite{chi2026familybloom}, where we varied device use within-subjects. \textchange{We designed FamilyBloom for two devices that represent contrasting ends between personal and shared device possibilities: smartwatches, representing personal, \textit{on-person devices}; and home display, representing \textit{ambient, shared, and peripheral devices} in the home.} Families alternated between three conditions that varied how family data were shared: smartwatch-only, home display–only, and a mixed condition combining both. We focused on mood and goal tracking and reflection, as these domains are central to self-regulation affecting health \cite{hennessy2020self}, require active interpretation, and can benefit from family coordination and support \cite{hennessy2020self,silva2023unpacking,stefanidi2025supporting}. These dynamics are relevant for families broadly \cite{Sanders2018}, and particularly relevant for families facing elevated challenges with everyday behavior regulation, such as those involving Attention-Deficit/Hyperactivity Disorder (ADHD) \cite{Classi2012,Cobb2013}. Accordingly, we recruited families with at least one ADHD member to examine how multi-device family informatics might support collaboration, awareness, and health behavior regulation in contexts where sustained engagement can be especially difficult.

Through our mixed-methods analysis of system logs and family interviews, we found that glanceable designs, and device form factor and home display placement, shaped not only how much families tracked, but how they integrated tracking and reflection into everyday life \textchange{towards mutual caregiving}. The additional presence of a home display was associated with significantly higher self-tracking engagement, with approximately 24\% more mood entries and 48\% more goal completion reporting, compared to family data on smartwatch-only use. At the same time, families developed complementary practices across devices: smartwatches supported opportunistic self-tracking and family awareness while apart and on the move, whereas home displays enabled ambient reminders during home navigation and deeper collaborative reflection during shared moments. Importantly, redundancy of family data across devices accommodated within-family differences in routines, preferences, and participation, allowing even members who avoided wearables or did not actively self-track (\textit{e.g.}, members who did not enroll in the study) to remain aware of and engaged with shared family data. Overall, we found a synergistic relationship of multi-device family informatics, with the smartwatch and home display being adopted with complementary uses. In summary, we contribute:
\begin{itemize}
    \item Quantitative evidence of family self-tracking engagement and the potential influence of glanceable displays on this behavior. Our log-based analysis shows that complementing smartwatch tracking with a home display was associated with more mood logging (24\%) and goal completion reporting (48\%) compared to smartwatch use alone, despite an overall decline in mood tracking volume of roughly 0.76\% decrease per day ($\sim$15\% over 20 days). Mood self-tracking engagement was also significantly higher in the afternoon compared to morning and nighttime periods, declined sharply on weekends (72\% reduction), and occurred far more often at home than away (2.6$\times$), with children tracking more consistently across contexts than adults and with significantly more entries (\textit{e.g.}, smartwatch-only condition: 34\% higher mood entry rate, 3.8$\times$ odds of reporting goal completion).

    \item An empirical understanding of how families perceive and distribute tracking, awareness, and reflection across devices and contexts. Our results indicate that smartwatches supported opportunistic, in-the-moment tracking and family awareness while members were apart; and that home displays enabled ambient reminders as well as collaborative reflection during home routines. Multi-device redundancy accommodated overall diversity of members' routines, mobility patterns within the home, and device preferences. This enabled data mediated collaboration even with members avoiding self-tracking, such as members that did not enroll in the study.

    \item A discussion on design opportunities for multi-device family informatics that balance engagement, coordination, and attention. Based on our findings, we reflect on supporting sustained engagement across devices, enabling mutuality in family connections, and managing the trade-offs introduced by glanceable family data, including attention demands, device maintenance, and family data visibility across contexts.
\end{itemize}
\section{RELATED WORK}
In this section, we situate our work in prior research on (1) family informatics and home displays for shared data review, (2) smartwatches and Ecological Momentary Assessments (EMAs) for health tracking, and (3) multi-device ecologies for tracking and reflection. Across these areas, prior work highlights the value of collaborative reflection, but leaves open how different \emph{multi-device designs and configurations} shape family self-tracking, awareness, reflection, and sensemaking across everyday contexts.

\subsection{Family Informatics \& Home Displays}
Family informatics extends personal informatics \cite{Epstein2015a,Li2010} by emphasizing that health and well-being management in families is inherently interconnected, rather than an individual activity performed in isolation~\cite{pina2017personal, Bell1979, Denham2015}. Family members often share routines, environments, constraints, and overlapping health goals or conditions, meaning that tracking a single person’s data can miss important context about what shapes collective behaviors in everyday life~\cite{pina2017personal, grevenstein2019better, berge2015all, bugelmayer2018family}. This becomes even more salient in families dealing with chronic conditions, where coordinating care requires aligning multiple roles, schedules, and informational needs~\cite{Nikkhah2022, Hong2016, Kaziunas2017, richards2021a, Richards2025FamilyCareRoutines, liangChi2026CollabCareWork}. For ADHD families, with ADHD being highly hereditary \cite{faraone2019genetics}, collaboration and coordination are crucial to sustaining behavior regulation between members \cite{Gulsrud2010,Gisladottir2017}. More broadly, families can be understood as dynamic systems in which members continuously shape each other’s routines and regulation through everyday interaction and coordination~\cite{Sanders2018,pina2020dreamcatcher,Skinner2000}.

A core focus in family informatics is how tracking can support collaborative sensemaking and reflection. Families can benefit from discussing data together because different members contribute complementary perspectives, fill gaps in each other’s understanding, and jointly explore what may underlie observed patterns~\cite{pina2020dreamcatcher, Lee2024FamilyScope, Grimes2009}. At the same time, shared data review can be difficult to sustain and remain superficial~\cite{saksono2019social, Kaziunas2017}. Situated \emph{home displays} can support this necessary family communication~\cite{lindley2009BubbleBoard, sellen2006whereaboutsClock, brown2007whereaboutsclock, sellen2006HomeNote,Neustaedter2009} and make data visible and available ~\cite{Bressa2022, Willett2017, Froehlich2012Display, Carman2006FamilyCalendar,Brush2008}. For example, Dreamcatcher combined sleep sensing from wristbands and self-reported mood on a tablet display to enable families to review each other’s data~\cite{pina2020dreamcatcher}. Similarly, Spaceship Launch used a shared gamified interface for parent-child engagement with physical activity data~\cite{saksono2015spaceship}. 

However, prior work highlights limitations of home displays for supporting continuous family tracking. Shared review often assumes co-location and simultaneous engagement, reducing usefulness when family members are apart and navigating different contexts during the day~\cite{pina2020dreamcatcher, silva2024Codesign}. Data sharing within families also poses challenges; family members may selectively disclose, negotiate privacy boundaries, or avoid sharing altogether depending on their situation~\cite{stefanidi2025supporting, pina2020dreamcatcher}. 
In addition, shared family tracking surfaces recurring tensions around autonomy, privacy, and control, especially when family members differ in roles and preferences~\cite{Kaziunas2017, Toscos2012, silva2023unpacking, stefanidi2025supporting, kaiwenSmartHome2026}. Therefore, we have limited understanding of how to support mutual, ubiquitous tracking for personal and family contexts across different roles and preferences.

\subsection{Smartwatches \& EMAs for Health Tracking}


Smartwatches, a prominent platform for everyday health tracking, are continuously available and usable across locations and activities~\cite{Armagan2011, Pizza2016, Le2025uEMA,gouveaGlanceable2016}. In family informatics, wearables have been used mostly as a means to capture data like physical activity~\cite{saksono2015spaceship, Saksono2020, Oygur2021}, sleep~\cite{pina2020dreamcatcher, shin2023BedtimePals}, and goal-related behaviors~\cite{silva2023unpacking}. Yet, much of this work positions the smartwatch primarily as a sensing endpoint, with reflection and interpretation happening elsewhere (\textit{e.g.}, on a phone or tablet) or left to caregivers, rather than enabling ongoing, shared family data access directly through the device~\cite{silva2023unpacking, Merel2025wearables}. 
This leaves open what \emph{role} a smartwatch can play in a multi-device family setting beyond a source of data capture.

Smartwatches are also increasingly used for Ecological Momentary Assessments (EMAs) and experience sampling, where users provide brief in-the-moment self-reports that would be hard to reconstruct later with memory alone~\cite{Le2025uEMA}. This is particularly relevant for mood and personal goals, where value depends on context and interpretation rather than passive sensing alone. At the same time, smartwatch-based tracking in families often depends on parental setup, maintenance, and interpretation, which can increase invisible labor and limit children’s ownership of the tracking practice~\cite{silva2023unpacking, oygur2020raising}. 
Prior systems provide early signals that wearables can mediate collaboration even when family members are not co-located (\textit{e.g.,} rewards, goals, and routines)~\cite{silva2023unpacking, RikuCalmReminder}, and that personalization can matter for children’s engagement and accessibility~\cite{charitos2025watch}. Still, we know less about EMA-style engagement for children and families in long-term use when the smartwatch is part of a broader device setup for shared awareness and reflection.

Overall, prior work establishes smartwatches as promising for capturing data across contexts and enabling brief, glanceable interactions, including for children. However, we know less about how to use smartwatch glanceability to support \emph{continuous, mutual} tracking and sharing within families, especially when aiming to connect \textit{in situ} everyday experiences to shared reflection opportunities at home. This gap becomes particularly visible for mood- and goal-related tracking, where interpretation and collaborative sensemaking are crucial~\cite{silva2024Codesign,hennessy2020self}.

\subsection{Multi-Device Ecologies}
Classic Ubicomp work on domestic technology shows that households use and share multiple connected devices across rooms and routines, suggesting opportunities for distributing glanceable information in everyday pathways rather than assuming a single primary device~\cite{brush2007yoursmineours, kawsar2013homecomputingunplugged}. Domestic ecologies also complicate identity and privacy, since devices and accounts are often shared and families use lightweight mechanisms that separate personal contexts when needed~\cite{egelman2008familyaccounts}. Research on multi-device interaction examines how tasks unfold across devices and how systems can support continuity when people switch between devices over time~\cite{Nebeling2017, Nebeling2016, Chi2015}. In contrast to tasks like browsing and searching, tracking and journaling health behaviors introduce different temporal dynamics: devices may act as ``collectors'' of data~\cite{OLeary2017}, but it remains unclear when people want to combine devices for logging (in real time, later the same day, or not at all), and how goals and context shape these choices.

Prior work emphasizes leveraging complementary device strengths while reducing friction in transitions, synchronization, and task continuity~\cite{Jokela2015, DiGeronimo2016}. In health and well-being tracking, researchers have explored such combinations, for example pairing smart speakers with mobile apps to provide complementary feedback during exercise~\cite{Luo2020}, or combining different input mechanisms for data access and exploration~\cite{Kim2021,ModEatMobileHCI}. Family informatics systems have also combined devices, like wristbands for capture with shared displays for review \cite{pina2020dreamcatcher, saksono2015spaceship}). 
However, many designs still centralize interaction on a single ``primary'' device and user, without considering the potential of redundant access across devices and how such configurations might shape tracking, awareness, and reflection over time. We would also benefit from understanding how such device redundancy influences coordination in family with heightened behavior regulation challenges, such as ADHD.
\section{METHODS}
In this section, we describe how we tested three FamilyBloom design configurations in a within-subject deployment study, including recruitment and enrollment, procedures, and data analysis methods. Our study was approved by our university's Institutional Review Board (IRB).

\subsection{Family Tracking Designs}
Our primary objective was to investigate how families perceive tracking and data sharing across multiple devices for wellness and collaboration. We adapted three designs of our FamilyBloom system \cite{chi2026familybloom} for mood and goal tracking, which are central to health behavior regulation \cite{hennessy2020self} and family functioning \cite{Skinner2000}. \textchange{In particular, we developed three variations of FamilyBloom corresponding to the study conditions (Section \ref{sec:3.3}): watch-only, home display-only, and mixed designs. The designs presented the same underlying mood and goal representations (\textit{e.g.}, identical visuals and content) while leveraging device's unique affordances (\textit{e.g.}, watchface for persistent visualization, home display's larger screen for viewing multiple days of data).}

\begin{figure*}[]
      \centering
      \subfloat[Watchface with personal moods and goals.\label{FB_watchface_personal}]{{\includegraphics[width=0.18\textwidth]{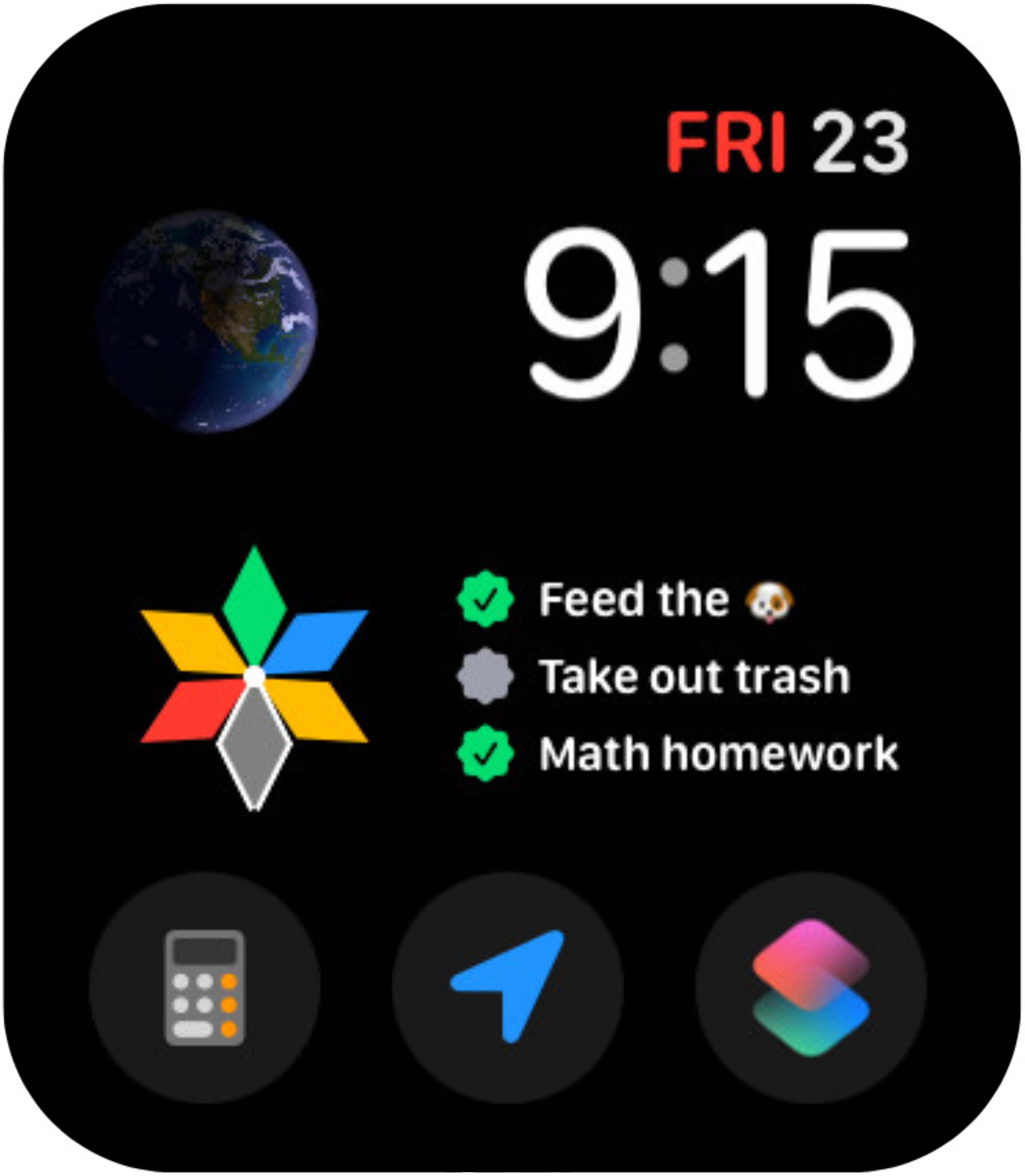} }}
      \hfil
      \subfloat[Menu navigation.\label{FB_personal_menu}]{{\includegraphics[width=0.18\textwidth]{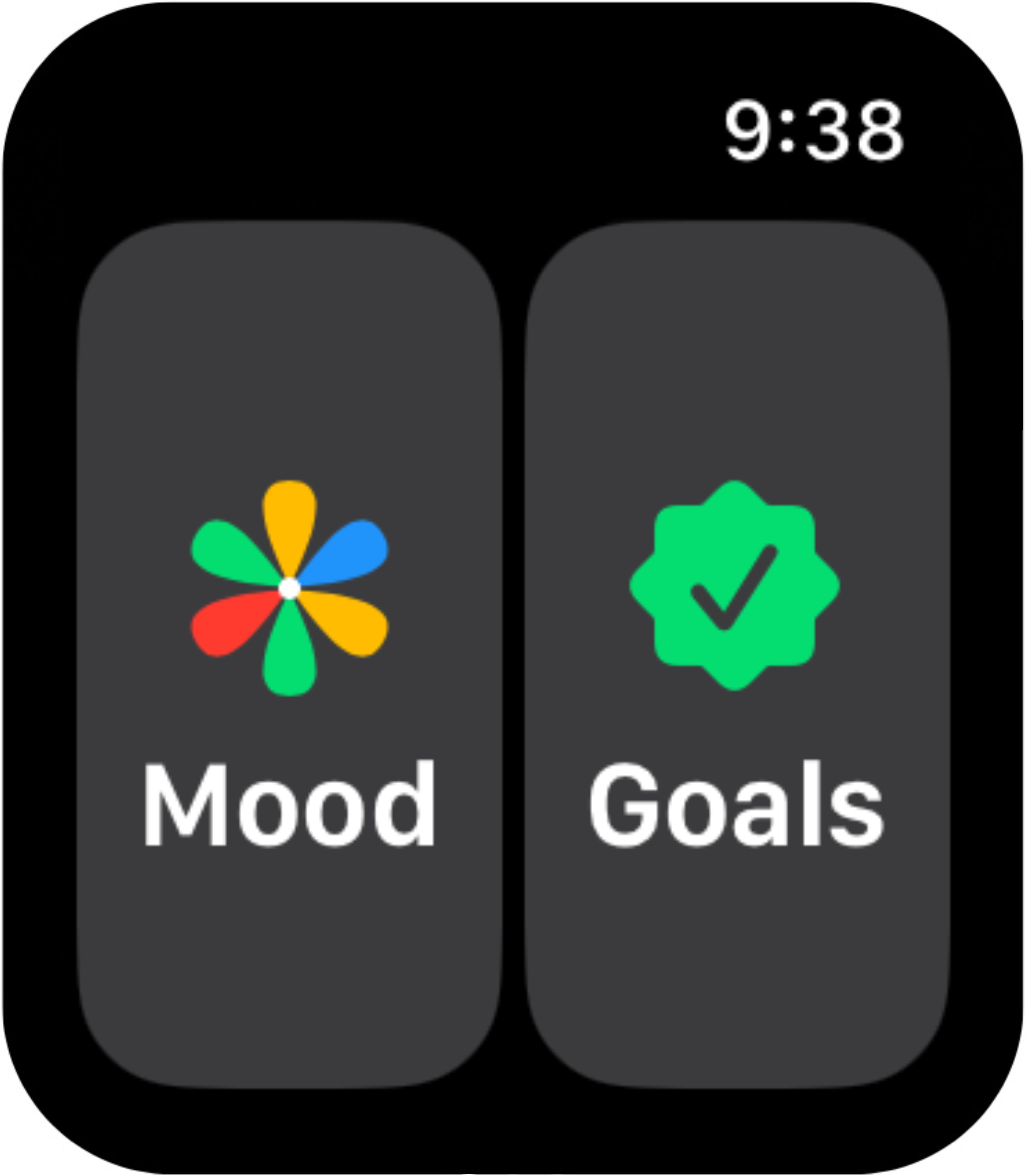} }}
      \hfil
      \subfloat[Goal management.\label{FB_goal}]{{\includegraphics[width=0.18\textwidth]{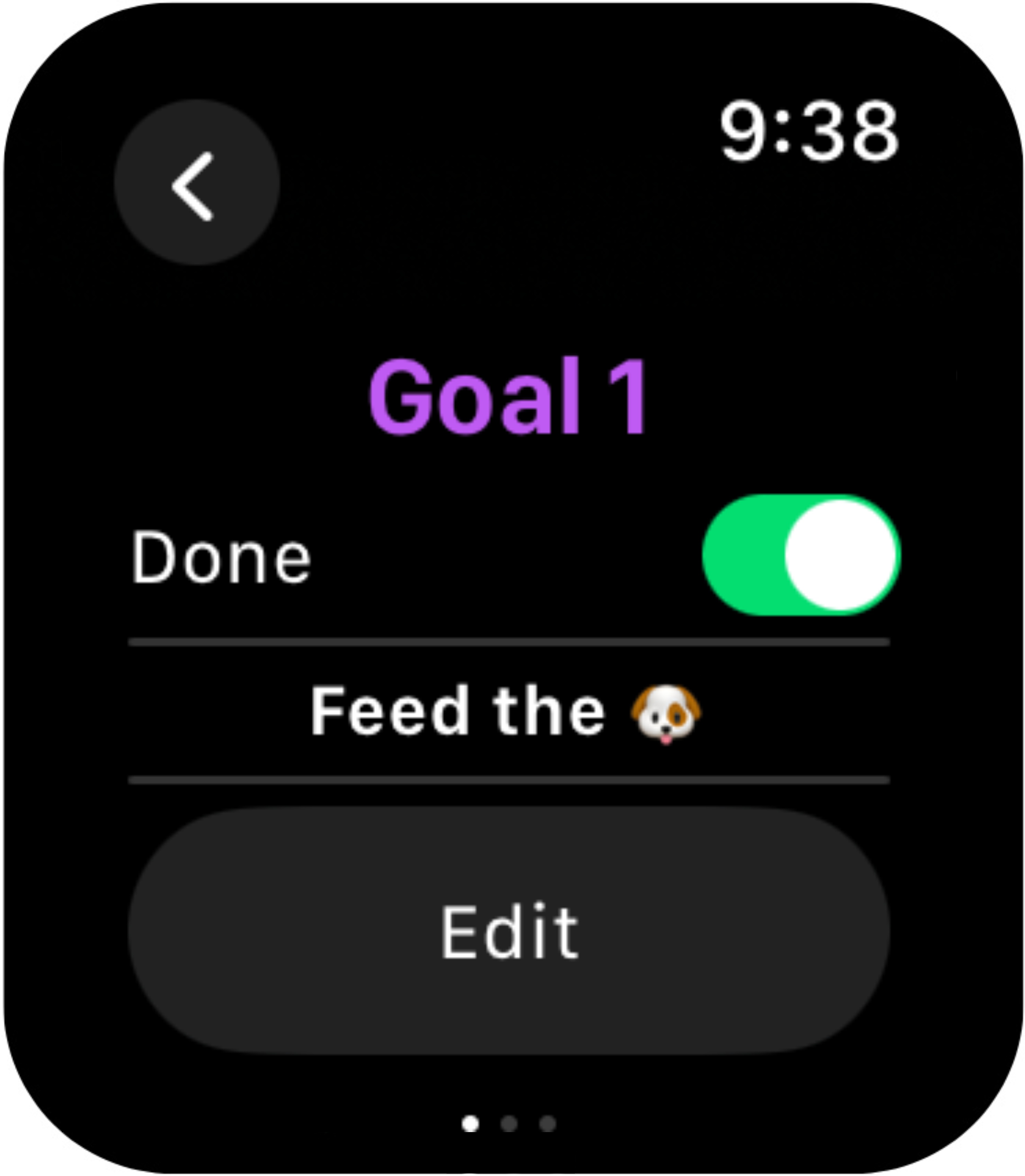} }}      
      \hfil
      \subfloat[Mood color selection.\label{FB_mood_selection}]{{\includegraphics[width=0.18\textwidth]{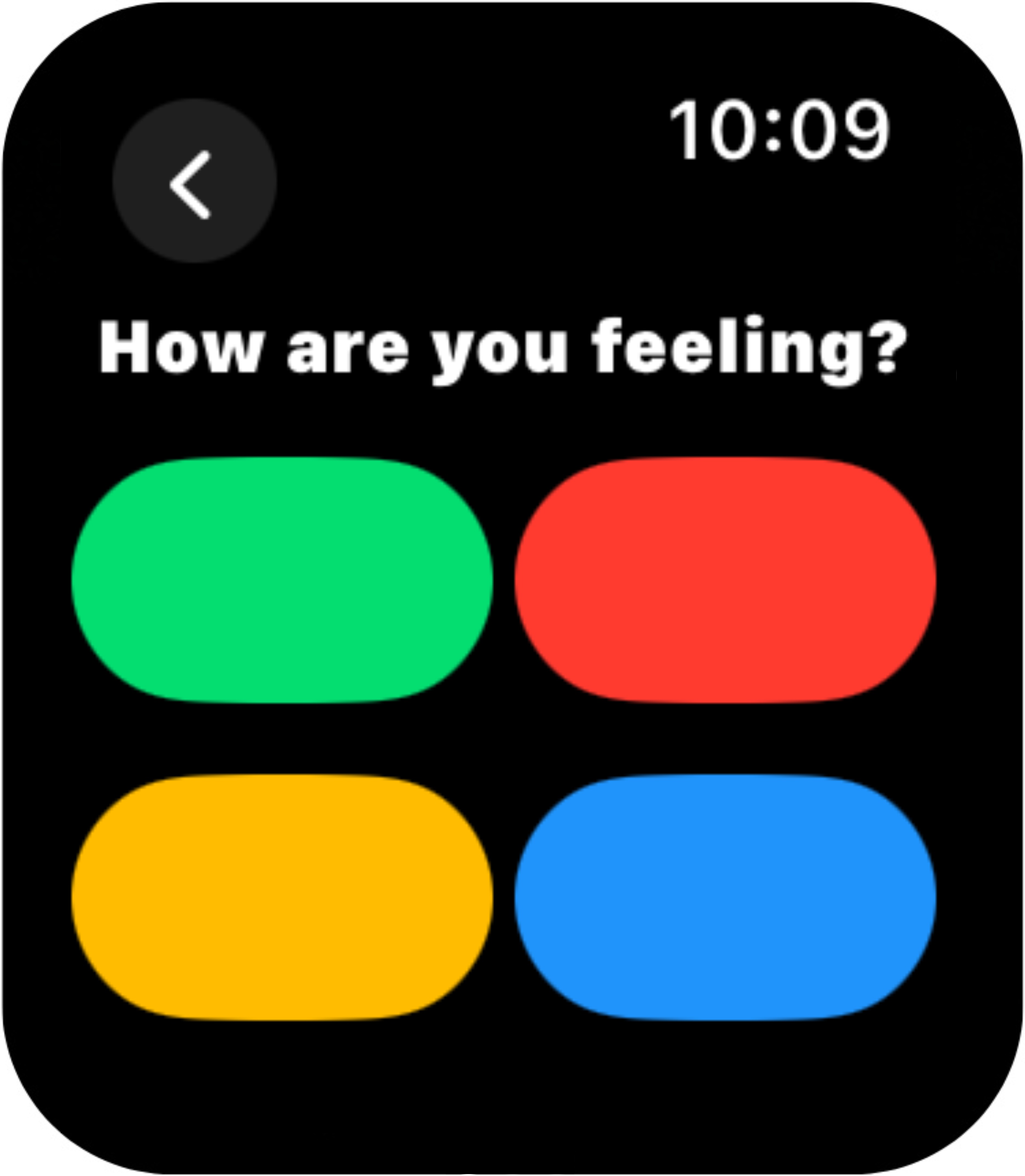} }} 
      \hfil
      \subfloat[Selection confirmation and optional notes.\label{FB_mood_notes}]{{\includegraphics[width=0.18\textwidth]{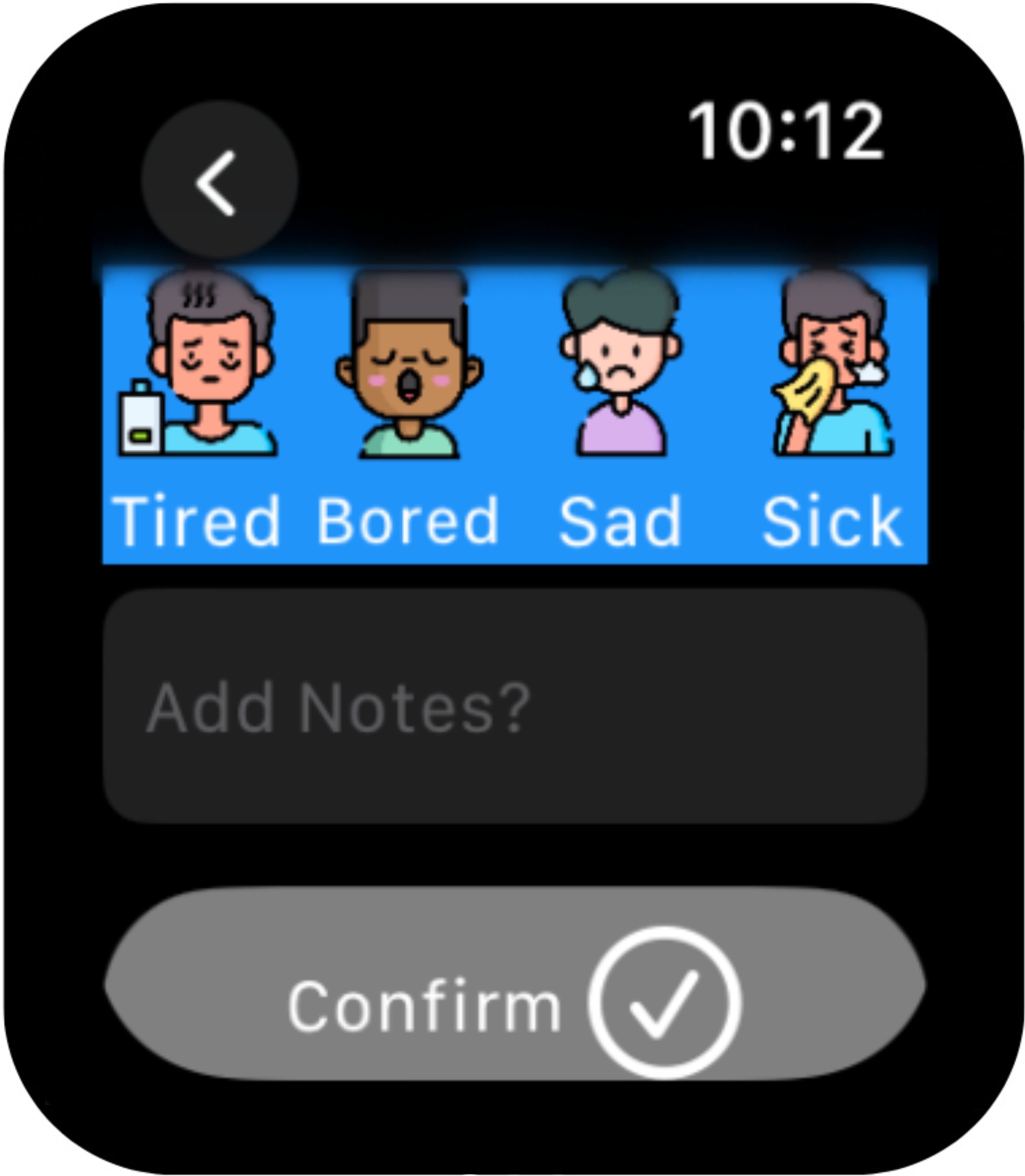} }}      
      \hfil
      
      \caption{FamilyBloom’s smartwatch app \textbf{for self-tracking}. These interfaces are available on the watch during the home display condition, with users having access only to their personal data.}
        \label{fig:FB_watch_self_tracking}
\end{figure*}

Self-tracking of moods and goals is implemented as an Apple Watch application (Figure \ref{fig:FB_watch_self_tracking}). Mood tracking is performed via color selection corresponding to the intensity of internal states (Figure \ref{FB_mood_selection}), with optional open-text notes for flexibility with speech-to-text input (Figure \ref{FB_mood_notes}). For goals, users can manage up to three personal goals represented as text descriptions, edit them via speech-to-text, and toggle each goal’s status as ``done'' or ``not done'' (Figure \ref{FB_goal}). Goal status resets daily to ``not done,'' though users may update goals and their status at any time.

To support personal data glanceability, we implemented watchface widgets for moods and goals (Figure \ref{FB_watchface_personal}). Mood widgets have individual petals or star points correspond to two-hour time blocks positioned like clock hands. The color of each component matches the most recent mood logged within its time window. The component at the 6 o’clock position aggregates moods logged after 6 pm, while the 8 o’clock position corresponds to morning routines and includes moods logged up to 10 am. The goal widget displays the three active goals; each marked with a gray icon or green checkmark to show completion.

\begin{figure*}[]
      \centering
      \subfloat[Watchface for glanceable family data.\label{FB_watchface_family}]{{\includegraphics[width=0.19\textwidth]{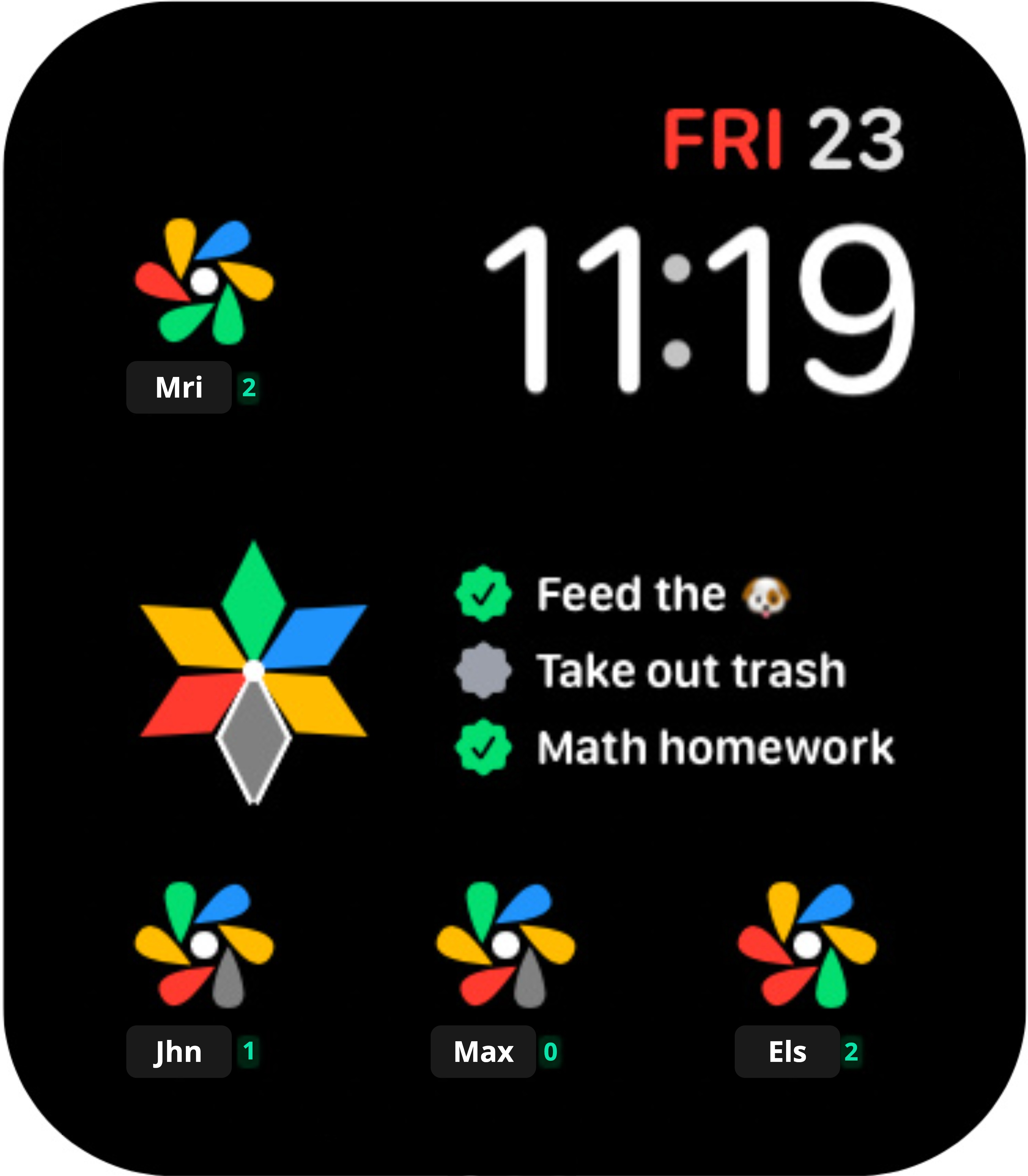} }}
      \hfil
      \subfloat[Menu view with family-data navigation.\label{FB_fam_menu}]{{\includegraphics[width=0.19\textwidth]{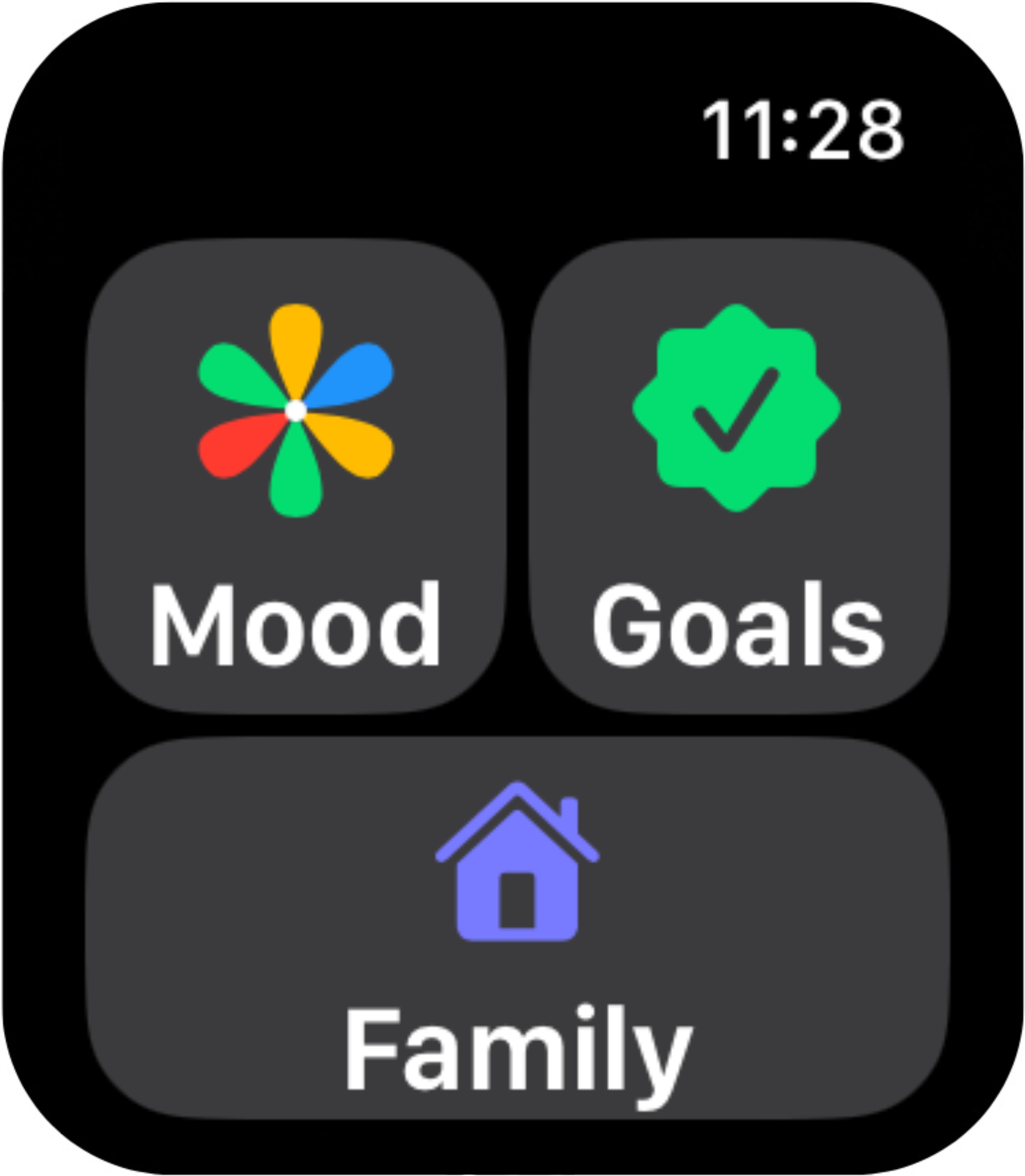} }}
      \hfil
      \subfloat[In-app view of a family member's data.\label{FB_fam_view}]{{\includegraphics[width=0.19\textwidth]{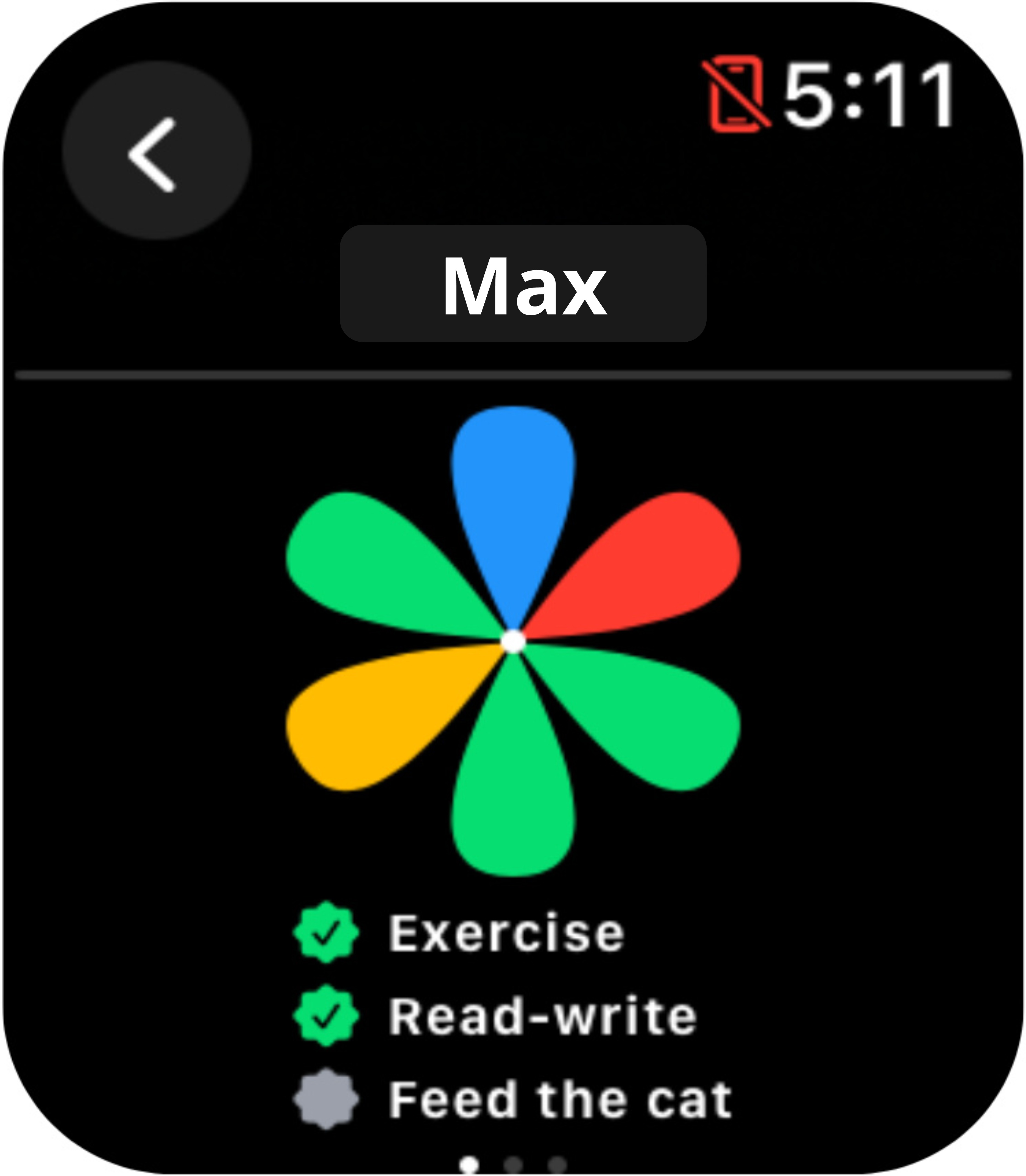} }}      
      \hfil
      \subfloat[\textchange{Example of device with glanceable family data.}\label{FB_eg_watchface}]{{\includegraphics[width=0.19\textwidth]{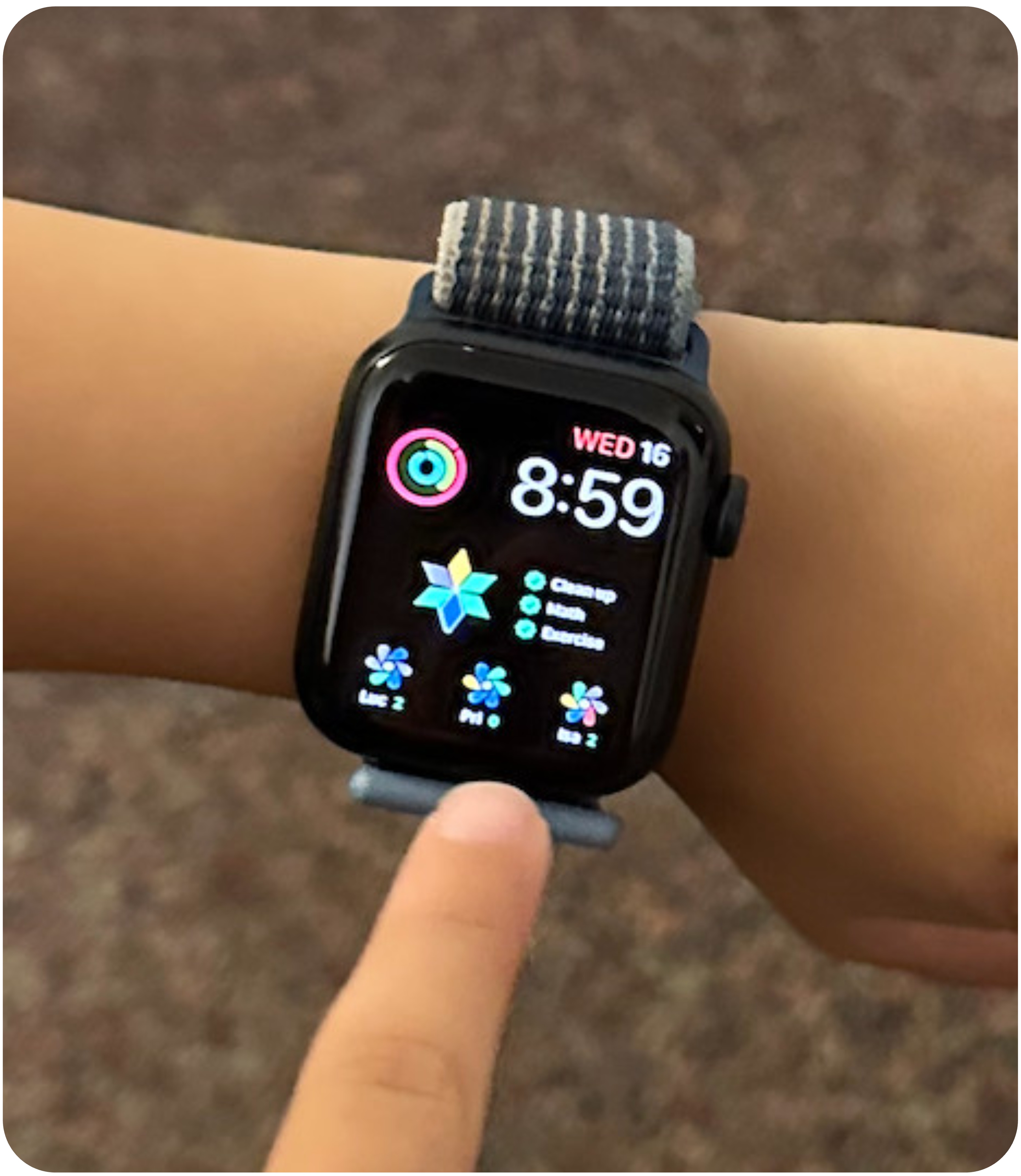} }}      
      \hfil
      \caption{\textbf{Smartwatch-only design} that supports both self-tracking and family data review. Family data is persistently available on the watchface as well as in app.}
        \label{fig:FB_watch_only_condition}
\end{figure*}

In the \textbf{watch-only design} (Figure \ref{fig:FB_watch_only_condition}), additional compact watchface widgets display other family members’ mood flowers and daily goal completion counts (up to four other members) via Apple Watch complications. Family members’ names are abbreviated to three letters alongside their goal completion count and are clickable for in-app details.

\begin{figure*}[]
      \centering  
      {\includegraphics[width=0.95\textwidth]{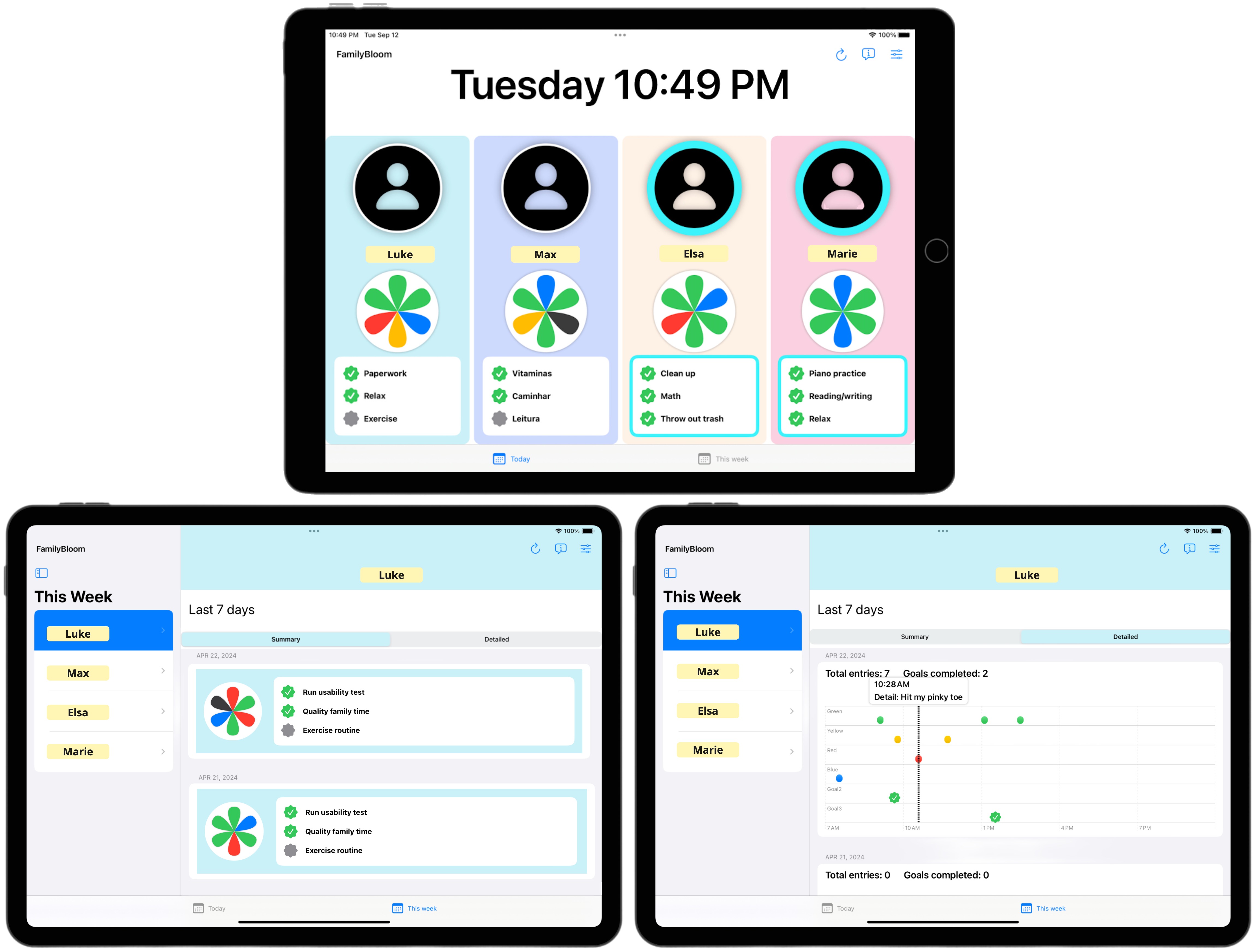} }
    
      \caption{\textbf{The home display-only design} supports family data glanceability on the always-on display (top). Granular data reviews can be accessed through secondary screens detailing the last seven days of member's tracking (bottom).}
        \label{fig:FB_home_display}
\end{figure*}

In the \textbf{home display design} (Figure \ref{fig:FB_home_display}), implemented for iPad, family data are presented on an always-on display synchronized across all members. Each family member can customize their background color and profile image. A secondary screen enables navigation through weekly data, either as a high-level summary or granular times and descriptions (Figure \ref{fig:FB_home_display}-bottom). During the home display-only condition, users only have access to the watch app described in Figure \ref{fig:FB_watch_self_tracking} for personal tracking.

In the \textbf{mixed-design condition}, family data is available both in the home display as well as on the watch (\textit{i.e.},~Figures \ref{fig:FB_watch_only_condition} and \ref{fig:FB_home_display} are both available).

\begin{table}[]
\caption{Participant demographics and device deployment details.}
\label{tab:participants}
\resizebox{0.9\columnwidth}{!}{%
\begin{tabular}{cllllc}
\hline
\begin{tabular}[c]{@{}c@{}}Family\\ ID\end{tabular} &
  \begin{tabular}[c]{@{}l@{}}ADHD \\ Children's\\ Gender, Age\end{tabular} &
  \begin{tabular}[c]{@{}l@{}}Non-ADHD \\ Children's\\ Gender, Age\end{tabular} &
  \begin{tabular}[c]{@{}l@{}}Caregiver\\ Participants\end{tabular} &
  \begin{tabular}[c]{@{}l@{}}Location of~\\ Situated Display\end{tabular} &
  \multicolumn{1}{l}{\begin{tabular}[c]{@{}l@{}}Assigned~\\ Group\end{tabular}} \\ \hline
1  & F, 10        & M, 6        & Father, Father & Dining/Living room   & 2 \\
2  & M, 10        & F, 13;M, 13 & Mother         & Kitchen              & 1 \\
3  & F, 10        & F, 7        & Mother, Father & Entrance             & 2 \\
4  & M, 9; F, 8   & -           & Mother         & Kitchen/Dinning room & 1 \\
5  & M,12         & F, 10       & Mother, Father & Kitchen/Dinning room & 1 \\
6  & F, 9         & F, 7        & Mother, Father & Dining/Living~room   & 1 \\
7  & F, 10        & -           & Mother, Father & Living~room          & 1 \\
8  & M, 11; F, 11 & -           & Mother, Father & Kitchen              & 2 \\
9  & M, 9         & M, 6        & Mother, Father & Entrance             & 1 \\
10 & M, 11        & F, 14       & Mother, Father & Kitchen/Dinning room & 2 \\
11 & M, 12        & F, 10       & Mother         & Kitchen             & 2 \\
12 & M, 10        & F, 9        & Mother         & Entrance             & 2 \\ \hline
\end{tabular}%
}
\end{table}
\subsection{Participants}

We enrolled 44 people (12 families) (Table \ref{tab:participants}) through collaboration with a local school, flyers, email, and word-of-mouth. Parents provided written consent, and children provided verbal assent. Given the focus on data related to behavior regulation in family settings, participation eligibility required families to have at least one child member with ADHD (aged 8-14). Additional children were eligible if they were at least 6 years old.

Overall, participating families included 14 children with ADHD, 10 neurotypical children, and 20 parents. Families resided across eight cities in a large metropolitan area in the United States. \textchange{Families were mostly middle- to high-income and varied in cultural background, although we did not formally collect related classifications from participants.} We use F\# to refer to a specific family, C\#[a-c] to reference a participating child ordered by appearance in Table 1 (\textit{i.e.}, ADHD diagnosis and age), and P\#[a-b] to reference a parent, also ordered by appearance in the table. Several parents (P05b, P08b, P09a, and P09b) reported a diagnosed or suspected ADHD condition. Seven parents owned smartwatches: P04, P05b, P06a\&b, P08a, P10b, P11. 

Not all family members participated: two children in family F06 were under age six, and fathers in families F02, F04, and F11 did not enroll. Family F11 participated in the full deployment (\textit{i.e.}, used the system for more than nine weeks) but did not complete the final two interviews due to a family crisis. We excluded their data from quantitative analysis but maintained interview analysis.

This deployment was part of a larger project and participants were compensated individually in USD and based on their activities: \$10 per interview, per survey, and per week of participation. Compensation ranged from \$120 to \$270 (M = \$151). Families received an iPad (9th gen) used as the home display if all eligible members completed all interviews.

\subsection{Study Deployment Procedures}\label{sec:3.3}

\begin{figure}[]
    \centering
    \includegraphics[width=\linewidth]{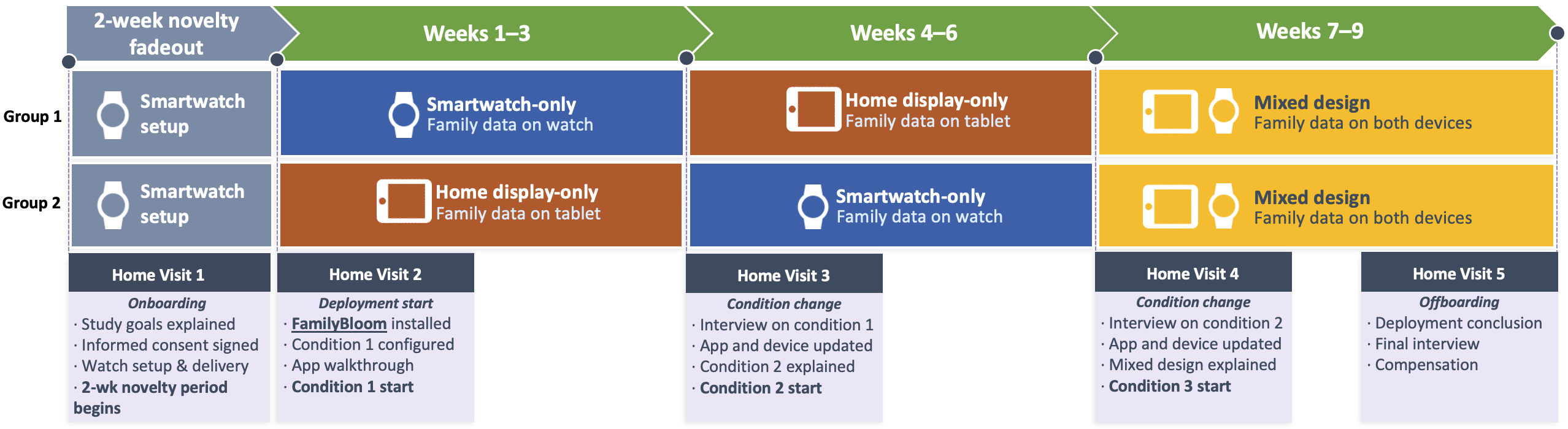}
    \caption{Study timeline and deployment procedure.  \textchange{To diminish novelty effect and help new users gain familiarity with using smartwatches, all families experienced an initial 2-week period with wearables without FamilyBloom.} The within-subject study included three conditions: smartwatch-only (family data visible on the watchface and app), home display (family data visible only on a shared tablet in the home), and mixed (family data visible on both devices). Families were randomly assigned to Group 1 or Group 2, which determined whether they began with the smartwatch or home display condition. The study included five home visits covering onboarding, condition deployments and transitions, interviews, and off-boarding across nine weeks.}
    \label{fig:study_timeline}
\end{figure}

We conducted a within-subjects, repeated measures deployment study in which device and FamilyBloom design conditions were varied every three weeks (Figure \ref{fig:study_timeline}): smartwatch-only (Figure \ref{fig:FB_watch_only_condition}), home display-only (Figure \ref{fig:FB_home_display}), and mixed-designs (both Figure \ref{fig:FB_watch_only_condition} and \ref{fig:FB_home_display}). \textchange{Before the first condition, participants experienced an initial 2-week fadeout period to get used to smartwatch's default apps and features, without FamilyBloom. The intention was to diminish novelty effects of wearables to new users, help children acclimate and explore default features, and account for any participants that might already have experience with smartwatches.} Family units were randomly assigned to Group 1 or Group 2, which determined the order in which conditions were experienced. Because we were concerned with families being deprived of a device as their last experience in the study, we opted to fix the mixed-design condition as the last condition in the deployment. This had important \textchange{order effect implications to results on mixed-design use} and limitations, which we account for in our report in the next sections. Throughout the study, participants were encouraged to explore the system in their daily routines and reflect on likes, dislikes, and design suggestions.


The study consisted of five home visits conducted by the first author:

\begin{itemize}
    \item \textbf{Home visit 1 - Onboarding:} After adults signed informed consent, we visited each home to assent children and setup smartwatches for participants who did not already own one. We explained the study’s goals and asked families to use their smartwatches for two weeks to adjust to its use and minimize any novelty effects. 
    \item \textbf{Home visit 2 - Deployment Start (weeks 1–3):} We installed a condition-specific version of FamilyBloom on participant’s devices. We configured the Group 1 (smartwatch condition) watches to display family data and explained how to use the app to track moods and goals. Group 2 began with the home display condition, in which family data were shown on a tablet placed in a location of their choice (Table 1). For Group 2, smartwatches served only for personal data tracking and visualization.
    \item \textbf{Home visit 3 - Condition Change (weeks 4–6):} We met families again to interview about their experiences and perspectives with the initial design condition. We then explained how FamilyBloom’s design and device configuration would be different for the next phase before reconfiguring setup per condition. 
    \item \textbf{Home visit 4 – Mixed Condition (weeks 7–9):} We again interviewed participants about their experiences and perspectives, asking them to reflect on both prior conditions. We then configured FamilyBloom to the mixed design condition, with family data on both smartwatches and home display. 
    \item \textbf{Home visit 5 – Offboarding:} At the final interview, we discussed the mixed designs and overall study participation. 
\end{itemize}

\subsection{Data Analysis}
\textchange{We deductively analyzed system entry logs quantitatively to compare and understand differences of engagement across conditions. We also qualitatively analyzed interview transcripts to understand participant's perspectives and experiences with the different devices and designs of each study condition.}

\textbf{Quantitative analysis:} We analyzed log-based outcomes of mood entry frequency using generalized linear mixed-effects models (GLMMs) \textchange{with a Poisson distribution and log link, and goal completion additionally using binomial GLMMs with a logit link. Models included random intercepts for participants nested within families (\textit{i.e.}, (1|FamID/UserID)) to account for repeated observations,} controlling for within-participant dependencies and shared family-level routines and contexts. Mood tracking engagement was operationalized as the count of mood entries and modeled using a Poisson distribution with a log link, appropriate for non-negative count data. Goal completion was modeled as a binary outcome (completed vs. not) using a binomial distribution with a logit link. \textchange{All models were fit using the R package lme4 (v1.1-35.5).}

Fixed effects included design condition (watch-only, home display, mixed), condition order, participant role (parent vs. child), day of the week (weekday vs. weekend), time of day, day relative to study or condition start, and location (home vs. out-of-home). Log entry location was classified as out-of-home if their GPS location was more than 0.5~km from the household’s address, but in practice entries at home typically fell within just a few meters of the address. 

Entries were grouped into six day-periods: early morning (before 9am), morning (9am–12pm), afternoon (12pm–4pm), evening (4pm–9pm), night (9pm–12am), and late night (12am–5am). Differences in self-tracking frequency across day periods were evaluated using GLMMs with Tukey-adjusted pairwise comparisons on the log scale. We also conducted an exploratory hour-level analysis in which time of entry was treated as a continuous variable to assess potential non-linear patterns across the day.


\begin{table}[]
\centering
\caption{\textchange{Example of GLMMs used in analysis. All models were fit using the lme4 package in R.}}
\label{tab:model_formulas}
\resizebox{0.9\columnwidth}{!}{%
\begin{tabular}{p{5.2cm} p{10cm}}
\toprule
\textbf{Outcome / Analysis} & \textbf{Model Formula} \\
\midrule

Mood tracking (condition effects) 
& \texttt{EntryCount $\sim$ Condition + (1 | FamID/UserID)} \\

Mood tracking (location effects) 
& \texttt{EntryCount $\sim$ isHome + (1 | FamID/UserID)} \\

Mood tracking \newline (condition $\times$ day period) 
& \texttt{EntryCount $\sim$ Day\_Period * Condition + (1 | UserID)} \\

Mood tracking (order/fatigue effects) 
& \texttt{EntryCount $\sim$ Condition + Order + (1|FamID/UserID) \newline DailyEntries $\sim$ Relative\_Day * Condition + (1|FamID/UserID)} \\ 

\midrule

Goal completion (condition effects) 
& \texttt{isDoneCount $\sim$ Condition + (1 | FamID/UserID)} \\

Goal completion odds (condition \newline  effects)
& \texttt{isDone $\sim$ Condition + (1 | FamID/UserID)} \\

Goal completion odds (child $\times$ \newline parents)
& \texttt{isDone $\sim$ isChild (1 | FamID/UserID) 
\newline isDone $\sim$ isChild * Condition + (1 | FamID/UserID)} \\

\bottomrule
\end{tabular}
}
\end{table}

\textbf{Qualitative analysis:} We used reflexive thematic analysis \cite{Braun2006} to understand families’ lived experiences with health tracking and sharing in everyday life, including their routines, device preferences, and system interactions. We also sought to understand families’ perspectives about the different FamilyBloom device and design configurations. We first reviewed and discussed interview memos as a team. Then, researchers independently open-coded two family interview transcripts before iteratively discussing codes and patterns. The team then used affinity diagramming with quotes to form initial higher and lower-level code clusters. The first author analyzed remaining transcripts and iterated with the team on the affinity diagram. Overall, our analysis led to eight higher-level themes and twenty-two sub-themes. For example, the higher-level theme ``multi-device design accommodating individual preferences'' had the sub-themes ``diversity in preferences'', ``watch and home display synergy'', ``maintenance burden'', and “involving non self-tracking members.'' We further iterated on themes while writing and in conversation with the quantitative findings.

\subsection{Limitations}

Our study includes limitations around both the similarity and diversity of participants. Seven parents were already smartwatch users, and there were tensions based on prior experiences that shaped preferences \textchange{(\textit{e.g.}, a preferred watchface, Section \ref{sec:4.2.2})} and may have influenced quantitative insights about the smartwatch-only design condition. \textchange{Still, we observed that the deployment's initial 2-week period prior to FamilyBloom use helped diminish some device novelty effect and equalize novel and experienced users' familiarity to smartwatch's affordances (\textit{e.g.}, reported exploring fitness tracking, Memoji avatar creation).  Overall, while some participants might have been more dexterous with smartwatches and technology than others, all had gained a practical understanding and experience with their smartwatches before study conditions.}

Although all families had 3 or 4 members enrolled and were balanced across conditions, variation in household structure may have affected how family data was interpreted and discussed. We included family-level analysis in our quantitative analysis as a means of accounting for differences of such structure. Some children were unable to wear or interact with the smartwatch for some school hours due to school policies or teacher discretion. Parents reported having resolved this restriction with the school, except for F06. While this limited continuous daytime data collection, it reflects real-world constraints on children’s device use.

Deploying with ADHD families provided valuable insights into multi-device ecosystems in contexts of heightened behavior regulation challenges, but families facing different health conditions or neurotypical families may experience different engagement patterns and challenges. For instance, observed system benefits and severity of distraction concerns may be elevated in ADHD contexts compared to other \textchange{(\textit{e.g.}, glanceability becoming distracting under some circumstances, see Section \ref{sec:4.2.2})}. Additionally, participants consisted primarily of middle- to high-income families, who likely had greater access to digital infrastructure, health resources, and supportive services. Thus, these findings may not generalize less-resourced environments, underscoring the need for future research with more diverse populations.

We inferred reflective engagement through qualitative insights and log analysis. We avoided potentially intrusive instrumentation (\textit{e.g.}, life camera measures of glance behavior \cite{Pizza2016}) to limit participant burden, particularly important when working with families and children with ADHD. Our mixed-methods, longitudinal approach allowed us to balance ecological validity with sustained participation. Additionally, our findings intertwine with the particular device and design choices of our study. It is possible that different visualization strategies, screen sizes, and device formats might inform different family collaboration practices or a lack thereof. Relatedly, we cannot rule out that the design's persistent shared visibility may itself have influenced individual members’ emotions, self-reporting practices, or behavioral patterns over time (\textit{e.g.}, social desirability, emotional contagion, avoidance). Exploring different data visualizations, interaction modalities, sharing/privacy controls, and devices for family caregiving are important opportunities for future work.

\section{FINDINGS}
Our analysis revealed varying perspectives on the utility, strengths, and weaknesses of smartwatches, situated home displays, and mixed designs for self-tracking and family data sharing. These perspectives were influenced by the family’s connection, individual daily routines, device and design preferences, and surrounding context. Having both designs for smartwatch and home display devices with glanceable data sharing was associated with increased self-tracking engagement, despite some drop-off from mood tracking burden over time. While family members are connected as a unit, we find they can have diverging device and design preferences impacting digital health sharing and collaboration adoption in practice. Multi-device system designs can accommodate individual differences in adopting ubiquitous family health data sharing. \textchange{Overall, multi-device glanceability for family data sharing prompted mutual caregiving practices that leveraged each device and design's strengths.}

In this section, we report on how deployment design conditions impacted participant’s self-tracking engagement (\ref{sec:4.1}), unpack unique patterns of engagement and perspectives on smartwatch’s (\ref{sec:4.2}) and home display’s (\ref{sec:4.3}) strengths and barriers, and describe synergistic benefits and limitations of having both devices (\ref{sec:4.4}).

\subsection{The Impact of Family Sharing Designs on Self-Tracking Engagement}\label{sec:4.1}

Across conditions, participants logged 7,403 mood entries and 5,286 goals, of which 1,270 were marked complete. Table \ref{tab:quant_summary} summarizes our analyses of self-tracking engagement over the deployment period.

\begin{table}[t]
\centering
\caption{Summary of main quantitative effects on mood tracking and goal completion reporting.}
\label{tab:quant_summary}
\resizebox{0.99\columnwidth}{!}{%
\begin{tabular}{p{3.0cm} p{6.8cm} p{3.2cm}}
\toprule
\textbf{Factor} & \textbf{Outcome} & \textbf{Effect Size} \\
\midrule
Home display \newline(vs.\newline Smartwatch-only) 
& Higher mood tracking frequency (24\%); higher goal completion reporting (48\%)
& Mood: \newline$\beta = 0.213$, $p < .001$ \newline
Goals: \newline $\beta = 0.39$, $p < .001$ \\[5pt]

Time of day 
& Mood entries most frequent in the afternoon compared to other day periods 
& Tukey-adjusted \newline$p < .05$ \newline\\[5pt]

Study duration fatigue 
& Decline in mood tracking over time ($\sim$15\% over 20 days); no association with goal logging 
& Mood: \newline$\beta = -0.0076$, $p = .009$
Goals: \newline $\beta \approx 0$, $p = 1.00$ \\[5pt]

Condition order \newline fatigue
& Fewer mood entries following each condition transition (every 3 weeks): $\sim$28\% reduction per transition \newline
& $\beta = -0.3236$, $p < .001$ \\[5pt]

Weekend \newline(vs.\ Weekday) 
& Substantially fewer mood entries on weekends \newline ($\sim$72\% reduction) \newline
& $\beta = -1.28$, $p < .001$
\\[5pt]

Location \newline (home vs.\ away) 
& Mood entries more likely at home than away (118–159\% higher rate); with a smaller location effect for children 
& Home: \newline $\beta = 0.78$-$0.95$, $p < .001$ \newline
Children: \newline$\beta = -0.34$, $p < .001$ \\[5pt]

Participant role \newline(child vs.\ adult) 
& Children logged more mood entries and goal completions overall; larger gap during smartwatch-only phase (34\% higher mood entry rate, 3.8$\times$ odds of reporting goal completion)
& Mood: \newline $\beta = 0.29$, $p < .001$ 
Goals: \newline $\beta = 1.33$, $p < .001$; \newline OR $\approx 3.8$\\
\bottomrule
\end{tabular}
}
\end{table}


\subsubsection{Home Displays Associated with Higher Personal Self-Tracking Engagement}

\begin{figure}
      \centering
      \subfloat[Mood tracking frequency per condition.\label{FB_moodCondition}]{{\includegraphics[width=0.45\linewidth]{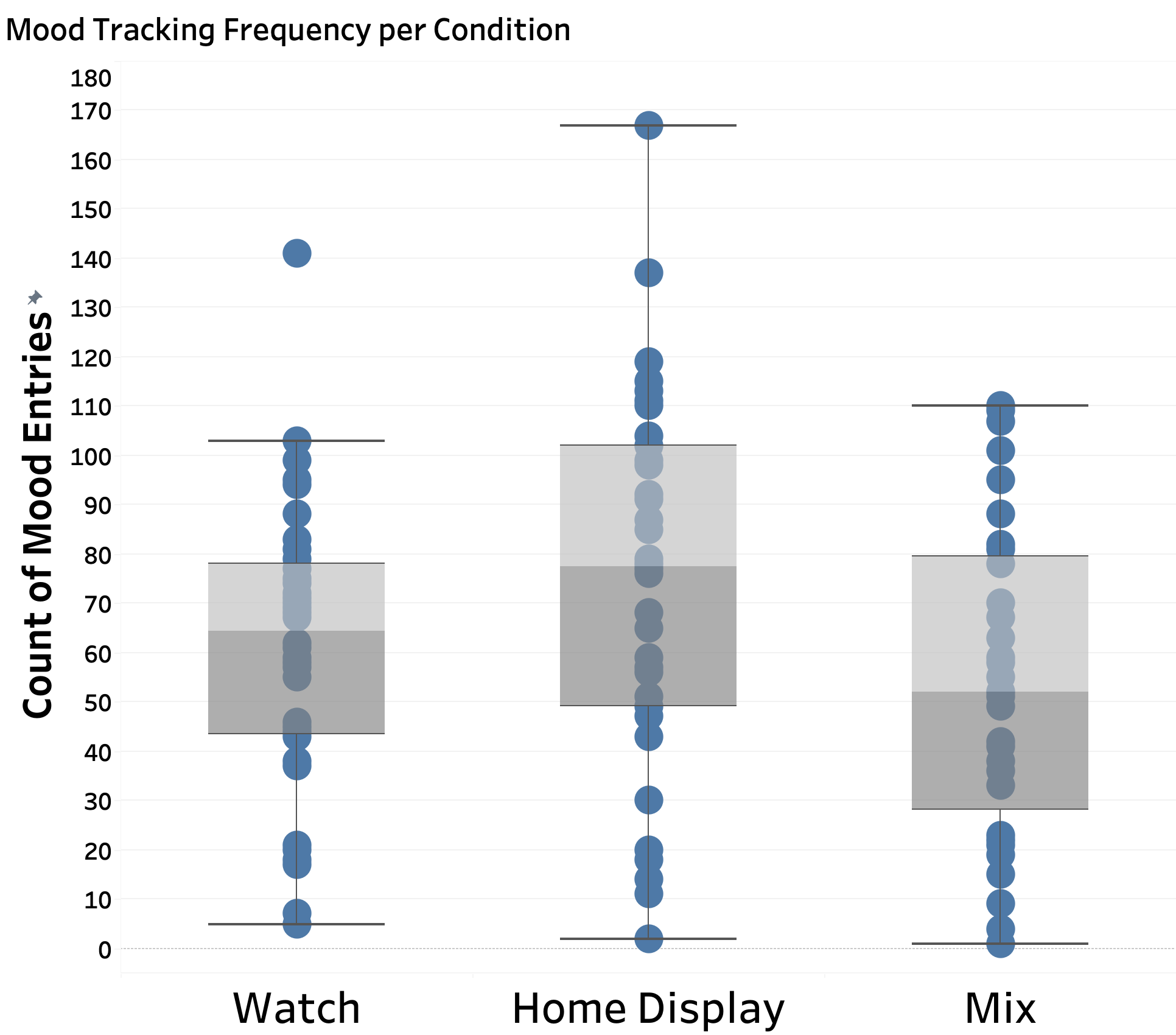} }}
      \hfil
      \subfloat[Goal completion reporting per condition.\label{FB_goalCondition}]{{\includegraphics[width=0.45\textwidth]{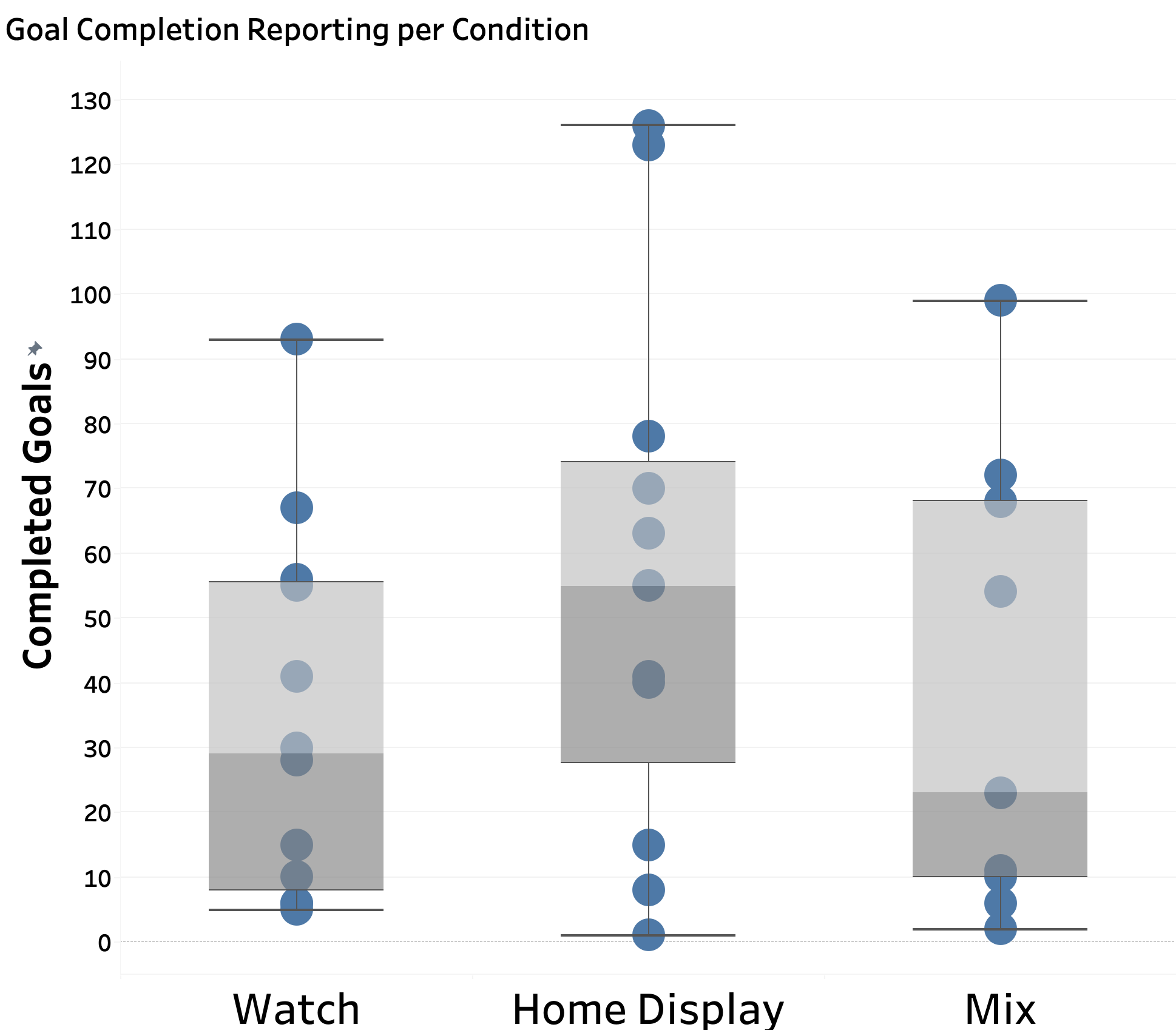} }}
      \caption{Home display design condition was associated with higher mood entry rate ($\beta = 0.213$, $p < .001$) and goal completion reporting ($\beta = 0.39$, $p < .001$) compared to smartwatch-only condition. Mixed-device condition was similarly higher when controlling for order (\textit{i.e.}, accounting for tracking fatigue; $\beta = 0.353$, $p < .001$).}
        \label{fig:FB_engagement_vsConditions}
    \Description{}
\end{figure}

When families had a home display available \textchange{rather than only wearable devices}, self-tracking engagement was significantly higher for mood inputs and goal completion reporting (Figure \ref{fig:FB_engagement_vsConditions}). In the mood logging model, compared to the watch-only condition, the home display condition was associated with a significantly higher entry rate ($\beta = 0.213$, $p < .001$, 24\% more mood entries). \textchange{While it might be expected that multi-device availability directly increases engagement, the mixed-design condition suffered from relatively lower engagement. We believe this was due to it being fixed as the last condition, and when controlling for condition order, it did show a significant increase in engagement compared to the watch-only condition} ($\beta = 0.353$, $p < .001$). \textchange{We further report on ordering effect impacting tracking drop-off in the next subsection and participants' perceived benefits in Section \ref{sec:4.4}.}

Goal completion reporting was significantly higher in conditions that included a home display compared to watch-only use. In condition-only models, the home display condition was associated with significantly higher numbers of completed goals compared to watch-only use ($\beta = 0.39$, $p < .001$), corresponding to approximately 48\% more completed goals. When accounting for \textchange{condition order}, both \textchange{the home display and mixed-design} remained significantly associated with higher goal completion counts, with the mixed-device condition associated with over twice as many completed goals compared to watch-only use ($\beta = 0.97$, $p < .001$). In addition to these insights on modeling volume of goal completion reporting, we analyzed likelihood of tracked goal being marked as complete (\textit{i.e.}, binomial mixed-effect modeling). The home display-only was associated with a 55\% increase in the odds of goal completion, and when accounting for order effects, both the home display (OR $= 1.63$, $p < .001$) and mixed-device conditions (OR $= 3.00$, $p < .001$) showed substantially higher completion probabilities compared to watch-only use.

Together, these results indicate that the presence of a home display for personal and family review \textchange{in addition to data on watchfaces} was associated with higher self-tracking engagement beyond when having watch-only family data, both for mood and goal domains.

\subsubsection{Engagement Patterns of Self-Tracking}

\begin{figure}
    \centering
    \includegraphics[width=0.95\linewidth]{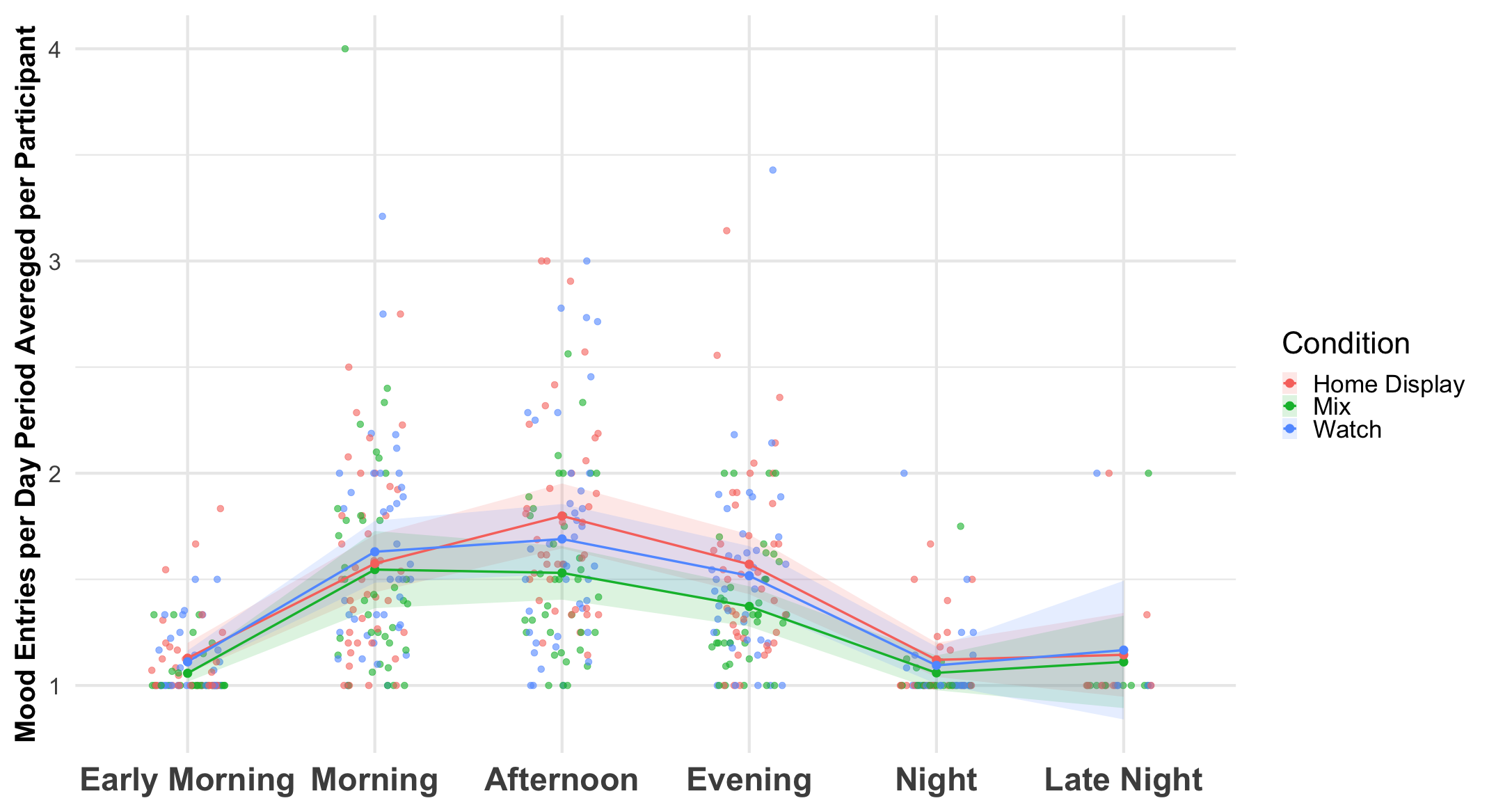}
    \caption{Average of mood tracking per day period per participant, with shaded regions representing $\pm95$\% confidence intervals across participants. Mood entries were most frequent during the afternoon, which was significantly higher than early morning, night, and late-night periods (Tukey-adjusted p < .001 for all) and morning and evening periods (Tukey-adjusted $p  < .05$). Design conditions did not \textit{significantly} alter \textit{when} participants typically self-tracked ($p = .454-.836$), although impacting volume of tracking.}
    \label{fig:dayperiod}
\end{figure}

\textbf{Self-tracking varied significant across the day:} Mood tracking frequency varied significantly by time of day ($p < .001$; Figure \ref{fig:dayperiod}). Mood entries were most frequent during the afternoon, which was significantly higher than early morning, night, and late-night periods (Tukey-adjusted $p < .001$ for all), and moderately higher than morning and evening periods (Tukey-adjusted $p  < .05$). All nighttime periods showed substantially lower engagement compared to daytime periods (Tukey-adjusted $p  < .001$). 

These temporal patterns were consistent across conditions and participant age groups, with no meaningful interactions between day period and condition or age group ($ps > .05$). Hour-level analysis did not reveal additional structure beyond these day-period effects. While the presence of a home display was associated with higher logging volume, it did not significantly alter when participants tracked across the day (day periods $\times$ condition interaction:  $ps = .454-.836$).

\textbf{Tracking moods showed \textchange{gradual} fatigue drop-off, but goal \textchange{completion} tracking did not:} While home displays were positively associated with engagement, we observed evidence of tracking fatigue over the course of the study. C03a explained: \textit{``It’s like brushing your teeth, it makes sense, but you end up not wanting to do it sometimes.''} F01 similarly said: \textit{``These last three weeks were really different from the first three. Maybe it was also fatigue, I don't know.''} (P01a); In examining daily mood logging as a function of days since study start, there was a significant negative association for number of mood entries ($\beta = -0.0076$, $p = .009$), corresponding to a 0.76\% decrease in daily mood entries per additional day in the study ($\sim$15\% over 20 days). In contrast, there was no significant association between study day and goal logging frequency ($\beta \approx 0$, $p = 1.00$), nor any interaction between study day and condition ($ps > .96$). Goal completion reporting could be more episodic than moods throughout the day. Still, goal completion reporting was low overall ($\sim$24\% of all goals) with substantial between-user and family variability (see Section \ref{sec:4.4}).

Considering the condition order as a fixed effect similarly revealed fatigue. Each successive condition order was associated with a significant reduction in mood entry counts ($\beta = -0.3236$, $p < .001$, $\sim$28\% fewer entries from first to last condition). Because the mixed-design condition was always last, it received the largest order-related penalty. However, when controlling for order, the mixed condition yielded higher engagement than the watch-only condition ($\beta = 0.35340$, $p < .001$), suggesting that the presence of a home display \textchange{combined with family data on the smartwatch} may have partially mitigated the effects of tracking fatigue.

While we observed indicators of fatigue, models showed engagement remained above zero throughout the nine-week deployment; families tracked even in their final weeks of participation\textchange{, and they reported tracking} helped maintain benefits. They described tracking interactions as \textit{``really easy to use''} (P10b) and \textit{``just a few clicks ...to get the job done''} (P10a), becoming \textit{``a normal habit of checking in''} (P09b) for some. F06 reflected on this routine baseline, saying, \textit{``it will be weird to be without it [after the study], we were all connecting with it''} (P6a); and \textit{``now it just is part of our routine, part of the day-to-day check-in''} (P06b). While tracking became less frequent, participants did not disengage entirely and continued to incorporate self-tracking into their routines over an extended period. \textchange{These qualitative insights indicate the complementary benefits of combining designs across devices, which help explain why families still valued the mixed condition despite reduced overall logging volume due to the order effect. We further expand qualitative insights for multi-device, mixed-designs in Section \ref{sec:4.4}.}

\textbf{Weekdays for tracking, weekends for relaxing:} We observed a quantitative and qualitative difference between weekdays and weekends, with weekends being seen by many as \textit{``day offs''} from engaging with tracking. Mood tracking was lower during a weekend day versus a regular weekday in all conditions, with substantially fewer mood entries ($\beta = -1.28$, $p < .001$), corresponding to a 72\% reduction in entries (IRR $= 0.28$) for the mixed design condition. Weekend-related reductions were slightly smaller in the tablet condition (IRR $\approx 0.31$; $\sim$69\% reduction), although not statistically significant, and smallest in the watch-only condition (IRR $\approx 0.35$; $\sim$65\% reduction), significant compared to mix condition ($\beta = 0.22$, $p = .003$). In contrast, goal completion did not differ significantly between weekends and weekdays overall ($\beta = 0.12$, $p = .45$).

Qualitative findings corroborate with these statistical patterns. For example, C01a said \textit{``I never do them [tracking inputs] on weekends''}, with P01a complementing \textit{``I agree, things are different in weekends''}. Similarly, C04a explained \textit{``on weekends, I’ll barely see it [FamilyBloom] because I'm usually in the house quickly and grab something [to eat] and I'm playing outside.''} Overall, participants saw the weekend as a different routine for their wellness for which the shared and self-tracking designs were not as relevant as weekday routines. While these differences align with some cultural norms surrounding weekdays (\textit{e.g.}, for work and school) versus weekends (\textit{e.g.}, for leisure), they suggest opportunities for systems to support alternative forms of engagement on weekends, such as family reflection or joint wellness activities grounded in weekday data.

\textbf{Participants tended to track more at home versus away:} Beyond temporal patterns, tracking location also influenced engagement. Across all conditions, mood entries were significantly more likely to be logged at home than away from home ($\beta = 0.78$--$0.95$, $p < .001$), with approximately 2.6 times as many entries when participants were at home. However, children showed a significantly smaller home--away difference ($\beta = 0.34$, $p < .001$), indicating they were relatively more likely than adults to engage in mood tracking across locations. Similarly, 91\% of goal completion reporting occurred at home, though entries during the smartwatch-only period were significantly less likely to be marked at home compared to conditions with a home display ($\beta = -0.48$, $p = .028$). Qualitative analysis related to these behavior patterns \textchange{revealed several reasons reported by participants that could have led to logging more at home versus away: attention focused on work (\textit{e.g.}, meetings) and school (\textit{e.g.}, engagement only during recess). Other explanations} ranged from device-specific affordances \textchange{and constraints, to personal preferences and challenges faced in diverse contexts of regular everyday} routines, which we detail in Sections \ref{sec:4.2}--\ref{sec:4.4}.

\subsection{Smartwatch: Benefits and Shortcomings}\label{sec:4.2}

Participants expressed largely positive perspectives on using the smartwatch for self-tracking and reflection on glanceable \textit{personal} data. Perceptions of the value of viewing family-level data on the watch were more mixed, with tensions around contextual use, attention/distraction, or preferences for different watchfaces.

\begin{figure}
      \centering
      \subfloat[Mood tracking frequency per adults and children.\label{FB_mood_age}]{{\includegraphics[width=0.443\linewidth]{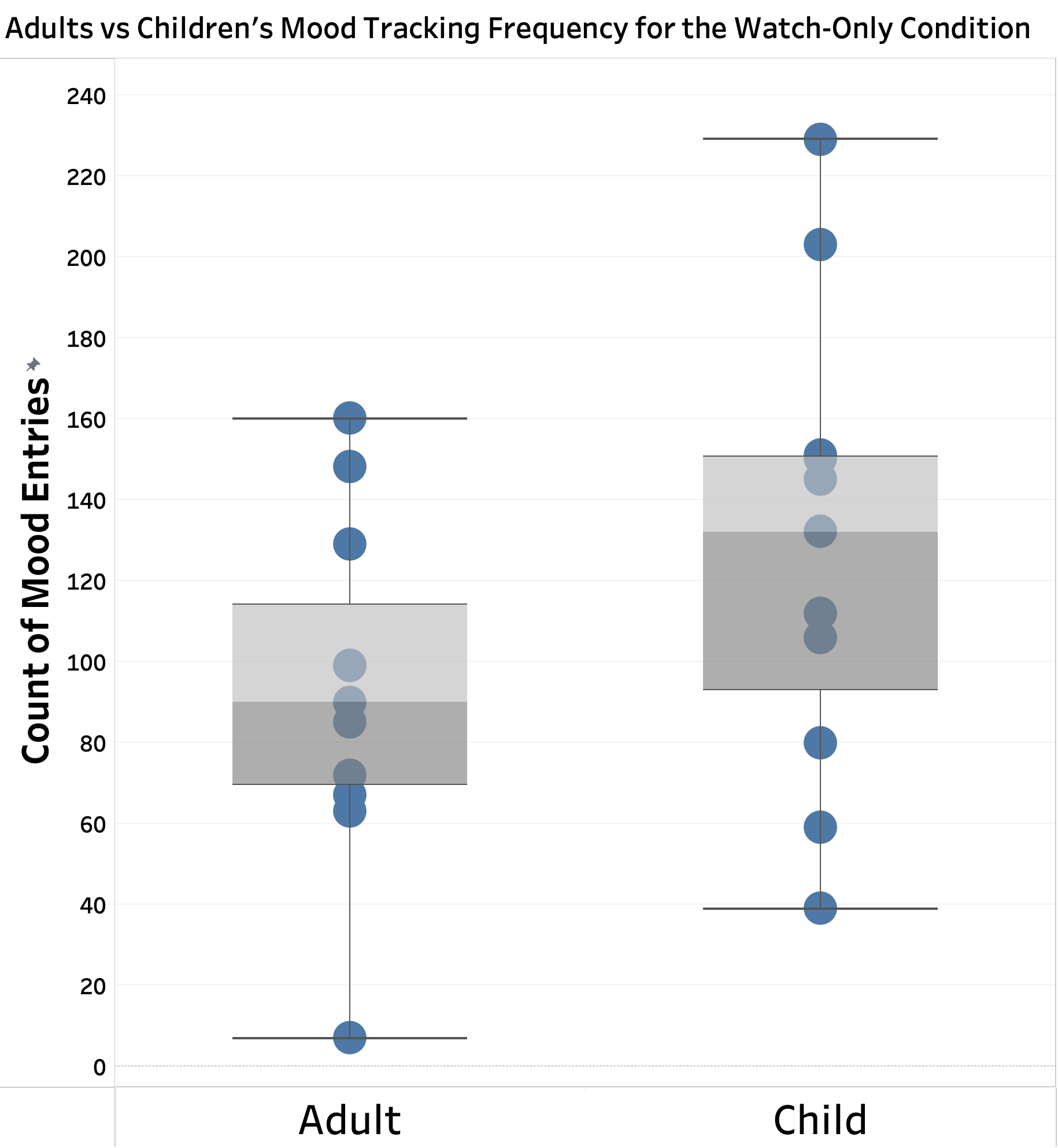} }}
      \hfil
      \subfloat[Goal completion reporting per adults and children.\label{FB_goal_age}]{{\includegraphics[width=0.45\textwidth]{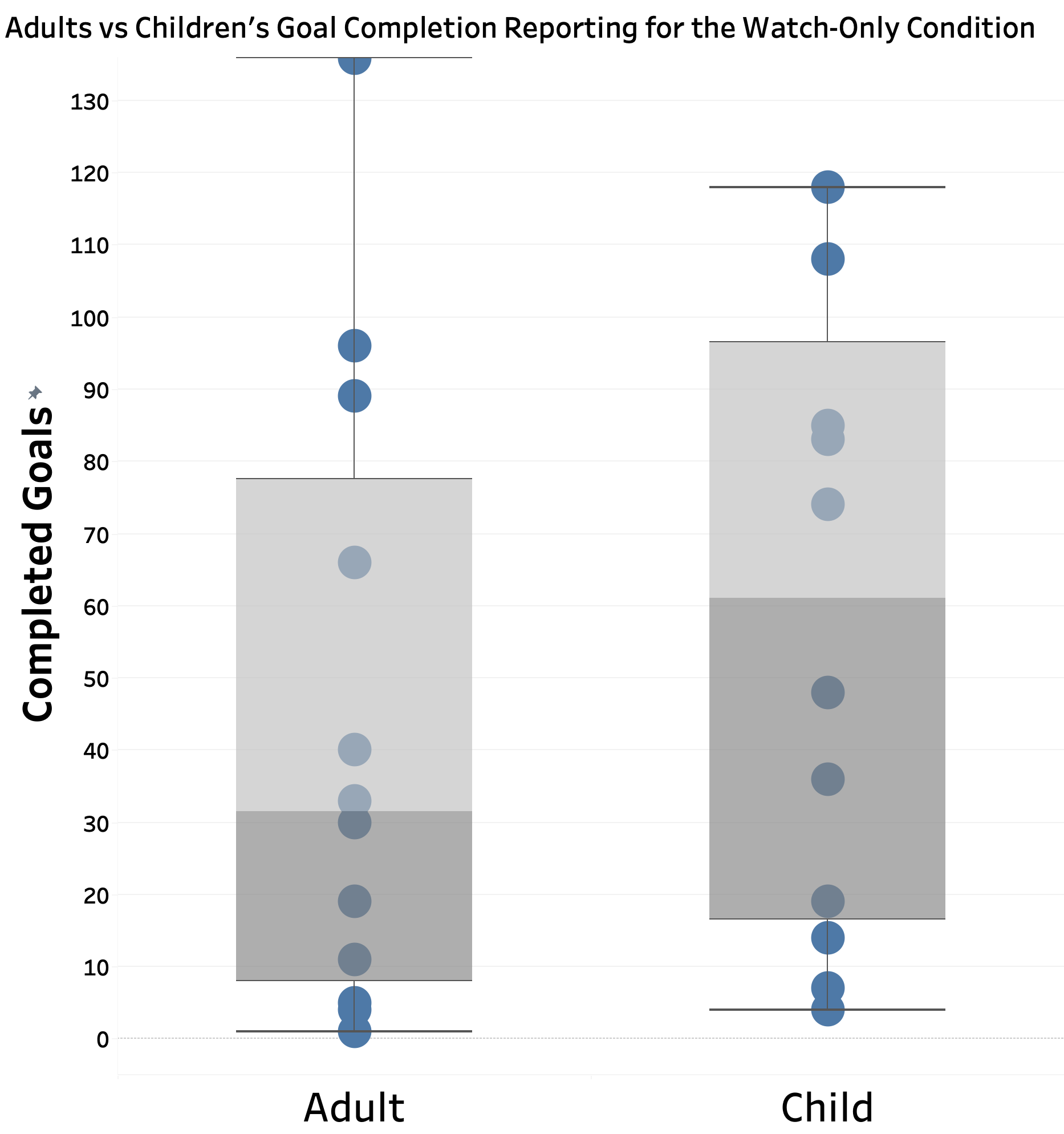} }}
      \caption{Children logged significantly more mood entries and goal completion than adults overall, with a larger gap during the smartwatch phase (mood, $\beta = 0.29$, $p < .001$) (goals, $\beta = 1.33$, OR $\approx 3.8$, $p < .001$).}
        \label{fig:FB_adults_vs_child}
    \Description{}
\end{figure}

Children consistently engaged more in self-tracking than adults \textchange{in all conditions}, with pronounced differences during the smartwatch-only condition. Children logged significantly more mood entries than adults during the smartwatch phase ($\beta = 0.29$, $p < .001$; Figure \ref{fig:FB_adults_vs_child}). Children were also significantly more likely than adults to mark goals as completed, nearly four times more during the smartwatch-only condition (OR $\approx 3.8$, $p < .001$), even after accounting for family and individual level clustering. These findings suggest some of the benefits from FamilyBloom on smartwatch were particularly meaningful for children and that these devices are a feasible platform for health behavior interventions for children as young as 6. In this section, we describe how the smartwatch-specific design appeared to provide benefits in availability for data input and reflection as well as challenges around distraction, competing with other watch apps, and maintenance burden.


\subsubsection{Benefits from Smartwatch for Personal and Family Data Availability}

\textbf{Glanceability and convenience promote engagement:} Participants reported that glanceability of mood tracking visuals stimulated moments of self-tracking, because \textit{``I look at the time, and then I see it. I remember''} (C03a). 
As P10b explained, \textit{``if it wasn't on my screen, I would probably not ever go in and do it.}'' P01a also compared with traditional non-glanceable apps: \textit{``...if it was buried in [an] app, I probably would do it half the time.''} F04 similarly explained the benefit of this persistent visualization on their wrist: \textit{``I like the flower and those [pointing at goals] because it just reminds me to do it, to check-in how I am feeling and see what time I put that feeling.''} (C04b) with the mother agreeing, \textit{``it helps you see how you are feeling''} (P04). Overall, opportunistic glances to the watchface led participants to often be reminded about mood tracking, reflect on their current state, and quickly interact with FamilyBloom.

For goals, persistent glanceability also worked as a subtle nudge for some participants to remember about incomplete goals under different daily contexts. For example, C04b reflected on the watch design versus home display as a reminder about his goals: \textit{``can I say something about the goals? They actually help me to do them [...] I prefer that on my watch because I can check them off whenever I go anywhere and I just look at my watch to see what I've done, what I need to do. And then I just don't need to come back home to see it.''} Likewise, P09b reported: \textit{``a lot of times I forget to check-in on the goals, but the watch actually reminded me like, `hey, you have three goals today, and here they are not checked'.''} P02a similarly said \textit{``I think if it wasn’t on the watch I wouldn’t think about it, so because it’s right there, I noticed it a lot more.''} Overall, participants reported that the persistent visualization of goal statuses prompted them to remember and reflect about their pending objectives.


\textbf{Opportunistic reflection about other family members:} Some participants found the watchface useful for staying connected with family members' moods and experiences throughout the day, especially when not together. For example, C12a said \textit{``I really like that I'm like at home, and my sister or my mom they leave to go somewhere. I could still see their moods because they registered everything.''} F06 appreciated this ubiquitous availability for family awareness and as a reminder to check-in on each other:

\begin{quote}
P06a: \textit{``I’m constantly moving, \textbf{so I liked being able to see everybody's moods on the watch}, because when I do get a moment to sit down I'm seeing how things are going throughout the day for them, and I don't get a chance to see the tablet [home display] often.''}

C06a: \textit{``I think the \textbf{watch reminded me more}, because it’s a quick look and like `oh my mom did this, I wonder why, I should ask my mom why'. It [watch] reminds me, `don’t forget to ask your parents questions', it kind of comes with it. If you use only the tablet you can forget, but if you use the watch, you can feel it on you. Takes one glance, and it shows they did this, [then] why? I should check in with them.''}
\end{quote}

During the home display-only as second condition (Group 2), some participants missed viewing family data on the watch, highlighting the potential for using the watch's pervasiveness and opportunistic visualization for family viewing. For example, C02a said \textit{``I wish the family stuff [widgets] was back on it [watchface]. So, I really like to know how they feel,''} demonstrating a strength of ubiquitous family data access.

Parents viewed watch-based sharing of goals as helpful for co-regulation. For example, P03a said \textit{``it was helpful for us to make sure that we're doing things within a reasonable timeframe. You don't want your kids doing homework with Kumon late at night.''} Some siblings also noticed each other’s goal accomplishment via the watch, such as C02b: \textit{``it’s too much to fit everyone’s goals [on the watchface] but the number is perfect, you can just see how many goals they’ve done. Because, I didn’t feel like [C02c] would do anything, he just plays all day, but he got all his goals done, that’s very cool to me.''} However, these positive experiences were not universal, as detailed next.


\subsubsection{Challenges with FamilyBloom on the Smartwatch}\label{sec:4.2.2}

Participants reported important barriers faced with using the wearable FamilyBloom in everyday life, including distraction, tensions between family and personal preferences, and inherent device limitations. These challenges experienced influenced how they used and viewed the system for personal and shared tracking of moods and goals on the watch.

\textbf{Contextual constraints and demands for focus can limit use:} Participants reported concerns in specific contexts where glanceability could be problematic, such as for ADHD children during intense self-regulation challenges. P06b, for example, explained that on some days \textit{``the watch was cutting in the middle of activities, there were times we had to take these away,''} though this was primarily due to \textit{``the other things [features] there [on the watch],''} leading [C06a]’s \textit{``use of the watch has lowered quite a bit.''} F06 later reported how routine went back to normal wear: \textit{``the distraction part is going a bit better. I think maybe just having the watch for longer.''} 

\begin{figure*}
      \centering
      \subfloat[P03b often changed his watchface towards personal domains instead of family data while away from home.\label{FB_p03_watchface}]{{\includegraphics[width=0.23\textwidth]{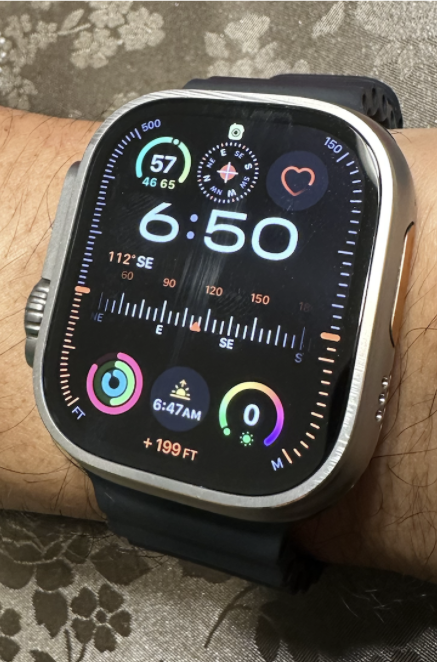} }}
      \hfil
      \subfloat[P08a wished to monitor different work-related time zones.\label{FB_tokyo}]{{\includegraphics[width=0.23\textwidth]{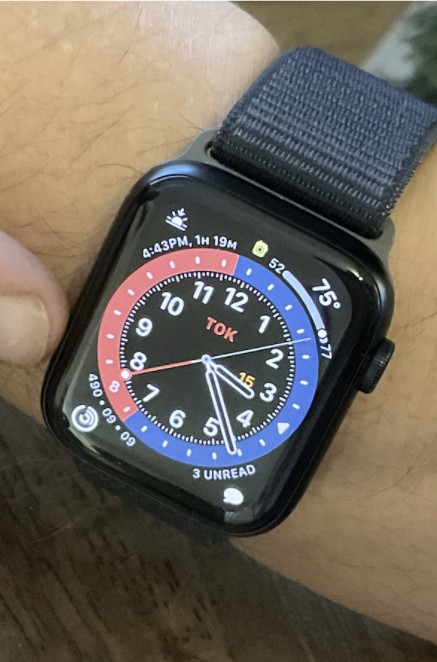} }}
      \hfil
      \subfloat[C06b, like P08b, sometimes switched the watchface for pictures of loved ones or pets.\label{FB_dogs}]{{\includegraphics[width=0.23\textwidth]{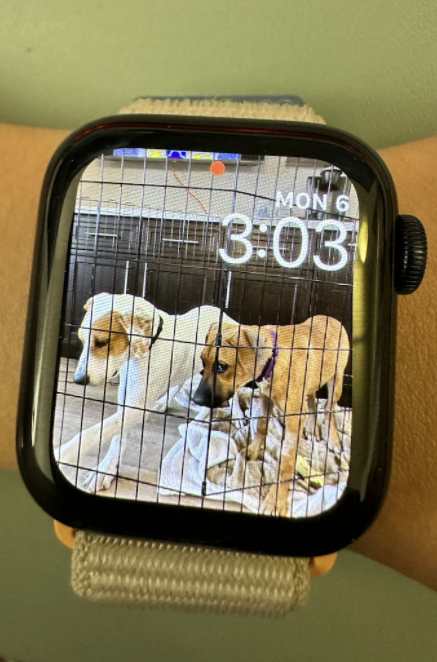} }}      
      \hfil
      \subfloat[C04b sometimes changed her watchface to ``something cute'' and diminish distraction tracking during class.\label{FB_c04_watchface}]{{\includegraphics[width=0.23\textwidth]{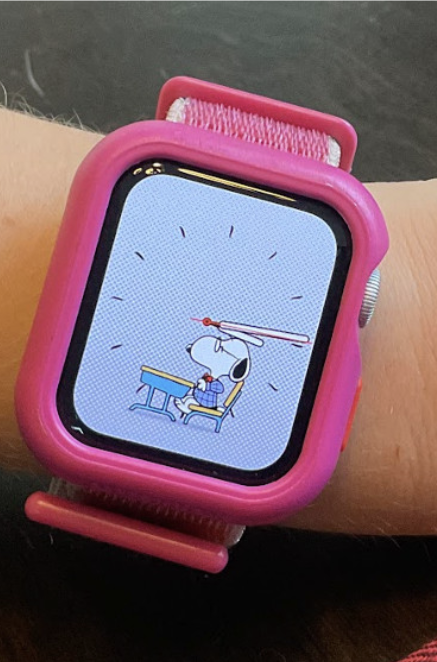} }}
      
      \caption{Some participants reported switching the watchface between family data views to display or track other personally relevant information (a-c) or to diminish distractions (d).}
        \label{fig:FB_watch_findings}
    \Description{Screenshots of smartwatch faces showing different data views for watchface, as described in the captions.}
\end{figure*}

Rather than removing the device entirely, some families developed strategies to manage potential distraction by changing the watchface during activities demanding focused attention, such as at school:
\begin{quote}
    C04b: \textit{``At school I get distracted a lot, so I change like this [changes watchface] so I don’t get distracted by having to do a [mood] check-in right then [...] I picked this one [Snoopy animated watchface], it does something [moving animation]''} (Figure \ref{FB_c04_watchface})
    
    P04: \textit{``But you didn’t get distracted by that [animation]?''}
    
    C04b: \textit{``It is less distracting. Because this one needs check-in, and if I am looking and need to check-in the teacher can see that and tell me to put the watch away in my backpack.''}
\end{quote}
This exchange illustrates how the very design features intended to prompt engagement could become counterproductive in moments requiring intense focus, especially for ADHD. However, concerns appeared to diminish over time (\textit{e.g.}, \textit{``At first [C10a] played with a lot of things in it, but after a while just used it for the quick [mood] check-in [...] it isn't distracting for me at work either.''} P10b) and glanceability was considered less disruptive than \textit{``dinging with notifications''} (P03a).

In contrast, stressful situations or the rush of activities led some participants to not pay attention to family data even when looking at the watchface. P06b noted \textit{``when I get really busy with work, I skip seeing their flowers, even when I look at my watch for time.''} P12 explained that \textit{``because this stressful situation [at work], I didn't even pay attention to their [children's] colors.''} C02b said \textit{``if I’m really into doing my homework or school work, I’m not distracted by it. Whenever I get the chance, I’ll look later.''} Similarly, P03a said \textit{``I am having to think about so many things throughout the day that I often don’t look at their flowers,''} and P09b reported \textit{``sometimes my job is back-to-back meetings. Then I have no time, I forget to go to the watch and look at their flowers.''} These examples indicate that cognitive load and competing attentional demands can limit even the quick glanceable interface.

Stressors and other demands on a person's cognition and attention can compete with family awareness in everyday life. The persistence of family data on the wrist does not guarantee attention to it when users are overwhelmed or focused on other priorities, suggesting limits to the ``always available'' advantage of smartwatches for family awareness.

\textbf{Competing for space on limited interfaces:} Some participants experienced tension between maintaining family-oriented widgets on their watchfaces and displaying personally relevant information. This tension was particularly relevant given that adults logged less often when away from home (mood: $\beta \approx 0.95$, $p < .001$; see Section 4.1.2), suggesting that out-of-home contexts created barriers to family-oriented tracking. This pattern reflected a broader shift in focus from collaborative family support to self-centered tasks and needs as participants moved between social contexts. Participants' practices indicate that such shifts typically occurred when they left the family environment for work or school, or even when working from home.

Several participants switched watchfaces in non-family contexts to highlight other self-tracking information, such as C05a focusing on \textit{``my [exercise] rings during the day;''} P08b \textchange{needing} \textit{``the GMT\footnote{Greenwich Mean Time, meaning he used to track different time zones.} a lot, because I work with [a company in] Japan,''} (Figure \ref{FB_tokyo}); and P06b \textchange{stating}, \textit{``the temperature, workout, and my schedule. Those are 3 things I check the most and I want it there [on the watch face] to not have to jump around other screens.''} P03b explained his watchface preference (Figure \ref{FB_p03_watchface}) when away from the family:
\begin{quote}
    \textit{``[I switch to] the watchface that had a lot more info. There's one [widget] that shows my heart rate. So, I think about my heart rate when I see it. Why was it so high earlier? Why was it low? So, there are other ones that I like to use. And I switched from that one to the family one.''}
\end{quote}

Some participants changed their watchface to center emotionally meaningful visualizations not related to self-tracking, such as C06b's  picture of the family dogs (Figure \ref{FB_dogs}) because \textit{``this one [watchface with dog pictures] reminds me of her whenever I look at it, I miss her.''} P08a used a picture of loved ones, \textit{``just a background picture and the time, simple''}, to limit distractions because focusing on data and others was \textit{``too much for me''} at the moment, with her husband adding \textit{``she has so many other things going on right now''.} P08a’s example showcases how the smartwatch’s glanceability might support nudges for personal wellbeing rather than data-centered reflection.

These findings demonstrate how people can adapt watchfaces towards context-specific needs, priorities, and attention. Participants switching between family-centered and other watchfaces throughout the day suggests the need for accommodating routine context shifts. This also points to the value of multi-device designs, where ambient displays maintain continuous family data available when watches are reconfigured for personal use.

\textbf{Inherent device limitations: wearing, charging, and display size:} Participants faced smartwatch maintenance challenges related to remembering to wear and charge their devices, similar to reports from prior work \cite{oygur2020raising,Oygur2021,Pizza2016}. P11 explained that \textit{``we tend to forget to put our watches on their chargers.''} Remembering to wear the watch was also problematic, as C01a said she \textit{``sometimes I forget''} with P01 adding that \textit{``she’s been wearing her watch like 70\% of the time, and [C02b], he’s more reliable, like 80, 90\%''.} Parents often took the burden of \textit{``reminding to charge and wear''} (P01b). As P12 explained, \textit{``if I'm stressed in the morning they're, I guess, more likely to forget their watches, too. I'd be driving and then `Oh no, you forgot to get the watch, I guess there will be no mood [tracking] today [laughs]'.''} Beyond maintenance, the smartwatch's limited screen size emerged as a barrier to family data review. C11a said that during the smartwatch-only phase \textit{``I never do the goals. They are just there, but they are small.''} Many participants \textit{``didn't notice''} (C01a) family members' goal achievement counts on the watchface and, because of the small size, did not \textit{``pay attention to the [goal] number, just when I went into the app. Maybe out of sight out of mind?''} (P04). P12 wished the family information \textit{``was a little bit bigger. Opening the app to see their flowers seems just more work for not as much more information.''} These reports highlight a tension between glanceability's continuous availability and the constrained space that limits depth of data to be communicated.

\subsection{Home Display: Situating Data Amid Regular Family Living}\label{sec:4.3}

Participants generally valued the home display for both its physical characteristics (\textit{e.g.}, \textit{``it's much bigger,''} C02a) and the shared data design. As C02b explained, \textit{``you can see many stuff [data] you missed, because you can see how you've improved, like emotionally throughout the week, or like gotten worse.''} Participants also perceived the home display as naturally situated within a family context, making it particularly useful for joint activities while remaining accessible to individuals as they moved through the home. Overall, as reported in Section \ref{sec:4.1}, the presence of the home display was associated with significantly higher engagement with mood and goal tracking and was generally perceived positively by families. In the following subsections, we detail participants' perspectives on the strengths and limitations of the home display, including differences they noted between the watch-only and home display designs.

\begin{figure*}[]
      \centering
      \subfloat[F03 positioned their display close to the home entrance and main corridor.\label{FB_F03_display}]{{\includegraphics[width=0.23\textwidth]{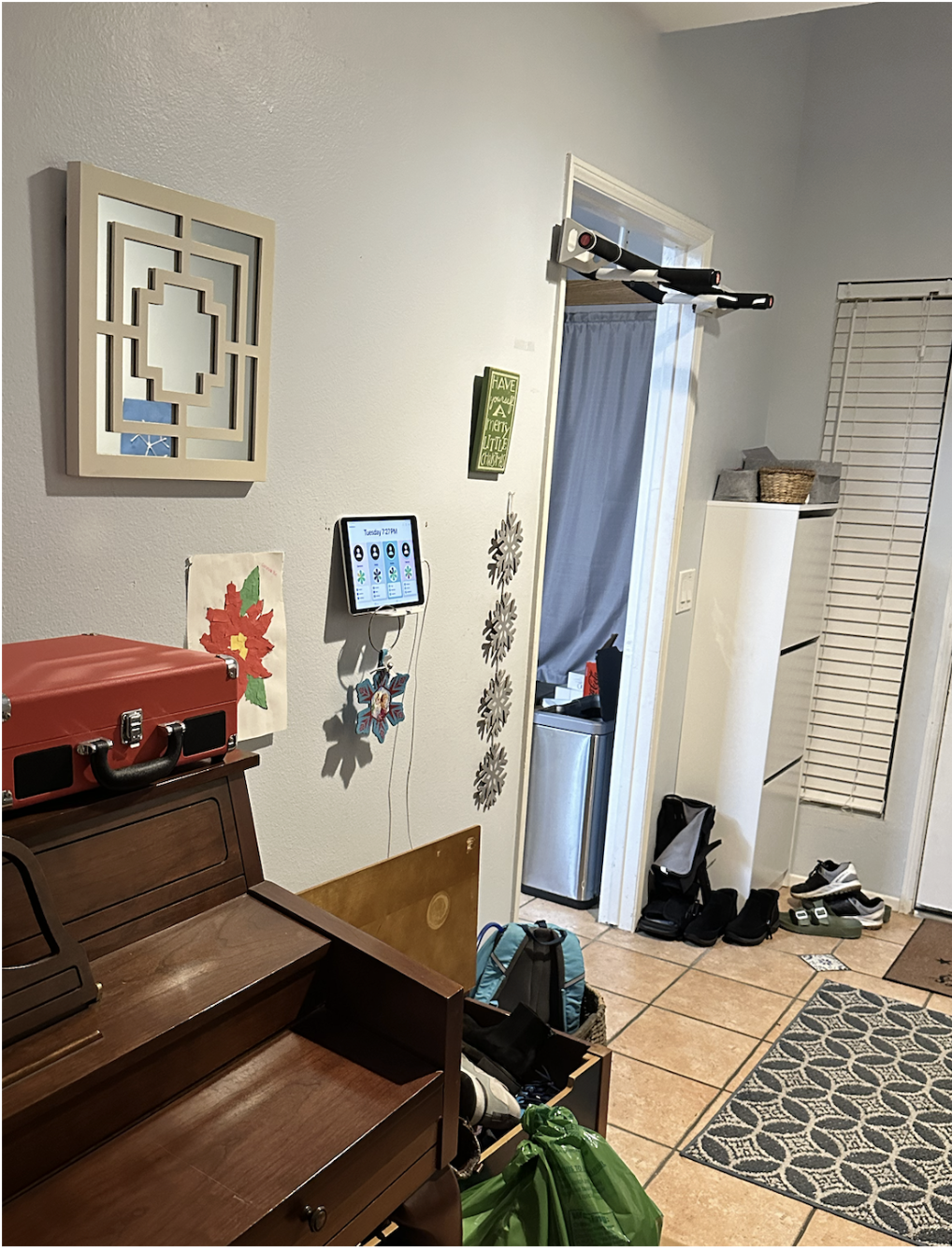} }}
      \hfil
      \subfloat[F01 had their home display next to the dinning area and used it for discussions during meals.\label{FB_dinner_display}]{{\includegraphics[width=0.46\textwidth]{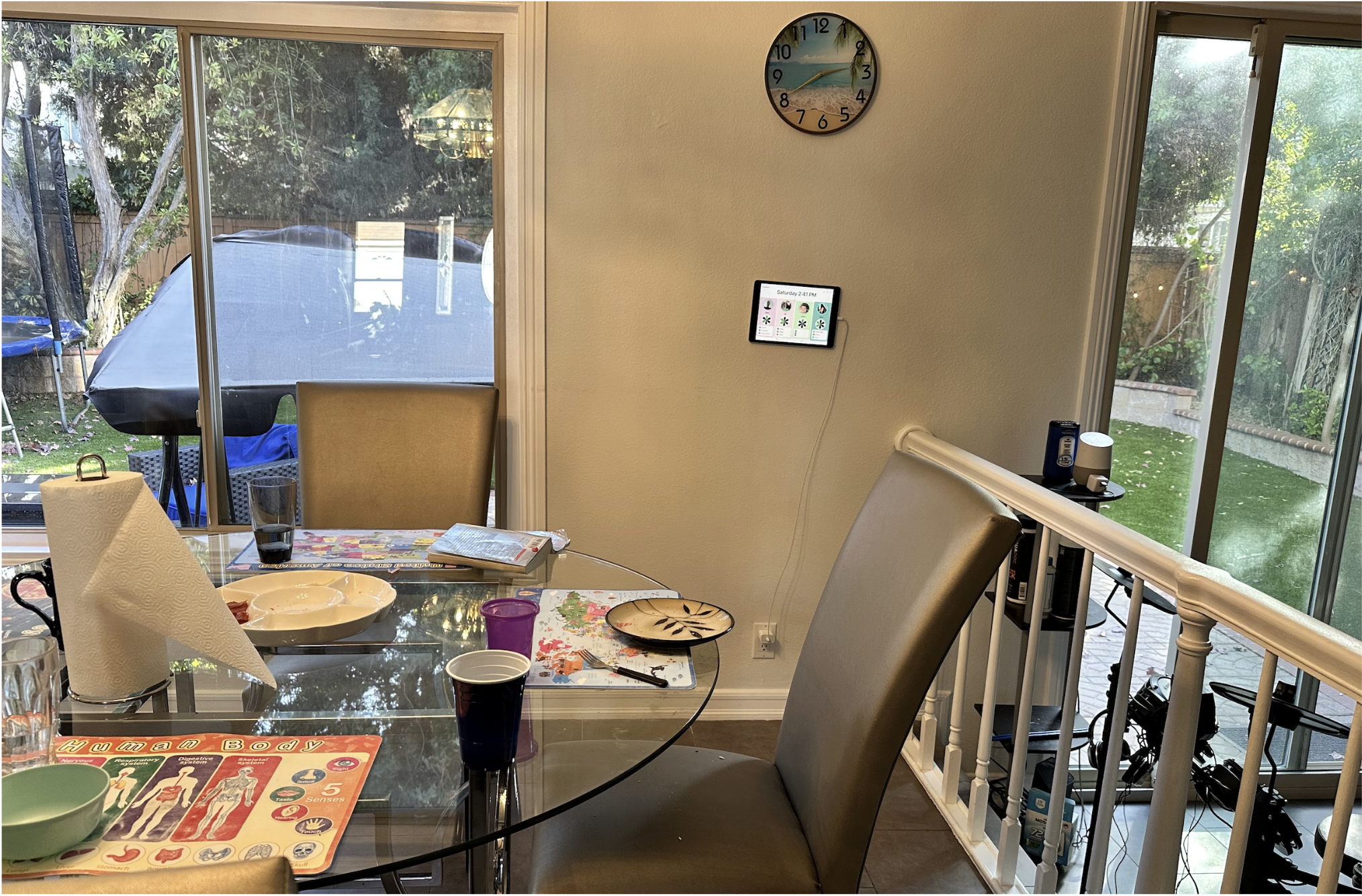} }}
      \hfil
      \subfloat[F09 moved their display to the living room to be closer to their main gatherings.\label{FB_living_room}]{{\includegraphics[width=0.23\textwidth]{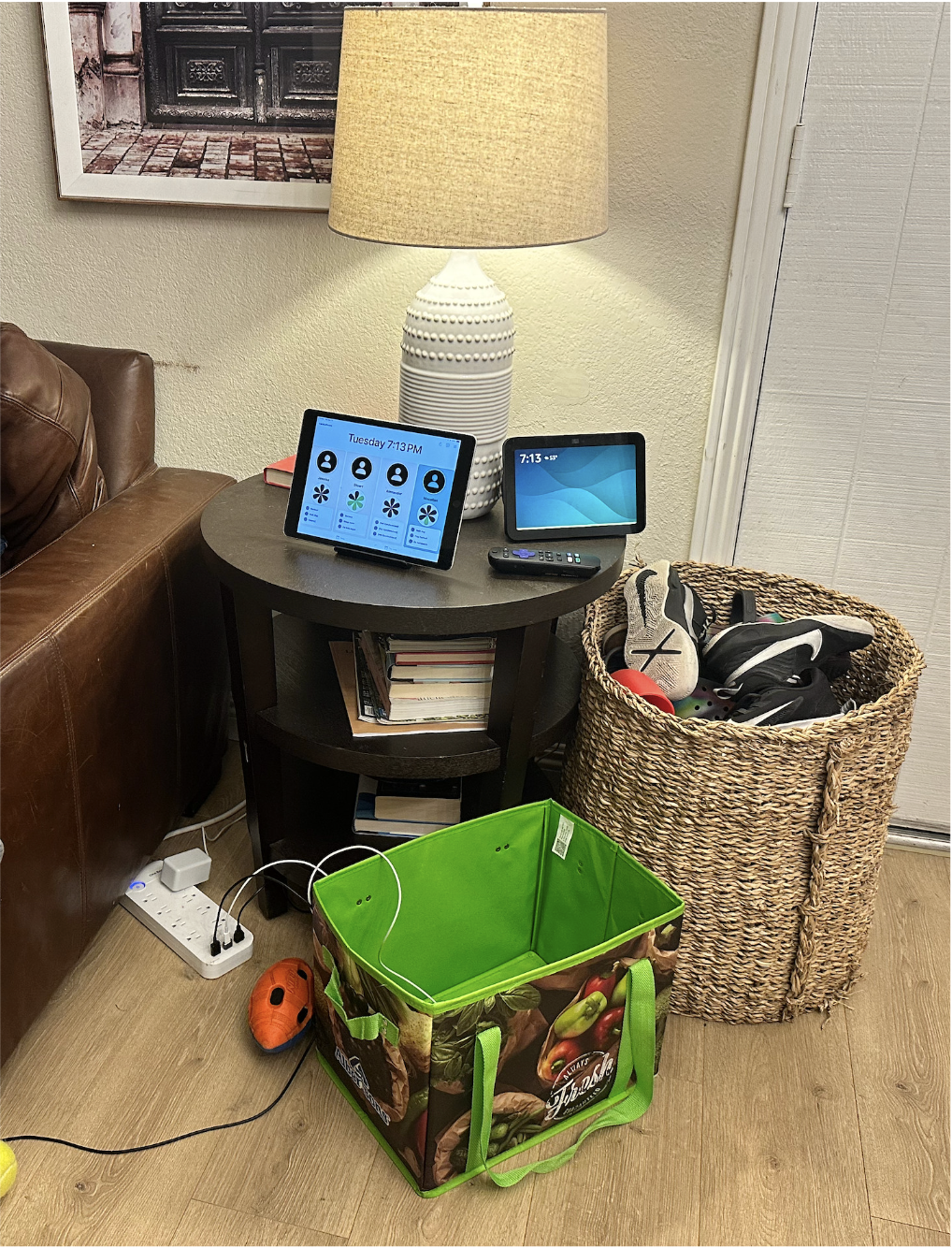} }}      
      \hfil
      \subfloat[F04 kept their home display in the kitchen, allowing the father (non-participant) to engage in family discussions.\label{FB_father_non_participant}]{{\includegraphics[width=0.39\textwidth]{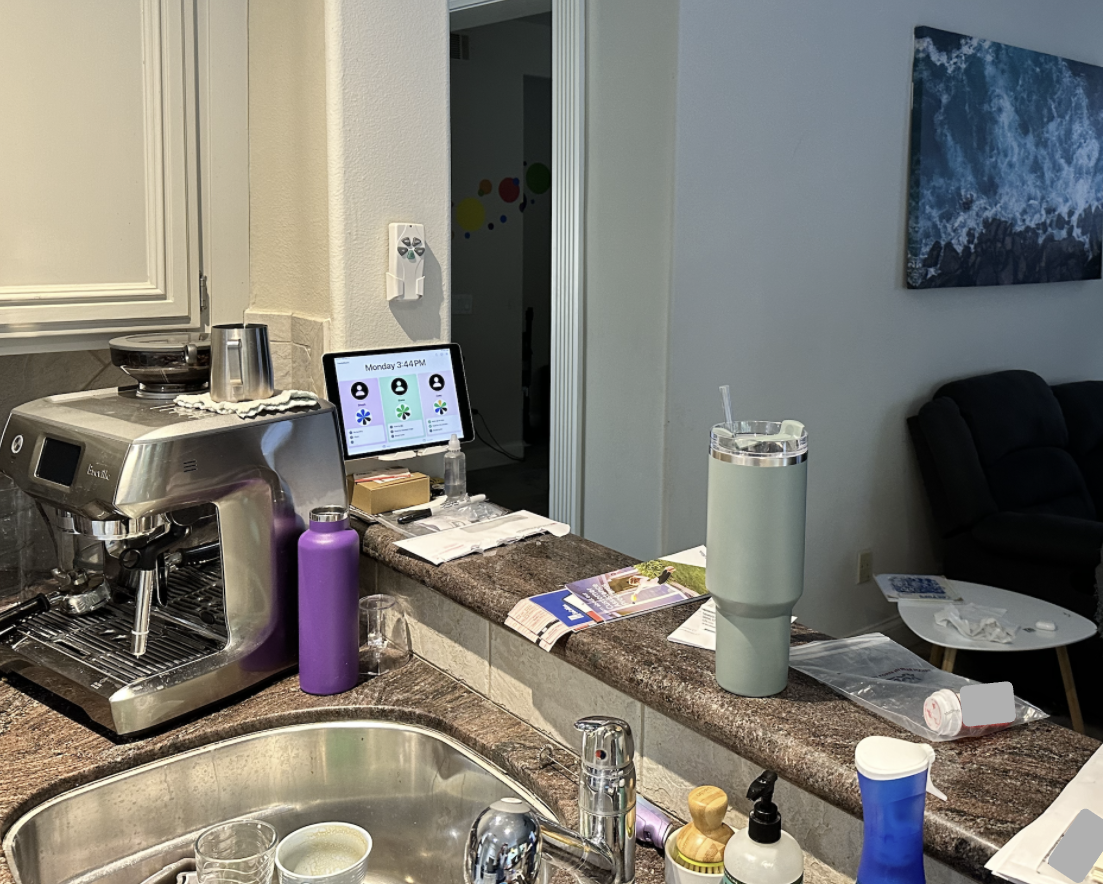} }} 
      \hfil
      \subfloat[P07b reported rarely seeing their home display due to its location at the back of the dining room.\label{FB_father_display_secluded}]{{\includegraphics[width=0.40\textwidth]{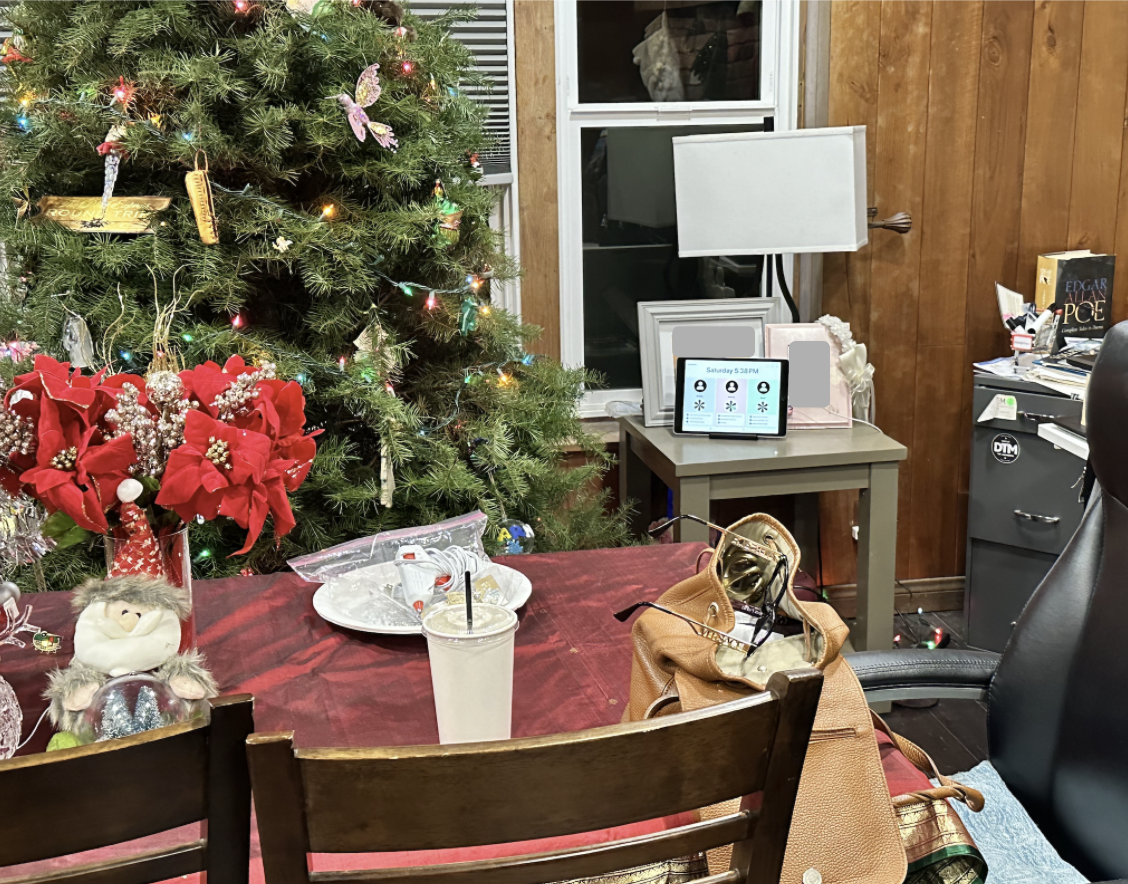} }}      
      \hfil
      
      \caption{Families typically positioned the situated display in their kitchen, living room, or in-between, such as a wall in a corridor or close to the front door. Families reported that opportunistically seeing the display in the home as they were going about the day reminded them to track and see how others were doing. During some joint moments, like dinner, it was also a useful reflection and discussion tool, inclusive even of members not enrolled in the study.}
        \label{fig:FB_home_display_findings}
\end{figure*}

\subsubsection{Benefits from Home Display Availability}

\textbf{Data review during opportunistic home navigation:} Participants positioned their FamilyBloom home displays in locations they frequently passed during everyday routines, such as the kitchen, living room, corridors, or near home entrance doors (Figure \ref{fig:FB_home_display_findings}, Table \ref{tab:participants}). These placements transformed data review into an opportunistic activity embedded within natural home navigation patterns. P03b explained: \textit{``I thought it was very impactful, like, I can just walk by and see it because we walk by iPad so often.''} P06b mentioned, \textit{``having the tablet there is like a very quick glance, I can check and see how they're at with their check-ins, it's a reminder. Like, I am concentrating on something and I come in to grab something from the kitchen, I'll see it on the tablet.''} C06b added, \textit{``It's like way bigger to see everything, and you can see goals better.''} The home display thus created multiple opportunities throughout the day for reflection on both personal and family data.

This occasional review of data supported some participants to go from family awareness to collaborative support. P10a described how the display's visibility prompted her to check in on her son: \textit{``I am home more and I see this [tablet]. So, I would notice something and go `hey bud, I notice you had this red and ask, what happened?'. Because it was at school and at the end of the day when we talk, he is typically generic about what happened, and this leads to something more specific.''} Similarly, P04 appreciated the ease of monitoring family goals for later providing nudges towards completing them: \textit{``it's easier to see if the kids have done their goals. It's nice to be able to walk past it at a glance. It's quick and easy and it's big and it's just there, versus like looking at it on the watch.''} Such opportunistic reflections served as important reminders to consider supportive actions towards others.

\textbf{Mediating shared use during family time for collaborative support:}
Beyond opportunistic individual glances, families consistently reported using the home display during shared moments in common spaces, particularly during meals or when gathering in the evening. The display's situatedness and constant visibility made it a convenient reference tool in family conversations about wellbeing.

Several families highlighted how the display facilitated discussion during meals, such as P11 emphasizing how this visibility generated engagement: \textit{``it just generates more engagement. Like when dinner is done and we are cleaning things up. If it had not been there, always on, it would never have been talked about. So, from zero to ten, we are at a five [score] right now, it would have been a zero [for family discussion] if it had not been there.''} The display's shared visibility created opportunities for more detailed conversations compared to viewing data on individual watchfaces. P06a explained:
\begin{quote}
    \textit{``Being able to see everybody on the watchface was almost like an immediate understanding of, okay, this is how things are. And then dinner times we're usually kind of in front of the tablet, and it's just a reminder for talking, like `oh wanna tell me about what that petal means?' And, we can look at the comments [mood notes] a little bit more about how she [C06a] was when she logged something.''}
\end{quote}
These examples demonstrate the home display's utility to mediate joint recollection and reflection. In addition, some considered that such shared use might have stimulated a higher motivation for self-tracking (Section \ref{sec:4.1}). For example, P03a said \textit{``I had the sense that she [C06a] would put the color there so that we would notice and talk about it at home.''} This kind of intentional communication was an unexpected and emergent mechanism for supporting family discussions and participation.


\subsubsection{Challenge: Individual Routines and Display Placement can Limit Usefulness}

While the home display's situated nature enabled opportunistic family data review, its fixed location created access limitations for family members whose daily routines kept them away from home or the display's location within the home. For example, extended absences from home naturally limited display access, with some members spending most of their time at school or work. P10b explained: \textit{``Yeah, sometimes I am away for work for like three days, or will arrive home really late, so seeing on the watch is better for me.''} C10a similarly noted: \textit{``For me, I get home late from [softball] practice and maybe go in the kitchen really quick, so I don't really see it.''} While many family members spend substantial time at home, others vary in their location routine and have less opportunity to casually view the~home~display.

Even when physically at home, some family members might not frequent the display's space. P07b noted: \textit{``I just don't walk over there all that often. Maybe if it was by the refrigerator, or something, where I might see it more often rather than having to go search for it''} (Figure \ref{FB_father_display_secluded}). C05b, who lived in a multi-story home, explained: \textit{``I am never down here [in the kitchen]. I stay in my room [upstairs].''} F09 had initially positioned their display in the kitchen, but later decided that moving it to the living room, close to the entrance door (Figure \ref{FB_living_room}), was more optimal for their regular engagement at home. Others similarly positioned their display in high-traffic corridors or entryways (\textit{e.g.}, Figure \ref{FB_F03_display}). These examples suggest that ``anchor points'' like main seating or high-traffic areas may be more effective, but optimal placement depends on individual and family routines.


\subsection{Multi-Device Synergies and Supporting Diverse Preferences Within Families}\label{sec:4.4}

Our analysis revealed substantial variety in personal preferences and individual differences alongside shared practices within family units. These differences highlight the potential for multi-device family informatics to accommodate diverse preferences for personal tracking, shared collaboration, and the shifting routines and contexts of families.

Our mixed-effect analysis revealed significant differences in self-tracking engagement depending on device and design availability but also that variance was explained differently for family-level dynamics. For moods, approximately 14.7\% of total variation was due to differences between family clusters (ICC $= 0.147$) while the remaining 85.3\% was due to within-family variation. This indicates that mood tracking frequency varied primarily due to individual differences rather than family dynamics. In contrast, goal tracking showed substantially more family-level clustering, with 39.2\% of total variance attributed to differences between families (ICC $= 0.392$). This suggests that family norms can explain a larger portion of variation in goal completion than for mood tracking.

Paired with our qualitative analysis, we find that individual and family-level variations were \textit{in part} because of overlap and tensions between family and individual preferences and contexts for mood tracking, and family differences for accountability and support towards personal goals. In the following subsections, we detail how the multi-device design accommodated some differences in preferences and practices within families.

\subsubsection{Synergistic Benefit of Smartwatch and Home Display Devices}

While maintaining both devices added some complexity to family routines (Section \ref{sec:4.2.2}), participants reported that the combination created complementary supports for more consistent engagement than either device alone. For example, several families described how the home display reminded them to wear their watches. P03b explained: \textit{``First thing in the morning we see the tablet there. Then the gray flowers means we forgot to put the watches on and we need to get them before going out the door.''} P01b elaborated on a similar new routine: \textit{``With the tablet we are getting breakfast and getting the backpacks, then realizing 'ok, we also need to go get the watches'. So, they also took the watches more,''} with P01a attributing this to increased engagement: \textit{``I guess this was more engaging than the previous period [watch only].''} During the watch-only condition, P04 reflected on missing this reminder function: \textit{``I've had to remind them to put them [watches] on more in the morning. Reminding them to actually get out the door with their watches on.''} The home display seemed to act as a reminder within families’ morning routines, mitigating the persistent challenge of remembering their wearable.

The display also reminded participants to self-track. P11 explained, \textit{``I am sure we wouldn't even have remembered to track if not the tablet there, even with the watch. It's super noticeable seeing the gray flowers there.''} P01a similarly said \textit{``having the tablet there also makes sure we actually fill it.''} For some, the display provided feedback for their self-tracking, such as C06a who said she \textit{``sometimes will check the tablet to see it upload''} during dinner.

In some families, the home display also helped coordinate smartwatch charging, another maintenance challenge. P10 had set up the charging station next to the home display and described how it prompted collective charging routines: \textit{``It reminds me to go 'hey you guys, we are starting up again, charge your watches!', at least the 3 of us got in the habit of charging our watches down here [kitchen]. It also helps me too in the morning to give them their watches because otherwise there will be a lot of missing colors [during the day].''} By prompting collective responsibility of device charging, this maintenance task became part of the routine, albeit often still needing parental coordination.

\subsubsection{Involving Non Self-Tracking Members in Family Collaboration}

The home display enabled participation in family tracking by members who did not wear smartwatches, extending collaborative benefits beyond active self-trackers. This inclusion proved value for non-participant family members. In F04, the father could not participate in self-tracking because \textit{``he can't wear a watch because of his [non-disclosed] job''} (P04), but the home display allowed him to remain engaged with his family's wellbeing:
\begin{quote}
C04b: \textit{``I like the tablet because dad can talk to us about it.''}

P04: \textit{``Yeah, he didn't say anything before [smartwatch-condition], just looked at their watch just to see what it was. What I like about the iPad [now] is that others who aren't doing, having a watch, can still be involved and participate and be more connected with the family. Like, he would see [C04b]'s moods or mine and be like `Oh, are you not feeling well today?' Like, he would see that she was mad at school halfway through the day, or whatever, and they would speak about it and find out what happened.''}
\end{quote}
F02 similarly appreciated that the home display enabled her husband's participation despite not being enrolled in the study: \textit{``it's different [from smartwatch-only condition], I mean it's the same information, but it's bigger and [husband’s name] wasn't part of the study, but he could see that. So, you know, he could participate even though he wasn't wearing the watch.''} C02s confirmed that their father started discussing their moods and goals.

Like these parents who did  not participate in our study, others might prefer to avoid wearables but still wish to be involved in family collaboration. P07b self-tracked with the watch but explained that \textit{``I typically don't wear watches, you know. There is a lot of people that don't want to,''} with P07a adding, \textit{``he doesn't like the actual feeling of things touching his body''} to which P07b responded, \textit{``It throws me out of balance, you know. Even the wedding ring was tough for me, I never wore a ring in my life. So, I make an effort for everyday things, like keys, my phone. I was using the watch, but I was making the effort to set my goals and my moods.''} Tactile sensitivity and other aversions to wearables might be a barrier to self-tracking, influencing family collaboration.

These accounts highlight how multi-device ecosystems can be important for all family members to participate in family health collaboration through awareness and support. The home display provided a shared access point that did not require personal device ownership, bodily wear, or engagement with self-tracking.

\subsubsection{Device and Design Redundancy Accommodate Diverse Routines and Preferences}

Even within families who shared tracking practices, individual members often expressed different preferences for which device and design they found most useful. These preferences typically aligned with their daily routines, such as how much time they spent at home and personal comfort with different interfaces. The availability of both devices allowed families to accommodate these differences, with each member leveraging the interface that best fit their circumstances.

When discussing the different designs, participants concluded that \textit{``having both''} (C02b) was ideal because it accommodated diverging views within their families. F12 exemplified this divergence:
\begin{quote}
C12b: \textit{``I like the tablet. When you are home and you look at the tablet, it is more convenient.''}

C12a: \textit{``What? But you just have to look at the watch. I like the watch, or having both [shows accessing family details in the watch app].''}

P12: \textit{``I still prefer the tablet. It’s my eyesight, you know. If there was a way to make them bigger [watch’s widgets], it would improve.''}

C12a: \textit{``But like, you can still see the colors. You can’t tell what it says, but you if you click on it, then it says what the things are [family’s goals and moods]. It’s just 2 clicks and that’s it, not that hard.''}
\end{quote}
This dialogue illustrates how multiple factors can influence personal choices within the family, such as location, eyesight, and interaction preferences.

Time at home emerged as a particularly important factor. F10 discussed how their family's preferences divided along differences in routine, with P10a concluding: \textit{``For me and [C10a], we are home often so we see the iPad. So, we prefer the iPad. [P10b] and [C10b] aren't here as much, so they use the watch to see the flowers. So, it's divided. The option to have it both places, that would be good, so whoever it suits best gets it the way they wanted, you know.''} Similarly, C01b explained: \textit{``I like looking at my watch instead of going downstairs to look at the tablet. So, I don't have to go downstairs, look at the tablet, and then go all the way up,''} to which C01a replied \textit{``So, I think having both [display and watch], especially for family because it is a bit crammed on my watch.''} 

Some participants speculated about integration of other devices based on their current technological ecosystem. P05a wished for smartring integration because \textit{``I don't really like the watch, I prefer this [pointing to Oura ring, a smartring], its more comfortable and it does a pretty good job at telling me about my stress.''} C10b wanted phone integration, saying \textit{``I can't always wear a watch so it would be nice to also have it on my phone. It could be one of those widgets on the phone.''} Others suggested voice assistants could support tracking during busy, multitasking moments. P08a explained: \textit{``you don't have time to sit down, you never stop moving, you know. I can guarantee a lot of the moms probably feel the same. If you could instead talk to Siri [voice assistant] about your moods while doing activities, it would be better.''} P11 similarly noted: \textit{``like, in the morning, when we are talking with Alexa, if it was [integrated] with Alexa, it would be better, because it has my daily stuff like tasks, calendar.''} These suggestions reveal how participants imagined FamilyBloom fitting into their broader personal device ecosystems.

While our study examined smartwatch and home display combinations, participants' reflections suggest that family informatics may benefit from even greater device diversity and interaction modalities to accommodate the heterogeneous preferences, routines, physical abilities, and existing technology practices found within families.

\subsubsection{\textchange{Challenge: Privacy Tensions with Pervasive Glanceable Visibility of Personal Data}}

\textchange{Participants generally perceived personal and family data on the watch as an inconspicuous visualization because \textit{``it looks just like a regular [watchface activity] ring''} (P07a) and because \textit{``anybody close enough to see the watch is somebody you trust''} (C02b). However, some participants had diverging perspectives about the home display's larger visibility of personal information. For some, it opened inquiry from visitors that could be unwanted. For example, P01a explained that if having visitors, he \textit{``might take it off the wall or close it... maybe it compromises our kids' privacy.''} F04 had a similar discussion:}
\begin{quote}
\textchange{(C04b): \textit{``My friends came [over] and they saw that [tablet] and asked me what it was. I said it was none of their business, because I really want to keep it private and away from them, because I don’t want them to spread it to anybody else.''}}

\textchange{\textit{(P04): ``If we hosted more [people], I would put it away. I don't want people to see that I got anxious.''}}
\end{quote}
\textchange{Yet, for others, the display was as conspicuous as the watch: \textit{``it's very similar to our photo frame out there [in living room] and no one comments on it''} (P10b); \textit{``We had people over for the [birthday] party. My friends asked about it, no big deal''} (C12a), \textit{``I had zero comments about what the app and the tablet was. No one really asked about it. Also, if they look closely I think people just assume it is a task manager''} (P12). These examples demonstrate a range of perspectives about the home display's potential to disclose information as well as hints at possible privacy preserving design choices depending on family's comfort with data disclosure to non-members.}

\textchange{Beyond concerns about non-family members, concerns about visibility and privacy within family sharing were generally tempered by the individual's control over whether to track their mood states and goal status. Because self-tracking was optional and manual, participants pragmatically and implicitly negotiated between resisting surveillance \cite{everydayResistance, minorResistance2025} and sharing as mediating care. Resistance strategies were intended to avoid discussions and typically took the form of either leaving data blank or inputting a default \textit{``green''} mood or completed goal. Participants also speculated that if the data were to be automatically input (\textit{e.g.}, \textit{``based on your heart beats''} P01a), then shared tracking could become \textit{``weird...creepy''} (C01a). Overall, these findings highlight how shared tracking systems leveraging glanceable interfaces across devices need to consider attitudes on personal data collection methods, visibility on and of devices, and intended versus unexpected viewers.}
\section{DISCUSSION}
Our results revealed several patterns of engagement with family health behavior tracking and preferences around multi-device family data reviews. Across design and device conditions, adding a shared home display to smartwatch-based tracking increased overall engagement, while also revealing consistent temporal and contextual patterns in when and where families tracked. Our findings highlight how multi-device configurations supported different forms of family awareness, coordination, and participation. They leveraged the smartwatch and home display in complementary and synergistic ways. At the same time, sustaining these device ecologies introduced maintenance work and raised questions about utility and attention. In the following sections, we reflect on opportunities for family \textchange{mutual caregiving} in light of these findings as well as prior work on multi-device and multimodal information systems\cite{OLeary2017,Jokela2015,DiGeronimo2016}. Our findings extend such work by showing that, in family tracking, devices also take on social and motivational roles supporting continuity of interaction, continuity of care, awareness, and participation across family members. 

\subsection{Supporting Mutuality in Health Collaboration through Ubiquitous Multi-Device Family Sharing}

Prior work has highlighted the benefits of shared family data for increasing awareness and accountability \cite{parkCollabTrackingReview,pina2017personal,saksono2015spaceship,schaefbauer2015snack}. However, these benefits are often concentrated around automated data collection and in structured, co-present review moments \cite{Sonne2016,pina2017personal}. While beneficial, such moments require temporal coordination and may miss accommodating diverse routines and attentional rhythms of multiple family members. Our findings extend this work by showing how continuous, mutual data sharing across devices can support collaboration beyond scheduled or co-present interactions. By distributing family data across both personal and shared devices, multi-device systems can surface moments of need, reassurance, or opportunity for support throughout the day. This expands family collaboration from expected shared interaction windows \cite{parkCollabTrackingReview,saksono2020storywell, silva2024Codesign} toward more flexible forms of coordination that align with everyday mobility and fragmented attention. Even during busy periods or traveling apart, participants described how pervasive family data access helped maintain awareness and connection.

Ubiquitous and integrated access to family data enabled continuous awareness and coordination that led to collaborative actions like behavior co-regulation. Brief encounters with the home display or glances at the smartwatch could surface cues about another family member’s mood or goal needs, prompting check-ins, co-regulation, or expressions of support when families reunited. In this way, multi-device family tracking supported mutual understanding and collective regulation that extended beyond expected co-located shared reviews, similar to prior work in situated displays \cite{Hardey2022, brown2007whereaboutsclock} and calendars \cite{Carman2006FamilyCalendar}. \textchange{Importantly, we observed mutual caregiving practices among multiple members of different roles, not only in parenting but also between spouses, siblings, and children-to-parents. Such mutual caregiving indicates potential for approaches like FamilyBloom to extend to other caregiving settings where bidirectional awareness and support are valued, such as older adult couples managing chronic conditions together, or adult children and aging parents seeking ways to stay mutually connected across distance.} Still, mutual caregiving system designs need to account for the uneven and often hidden collaboration work that families negotiate among members \cite{liangChi2026CollabCareWork}. Looking forward, multi-device family informatics can support pervasive and ongoing family awareness under different contexts, where data remains present enough to be noticed, but lightweight enough to fit into everyday attention. 

In addition to supporting collaboration, designing for mutual family data access \textchange{can influence individuals to more continuously self-track over time. In our findings, although personal} \textit{input} interfaces remained unchanged across conditions, participants logged more frequently when each other's data were made pervasively available across devices. \textchange{Our findings showed that despite overall mood tracking fatigue, participants continued tracking throughout the nine weeks, and the presence of family data across devices appeared to partially mitigate disengagement.} Participants described being reminded to track when noticing other people's logged entries and feeling motivated to contribute to family awareness and support. Shared family tracking can act as a social catalyst for personal engagement, where motivation to persist in self-tracking is socially influenced, resonating with prior work surfacing social dynamics of personal informatics \cite{saksono2024SocioCog, pina2017personal} and selective data sharing \cite{snapi,Dennis2024}. With equal sharing among family members \cite{parkCollabTrackingReview}, family-oriented designs \textchange{may be particularly valuable for sustaining engagement over time }in personal informatics practices, with pervasive sharing positioning self-tracking as a contribution to collective understanding alongside personal health objectives. \textchange{Future designs could intentionally leverage such social accountability mechanisms to support long-term engagement. At the same time, it is possible that shared visibility influences not only sustained engagement but also what individuals express (\textit{e.g.}, self-reported moods converging or shifting in response to repeated exposure to others' data), a dynamic worth investigating in future work.}


\subsection{Supporting Inclusive and Preferred Participation}


Our findings highlight how multi-device family informatics can support more inclusive participation by accommodating diverse preferences, constraints, and comfort levels around tracking. Added to differences in daily routines, individual aversions and device-specific limitations shaped how family members engaged with tracking and family data. While children, including those with ADHD, particularly appropriated smartwatch uses, some participants faced challenges with the device's form factor and others did not enroll in the study due to tactile sensitivity, work-related restrictions, or discomfort with wearables, excluding them from direct data production.

Redundancy of data across context and location via multiple devices helped increase data access to family members. Several members valued such family data access independently from their own self-tracking \textchange{engagement preferences as it enabled family participation}. Still, several participants expressed interest in integrating additional devices and modalities they were familiar with and enjoyed (\textit{e.g.}, smartphones, smart rings, and smart speakers). These preferences were grounded in familiar considerations such as multitasking, form factor, comfort, and context of use, aligning with prior work on device preference and appropriation \cite{Jokela2015,Kawsar2013,ModEatMobileHCI}. At the same time, these requests reflect a broader expectation that family-centered health tracking should integrate seamlessly into existing personal technology ecologies rather than exist as an isolated system.

Multi-device ecosystems for family tracking could expand to other devices and modalities to accommodate family members' \textchange{characteristics (\textit{e.g.}, age, literacy, role), and} affinities and interaction preferences in their current device ecosystems. \textchange{Our study examined designs for two devices at contrasting points in the spectrum of personal and shared device options. However, a broader range of form factors and interaction modalities exists beyond smartwatches and home displays. Some of our participants explained their preferences for phones and widgets on their home screens, others expressed a desire to interact with voice agents about their personal and family data, and others disclosed preferences for different tracker formats (\textit{e.g.}, smart rings). Such preferences reflect a broader need for} flexibility in defining data access and interactions for tracking, which can improve agency and tailoring for health management \cite{Kim2017,Ayobi2018}. \textchange{Multi-device supports for such flexibility} already exist for personal exercise \cite{Luo2020} and diet \cite{ModEatMobileHCI}. We see that family \textchange{caregiving could similarly leverage different} devices to be inclusive of mutual shared tracking, enabling non-trackers \textchange{or those with particular device aversions to benefit from and contribute to family collaboration through alternative devices in the ecosystem.} A critical challenge is the technical and social burden increased device ecosystems might bring to families, as we discuss next.


\subsection{Costs of Multi-Device Ecosystems: Maintenance Burden and Attention}

While participants valued the flexibility afforded by using multiple devices for family mood and goal tracking, our findings also surface the costs of multi-device systems. Expanding family informatics across multiple devices comes with increased technical and social complexity. Beyond challenges of data integration and use, families must contend with ongoing maintenance work, including charging, wearing, updating, and troubleshooting devices (\textit{e.g.}, maintaining internet connectivity). Consistent with prior studies of family and multi-device systems \cite{pina2020dreamcatcher,Silva2022,oygur2020raising, kaiwenSmartHome2026}, this maintenance work was often unevenly distributed, frequently burdening one parent (\textit{e.g.}, typically a caregiver with more intense charge of family wellness or more technically inclined). Wearable-specific constraints, such as limited battery life, further amplified these challenges \cite{silva2023unpacking, oygur2020raising}.

Our findings also reveal a tension around attention and device boundaries, particularly for smartwatches with both personal and shared purposes. The home display was widely perceived as a single, situated purpose, but less useful for people that did not frequent the surrounding space often or at all. Smartwatches had different attention challenges as a communal and personal data device. Some participants described wanting to switch between self-focused and family-focused watch modes depending on context, highlighting how personal devices become sites of negotiation when becoming a design space for shared family awareness. This echoes broader challenges in family informatics where personal data practices intersect, overlap, or conflict with collective goals \cite{parkCollabTrackingReview,Kaziunas2017,Toscos2012}, requiring ongoing negotiation of visibility, focus, and control.

Despite these challenges, families developed practical strategies to sustain their multi-device setups, such as charging stations and embedding reminders in existing routines. In these cases, the home display itself became a coordinating artifact, illustrating how one device in an ecosystem can scaffold the maintenance of others. These practices suggest opportunities for future designs to not only focus on classic personal/family informatics stages \cite{Epstein2015a, pina2017personal, Li2010} but also explicitly support the coordination of device upkeep. For example, systems could visualize device battery levels, connectivity, and wearing status at the family level or provide reminders, such as in glanceable widgets in the home display, for shared responsibility for maintenance alongside use. Overall, while multi-device designs for family data is the focal point towards collaborative health behaviors, practicality of sustaining the distributed family device ecosystem is foundational for family-centered caregiving systems.

\subsection{Smartwatch Glanceability as a Social Design Opportunity and Tension}
Our findings indicate that glanceable family tracking functioned as a pervasive social signal, but making family data persistently visible on a personal device introduced new tensions. The smartwatch is used differently in personal and family roles and across shifting contexts, in part represented by changing watchfaces or widgets as well as differing levels of use in these contexts. Prior research has shown that smartwatch users frequently customize and change watchfaces to reflect shifting goals, aesthetics, and social contexts throughout the day, using customization as a way to manage competing informational and identity needs \cite{charitos2025watch, Gouveia2023}. Broader personal informatics work advocates for flexible, self-defined tools and interfaces to better support self-tracking goals \cite{Ayobi2020, Kim2017}. Our findings extend this work by showing that, in family informatics, customization also becomes a mechanism for managing social role and attention transitions, from and between family-centered participation and self-centered health practices. Some ADHD children especially invested effort in watchface changes for attention management. Unlike personal tracking alone, family-oriented glanceable systems must therefore balance personal relevance with shared participation, introducing new design tensions around how and when family data should remain visible.

These tensions point to opportunities for \textbf{context-adaptive family tracking interfaces}. Beyond manually tailoring data sources on watchfaces, which can be burdensome, future systems could support visual adaptations based on inferred context. People might be generally willing to customize their self-tracking tools \cite{Kim2017} and watchfaces \cite{charitos2025watch,Gouveia2023}, but our findings indicate the opportunity for increased scaffolding when such changes are repetitive and expected over established routines. Instead of on/off family data views, systems could offer a range of foregrounding, diminishing, or removing family data. For example, systems could display each family member's data per individual widgets during regular moments, but aggregate or selectively display family data to smaller widgets when other information is more relevant, or completely receding family data into the background when personal focus is prioritized. Such designs could preserve the social benefits of glanceable family awareness while respecting users’ need to manage attention, priorities, and role boundaries. More broadly, our findings suggest that glanceability in family informatics should be understood not only as a design strategy for reducing interaction efforts, but as a socially situated mechanism for mediating care, accountability, and collaboration across everyday contexts.


\subsection{\textchange{Rhythms of Mobility, Temporality, and Fatigue}}


\textchange{Our quantitative findings revealed consistent temporal and spatial patterns in family self-tracking that carry implications for designing sustainable ubiquitous health systems. Mood tracking clustered in afternoons, dropped sharply on weekends, and occurred far more frequently at home than away. We see these patterns as signals of naturally occurring rhythms in family life that future systems could be designed around, but cultural and family norms outside the groups we studied should be considered. Systems could surface family data more prominently during high-engagement periods, which could be measured and assessed as in our study or understood as a personalized, societal, or cultural practice. For families like our participants, weekends could shift from active tracking toward collaborative reflection and goal-setting modes \cite{silva2024Codesign}, maintaining engagement while respecting the different social meanings families attach to weekend time. In other spaces, a different pattern might emerge and need to be supported.}

\textchange{Beyond weekly rhythms, we observed a gradual decline in mood tracking frequency over the nine-week deployment, while no significant decline was observed for goal tracking. The relative stability of goal tracking may reflect its more episodic and outcome-oriented nature, compared to mood tracking which requires repeated, open-ended self-assessment throughout the day. Prior work in self-tracking across other domains have highlighted the increased likelihood of health tracking abandonment, often before people reap benefits \cite{Lazar2015, Epstein2016b, Cordeiro2015a}. Still, our findings show that participants consistently tracked moods and goals by the end of the study, albeit less intensively for moods. Our findings also suggested that the synergistic combination of multiple devices alongside social support helped mitigate some self-tracking fatigue by leveraging different types of glanceable reminders.}

\textchange{Multi-device redundancy may itself be a sustainability mechanism. The presence of a home display was associated with higher engagement even when controlling for condition order, suggesting ambient visibility partially mitigated decline. Future designs could deliberately leverage cross-device coordination to sustain engagement. For example, surfacing and prompting for missing data on the home display towards specific members as they pass by (\textit{e.g.}, using recognition with a camera), or adapting watchface visualizations to signal tracking opportunities throughout the day depending on location, arm movement indicating glances, and with facilitated input (\textit{e.g.}, suggested input options based on inferences with sensor). Beyond individual prompting, shared devices could facilitate gentle social accountability, where family members are nudged to notice and respond to each other's missing or incomplete data. Ultimately, however, sustained tracking may depend heavily on whether families find meaningful collaboration opportunity with using each other's data. This highlights the importance of not only designing to facilitate data collection and sharing, but for the reflection \cite{pina2017personal, saksono2024SocioCog} and co-regulation \cite{silva2023unpacking,silva2024Codesign} processes that give family tracking its value.}
\section{CONCLUSION}
 Our results highlight how the combined designs across home display and smartwatches devices supported family engagement with self-tracking and collaborative reflection of moods and goals. While tracking fatigue for moods surfaced as the study progressed, home display presence somewhat mitigated the drop in engagement. Devices functioned synergistically, where home displays reminded families to charge and wear watches, and smartwatches enabled opportunistic tracking and awareness when apart. Family members' preference for device and design diverged based on individual routines, mobility within and outside the home, and comfort with wearables. Our findings indicate that multi-device redundancy can support collective participation in health tracking even for those not actively self-tracking. However, sustaining these device ecologies introduces maintenance burdens and requires ongoing negotiation of attentional boundaries, particularly as personal devices shifted between family-centered and self-centered roles across daily contexts. Our findings suggest opportunities for adaptive, context-aware family informatics that balance ubiquitous mutual awareness with individual preferences, while acknowledging the costs of expanding device ecosystems for collaborative health management.

\begin{acks}
We thank our participants for their long involvement in our deployment and willingness to open their door to us. We also thank Priscila, Elisa, and Natan Silva for piloting the system and providing early feedback. ChatGPT and Claude LLMs were used to provide writing feedback, syntax/grammar checking, and format tables. This work was partially supported by AHRQ under award numbers 1R21HS028871-01 and 5R21HS028871-02, by the National Science Foundation under award IIS-2237389, and the CERES network.
\end{acks}

\bibliographystyle{ACM-Reference-Format}
\bibliography{references}

@String{Computing = "Computing" }

@String{Computer = "{IEEE} Computer" }

@String{Springer = "Springer-Verlag" }

@inproceedings{Luo2020,
address = {New York, NY, USA},
author = {Luo, Yuhan and Lee, Bongshin and Choe, Eun Kyoung},
booktitle = {Proceedings of the SIGCHI Conference on Human Factors in Computing Systems (CHI 2020)},
doi = {10.1145/3313831.3376616},
isbn = {9781450367080},
month = {apr},
pages = {1--13},
publisher = {ACM},
title = {{TandemTrack: Shaping Consistent Exercise Experience by Complementing a Mobile App with a Smart Speaker}},
url = {https://doi.org/10.1145/3313831.3376616 https://dl.acm.org/doi/10.1145/3313831.3376616},
year = {2020}
}

@inproceedings{Hong2016,
address = {New York, NY, USA},
author = {Hong, Matthew K and Wilcox, Lauren and Machado, Daniel and Olson, Thomas A and Simoneaux, Stephen F},
booktitle = {Conference on Human Factors in Computing Systems - Proceedings},
doi = {10.1145/2858036.2858508},
isbn = {9781450333627},
pages = {5337--5349},
publisher = {ACM},
title = {{Care partnerships: Toward technology to support teens' participation in their health care}},
url = {http://dx.doi.org/10.1145/2858036.2858508},
year = {2016}
}

@article{Skinner2000,
author = {Skinner, Harvey and Steinhauer, Paul and Sitarenios, Gill},
doi = {10.1111/1467-6427.00146},
issn = {01634445},
journal = {Journal of Family Therapy},
number = {2},
pages = {190--210},
title = {{Family Assessment Measure (FAM) and Process Model of Family Functioning}},
volume = {22},
year = {2000}
}

@inproceedings{Cordeiro2015a,
address = {New York, NY, USA},
author = {Cordeiro, Felicia and Epstein, Daniel A. and Thomaz, Edison and Bales, Elizabeth and Jagannathan, Arvind K. and Abowd, Gregory D. and Fogarty, James},
booktitle = {Proceedings of the ACM Conference on Human Factors in Computing Systems (CHI 2015)},
doi = {10.1145/2702123.2702155},
isbn = {9781450331456},
month = {apr},
pages = {1159--1162},
publisher = {ACM},
title = {{Barriers and Negative Nudges}},
url = {https://dl.acm.org/doi/10.1145/2702123.2702155},
volume = {2015-April},
year = {2015}
}

@article{Gulsrud2010,
author = {Gulsrud, Amanda C. and Jahromi, Laudan B. and Kasari, Connie},
doi = {10.1007/s10803-009-0861-x},
issn = {0162-3257},
journal = {Journal of Autism and Developmental Disorders},
month = {feb},
number = {2},
pages = {227--237},
pmid = {19714458},
publisher = {Springer},
title = {{The Co-Regulation of Emotions Between Mothers and their Children with Autism}},
url = {https://link.springer.com/article/10.1007/s10803-009-0861-x http://link.springer.com/10.1007/s10803-009-0861-x},
volume = {40},
year = {2010}
}

@inproceedings{Kawsar2013,
author = {Kawsar, Fahim and Brush, J. Bernheim},
booktitle = {UbiComp 2013 - Proceedings of the 2013 ACM International Joint Conference on Pervasive and Ubiquitous Computing},
doi = {10.1145/2493432.2493494},
isbn = {9781450317702},
pages = {627--636},
title = {{Home computing unplugged: Why, where and when people use different connected devices at home}},
url = {http://dx.doi.org/10.1145/2493432.2493494},
year = {2013}
}

@article{Seiderer2017,
author = {Seiderer, Andreas and Flutura, Simon and Andr{\'{e}}, Elisabeth},
doi = {10.1145/3141788.3141790},
isbn = {9781450355568},
title = {{Development of a Mobile Multi-device Nutrition Logger}},
url = {https://doi.org/10.1145/3141788.3141790},
year = {2017}
}

@article{Hardey2022,
author = {Hardey, Mariannn (Maz)},
doi = {10.1177/20552076221093131},
issn = {2055-2076},
journal = {DIGITAL HEALTH},
month = {jan},
pages = {205520762210931},
title = {{Tracking the trackers: Self-tracking in households as social practice}},
url = {https://us.sagepub.com/en-us/nam/open-access-at-sage http://journals.sagepub.com/doi/10.1177/20552076221093131},
volume = {8},
year = {2022}
}

@inproceedings{Lazar2015,
author = {Lazar, Amanda and Koehler, Christian and Tanenbaum, Joshua and Nguyen, David H},
booktitle = {UbiComp 2015 - Proceedings of the 2015 ACM International Joint Conference on Pervasive and Ubiquitous Computing},
doi = {10.1145/2750858.2804288},
isbn = {9781450335744},
pages = {635--646},
title = {{Why we use and abandon smart devices}},
url = {http://dx.doi.org/10.1145/2750858.2804288},
year = {2015}
}

@inproceedings{Figueiredo2021,
author = {Figueiredo, Mayara and Chen, Yunan},
booktitle = {Conference on Human Factors in Computing Systems - Proceedings},
doi = {10.1145/3411764.3445189},
isbn = {9781450380966},
pages = {17},
publisher = {ACM},
title = {{Health data in fertility care: An ecological perspective}},
url = {https://doi.org/10.1145/3411764.3445189},
year = {2021}
}

@inproceedings{Kaziunas2017,
address = {New York, NY, USA},
author = {Kaziunas, Elizabeth and Ackerman, Mark S and Lindtner, Silvia and Lee, Joyce M},
booktitle = {Proceedings of the ACM Conference on Computer Supported Cooperative Work (CSCW 2017)},
doi = {10.1145/2998181.2998303},
isbn = {9781450343350},
pages = {2260--2272},
publisher = {ACM},
title = {{Caring through data: Attending to the social and emotional experiences of health datafication}},
url = {http://dx.doi.org/10.1145/2998181.2998303},
year = {2017}
}

@inproceedings{Nebeling2016,
address = {New York, NY, USA},
author = {Nebeling, Michael and Dey, Anind K},
booktitle = {Proceedings of the SIGCHI Conference on Human Factors in Computing Systems (CHI 2016)},
doi = {10.1145/2858036.2858048},
isbn = {9781450333627},
month = {may},
pages = {5494--5505},
publisher = {ACM},
title = {{XDBrowser}},
url = {http://dx.doi.org/10.1145/2858036.2858048 https://dl.acm.org/doi/10.1145/2858036.2858048},
year = {2016}
}

@article{Braun2006,
author = {Braun, Virginia and Clarke, Victoria},
doi = {10.1191/1478088706qp063oa},
issn = {14780887},
journal = {Qualitative Research in Psychology},
number = {2},
pages = {77--101},
title = {{Using thematic analysis in psychology}},
url = {https://www.tandfonline.com/action/journalInformation?journalCode=uqrp20},
volume = {3},
year = {2006}
}

@inproceedings{Nebeling2017,
address = {New York, NY, USA},
author = {Nebeling, Michael},
booktitle = {Proceedings of the SIGCHI Conference on Human Factors in Computing Systems (CHI 2017)},
doi = {10.1145/3025453.3025547},
isbn = {9781450346559},
month = {may},
pages = {4574--4584},
publisher = {ACM},
title = {{XDBrowser 2.0: Semi-Automatic Generation of Cross-Device Interfaces}},
url = {http://dx.doi.org/10.1145/3025453.3025547 https://dl.acm.org/doi/10.1145/3025453.3025547},
volume = {2017-May},
year = {2017}
}

@inproceedings{Luo2021,
author = {Luo, Yuhan},
doi = {10.1145/3468002.3468232},
isbn = {9781450385596},
pages = {11--15},
publisher = {Virtual Event, USA. ACM},
title = {{Designing Multimodal Self-Tracking Technologies to Promote Data Capture and Self-Reflection}},
url = {https://doi.org/10.1145/3468002.3468232},
year = {2021}
}

@article{Kim2021,
archivePrefix = {arXiv},
arxivId = {2101.06283},
author = {Kim, Young-Ho and Lee, Bongshin and Srinivasan, Arjun and Choe, Eun Kyoung and {Young-Ho Kim} and {Bongshin Lee} and {Arjun Srinivasan} and {Eun Kyoung Choe}},
doi = {10.1145/3411764.3445421},
eprint = {2101.06283},
isbn = {978-1-4503-8096-6},
journal = {Proceedings of the SIGCHI Conference on Human Factors in Computing Systems (CHI 2021) (CHI '21)},
month = {jan},
title = {{Data@Hand: Fostering Visual Exploration of Personal Data on Smartphones Leveraging Speech and Touch Interaction}},
url = {https://doi.org/10.1145/3411764.3445421 http://arxiv.org/abs/2101.06283 http://dx.doi.org/10.1145/3411764.3445421},
volume = {21},
year = {2021}
}

@inproceedings{Epstein2016b,
address = {New York, NY, USA},
author = {Epstein, Daniel A and Caraway, Monica and Johnston, Chuck and Ping, An and Fogarty, James and Munson, Sean A},
booktitle = {Proceedings of the ACM Conference on Human Factors in Computing Systems (CHI 2016)},
doi = {10.1145/2858036.2858045},
isbn = {9781450333627},
month = {may},
pages = {1109--1113},
publisher = {ACM},
title = {{Beyond Abandonment to Next Steps}},
url = {http://dx.doi.org/10.1145/2858036.2858045 https://dl.acm.org/doi/10.1145/2858036.2858045},
year = {2016}
}

@inproceedings{Jokela2015,
address = {New York, NY, USA},
author = {Jokela, Tero and Ojala, Jarno and Olsson, Thomas},
booktitle = {Proceedings of the ACM Conference on Human Factors in Computing Systems (CHI 2015)},
doi = {10.1145/2702123.2702211},
isbn = {9781450331456},
month = {apr},
pages = {3903--3912},
publisher = {ACM},
title = {{A Diary Study on Combining Multiple Information Devices in Everyday Activities and Tasks}},
url = {http://dx.doi.org/10.1145/2702123.2702211 https://dl.acm.org/doi/10.1145/2702123.2702211},
volume = {2015-April},
year = {2015}
}

@article{Yuan2022,
address = {New York, NY, USA},
author = {Yuan, Ye and Riche, Nathalie and Marquardt, Nicolai and Nicholas, Molly Jane and Seyed, Teddy and Romat, Hugo and Lee, Bongshin and Pahud, Michel and Goldstein, Jonathan and Vishkaie, Rojin and Holz, Christian and Hinckley, Ken},
doi = {10.1145/3491102},
isbn = {9781450391573},
journal = {CHI Conference on Human Factors in Computing Systems},
pages = {22},
publisher = {ACM},
title = {{Understanding Multi-Device Usage Patterns: Physical Device Configurations and Fragmented Workflows}},
url = {https://doi.org/10.1145/3491102.3517702},
year = {2022}
}

@article{Kim2017,
author = {Kim, Young-Ho and Jeon, Jae Ho and Lee, Bongshin and Choe, Eun Kyoung and Seo, Jinwook},
doi = {10.1145/3130930},
issn = {2474-9567},
journal = {Proceedings of the ACM on Interactive, Mobile, Wearable and Ubiquitous Technologies (IMWUT 2017)},
number = {3},
pages = {1--28},
title = {{OmniTrack: A Flexible Self-Tracking Approach Leveraging Semi-Automated Tracking}},
volume = {1},
year = {2017}
}

@inproceedings{Grimes2009,
address = {New York, New York, USA},
author = {Grimes, Andrea and Tan, Desney and Morris, Dan},
booktitle = {GROUP'09 - Proceedings of the 2009 ACM SIGCHI International Conference on Supporting Group Work},
doi = {10.1145/1531674.1531721},
isbn = {9781605585000},
pages = {311--320},
publisher = {ACM Press},
title = {{Toward Technologies that Support Family Reflections on Health}},
year = {2009}
}

@article{Choe2017,
author = {Choe, Eun Kyoung and Abdullah, Saeed and Rabbi, Mashfiqui and Thomaz, Edison and Epstein, Daniel A. and Cordeiro, Felicia and Kay, Matthew and Abowd, Gregory D. and Choudhury, Tanzeem and Fogarty, James and Lee, Bongshin and Matthews, Mark and Kientz, Julie A.},
doi = {10.1109/MPRV.2017.18},
issn = {1536-1268},
journal = {IEEE Pervasive Computing},
month = {jan},
number = {1},
pages = {74--84},
publisher = {IEEE},
title = {{Semi-Automated Tracking: A Balanced Approach for Self-Monitoring Applications}},
url = {http://ieeexplore.ieee.org/document/7807194/},
volume = {16},
year = {2017}
}

@inproceedings{Pizza2016,
address = {New York, NY, USA},
author = {Pizza, Stefania and Brown, Barry and McMillan, Donald and Lampinen, Airi},
booktitle = {Proceedings of the SIGCHI Conference on Human Factors in Computing Systems (CHI 2016)},
doi = {10.1145/2858036.2858522},
isbn = {9781450333627},
month = {may},
pages = {5456--5469},
publisher = {ACM},
title = {{Smartwatch in vivo}},
url = {http://dx.doi.org/10.1145/2858036.2858522 https://dl.acm.org/doi/10.1145/2858036.2858522},
year = {2016}
}

@article{Neustaedter2009,
author = {Neustaedter, Carman and Brush, A. J. Bernheim and Greenberg, Saul},
doi = {10.1145/1502800.1502806},
issn = {1073-0516},
journal = {ACM Transactions on Computer-Human Interaction},
month = {apr},
number = {1},
pages = {1--48},
publisher = {ACM PUB27 New York, NY, USA},
title = {{The Calendar is Crucial: Coordination and Awareness through the Family Calendar}},
url = {https://dl.acm.org/doi/10.1145/1502800.1502806},
volume = {16},
year = {2009}
}

@inproceedings{Toscos2012,
author = {Toscos, Tammy and Connelly, Kay and Rogers, Yvonne},
title = {Best Intentions: Health Monitoring Technology and Children},
year = {2012},
isbn = {9781450310154},
publisher = {Association for Computing Machinery},
address = {New York, NY, USA},
url = {https://doi.org/10.1145/2207676.2208603},
doi = {10.1145/2207676.2208603},
booktitle = {Proceedings of the SIGCHI Conference on Human Factors in Computing Systems},
pages = {1431–1440},
numpages = {10},
location = {Austin, Texas, USA},
series = {CHI '12}
}

@article{Denham2015,
author = {Denham, Sharon and Eggenberger, Sandra K. and Young, Patricia and Krumwiede, Norma},
journal = {FA Davis},
title = {{Family-Focused Nursing Care}},
year = {2015}
}

@inproceedings{yamashita2017,
author = {Yamashita, Naomi and Kuzuoka, Hideaki and Hirata, Keiji and Kudo, Takashi and Aramaki, Eiji and Hattori, Kazuki},
title = {Changing Moods: How Manual Tracking by Family Caregivers Improves Caring and Family Communication},
year = {2017},
isbn = {9781450346559},
publisher = {Association for Computing Machinery},
address = {New York, NY, USA},
url = {https://doi.org/10.1145/3025453.3025843},
doi = {10.1145/3025453.3025843},
booktitle = {Proceedings of the 2017 CHI Conference on Human Factors in Computing Systems},
pages = {158–169},
numpages = {12},
location = {Denver, Colorado, USA},
series = {CHI '17}
}

@inproceedings{Yamashita2018,
address = {New York, NY, USA},
author = {Yamashita, Naomi and Kuzuoka, Hideaki and Kudo, Takashi and Hirata, Keiji and Aramaki, Eiji and Hattori, Kazuki},
booktitle = {Proceedings of the SIGCHI Conference on Human Factors in Computing Systems (CHI 2018)},
doi = {10.1145/3173574.3173796},
isbn = {9781450356206},
month = {apr},
pages = {1--13},
publisher = {ACM},
title = {{How Information Sharing about Care Recipients by Family Caregivers Impacts Family Communication}},
url = {https://doi.org/10.1145/3173574.3173796 https://dl.acm.org/doi/10.1145/3173574.3173796},
year = {2018}
}

@inproceedings{OLeary2017,
address = {New York, NY, USA},
author = {O'Leary, Katie and Dong, Tao and Haines, Julia Katherine and Gilbert, Michael and Churchill, Elizabeth F and Nichols, Jeffrey},
booktitle = {DIS 2017 - Proceedings of the 2017 ACM Conference on Designing Interactive Systems},
doi = {10.1145/3064663.3064768},
isbn = {9781450349222},
month = {jun},
pages = {309--320},
publisher = {ACM},
title = {{The Moving Context Kit: Designing for Context Shifts in Multi-device Experiences}},
url = {http://dx.doi.org/10.1145/3064663.3064768 https://dl.acm.org/doi/10.1145/3064663.3064768},
year = {2017}
}

@article{Bell1979,
author = {Bell, Richard Q.},
doi = {10.1037/0003-066X.34.10.821},
issn = {0003066X},
journal = {American Psychologist},
month = {oct},
number = {10},
pages = {821--826},
title = {{Parent, child, and reciprocal influences}},
volume = {34},
year = {1979}
}

@inproceedings{DiGeronimo2016,
author = {{Di Geronimo}, Linda and Husmann, Maria and Norrie, Moira C},
booktitle = {PerDis 2016 - Proceedings of the 5th ACM International Symposium on Pervasive Displays},
doi = {10.1145/2914920.2915028},
isbn = {9781450343664},
pages = {220--227},
title = {{Surveying personal device ecosystems with cross-device applications in mind}},
url = {http://dx.doi.org/10.1145/2914920.2915028},
year = {2016}
}

@article{Willett2017,
author = {Willett, Wesley and Jansen, Yvonne and Dragicevic, Pierre},
doi = {10.1109/TVCG.2016.2598608},
issn = {10772626},
journal = {IEEE Transactions on Visualization and Computer Graphics},
month = {jan},
number = {1},
pages = {461--470},
pmid = {27875162},
publisher = {IEEE Computer Society},
title = {{Embedded Data Representations}},
url = {https://ieeexplore.ieee.org/document/7539328},
volume = {23},
year = {2017}
}

@article{Li2010,
author = {Li, Ian and Dey, Anind and Forlizzi, Jodi},
doi = {10.1145/1753326.1753409},
isbn = {9781605589299},
journal = {Conference on Human Factors in Computing Systems - Proceedings},
pages = {557--566},
title = {{A Stage-Based Model of Personal Informatics Systems}},
volume = {1},
year = {2010}
}

@inproceedings{Ayobi2020,
address = {New York, NY, USA},
author = {Ayobi, Amid and Marshall, Paul and Cox, Anna L.},
booktitle = {Proceedings of the SIGCHI Conference on Human Factors in Computing Systems (CHI 2020)},
doi = {10.1145/3313831.3376809},
isbn = {9781450367080},
month = {apr},
pages = {1--15},
publisher = {ACM},
title = {{Trackly: A Customisable and Pictorial Self-Tracking App to Support Agency in Multiple Sclerosis Self-Care}},
url = {https://dl.acm.org/doi/10.1145/3313831.3376809},
year = {2020}
}

@article{Cobb2013,
author = {Cobb, Stuart R. and Davies, Ceri H.},
doi = {10.1016/j.neuropharm.2013.02.001},
issn = {00283908},
journal = {Neuropharmacology},
month = {may},
pages = {1},
pmid = {23402709},
publisher = {American Psychiatric Association},
title = {{Neurodevelopmental disorders}},
url = {https://psychiatryonline.org/doi/10.1176/appi.books.9780890425596.dsm01 https://linkinghub.elsevier.com/retrieve/pii/S0028390813000427},
volume = {68},
year = {2013}
}

@inproceedings{Li2020,
address = {New York, NY, USA},
author = {Li, Qingyang and Caldeira, Clara and Epstein, Daniel A and Chen, Yunan},
booktitle = {Proceedings of the 14th EAI International Conference on Pervasive Computing Technologies for Healthcare},
doi = {10.1145/3421937.3422018},
isbn = {9781450375320},
month = {may},
number = {2020},
pages = {1--10},
publisher = {ACM},
title = {{Supporting Caring among Intergenerational Family Members through Family Fitness Tracking}},
url = {https://dl.acm.org/doi/10.1145/3421937.3422018},
volume = {14},
year = {2020}
}

@incollection{Sanders2018,
address = {Cham},
author = {Sanders, Matthew R and Turner, Karen M T},
booktitle = {Handbook of Parenting and Child Development Across the Lifespan},
doi = {10.1007/978-3-319-94598-9_1},
isbn = {9783319945989},
month = {dec},
pages = {3--26},
publisher = {Springer International Publishing},
title = {{The Importance of Parenting in Influencing the Lives of Children}},
url = {https://link.springer.com/chapter/10.1007/978-3-319-94598-9{\_}1 http://link.springer.com/10.1007/978-3-319-94598-9{\_}1},
year = {2018}
}

@inproceedings{Sonne2016,
address = {New York, NY, USA},
author = {Sonne, Tobias and M{\"{u}}ller, J{\"{o}}rg and Marshall, Paul and Obel, Carsten and Gr{\o}nb{\ae}k, Kaj},
booktitle = {Proceedings of the SIGCHI Conference on Human Factors in Computing Systems (CHI 2016)},
doi = {10.1145/2858036.2858157},
isbn = {9781450333627},
month = {may},
pages = {152--164},
publisher = {ACM},
title = {{Changing Family Practices with Assistive Technology: MOBERO Improves Morning and Bedtime Routines for Children with ADHD}},
url = {http://dx.doi.org/10.1145/2858036.2858157 https://dl.acm.org/doi/10.1145/2858036.2858157},
year = {2016}
}

@inproceedings{Oygur2021,
address = {New York, NY, USA},
author = {Oyg{\"{u}}r, Iþil and Su, Zhaoyuan and {A. Epstein}, Daniel and Chen, Yunan},
booktitle = {Proceedings of the SIGCHI Conference on Human Factors in Computing Systems (CHI 2021)},
doi = {10.1145/3411764.3445376},
isbn = {9781450380966},
month = {may},
pages = {1--12},
publisher = {ACM},
title = {{The Lived Experience of Child-Owned Wearables: Comparing Children's and Parents' Perspectives on Activity Tracking}},
url = {https://doi.org/10.1145/3411764.3445376 https://dl.acm.org/doi/10.1145/3411764.3445376},
year = {2021}
}

@article{Silva2022,
author = {Silva, Lucas M. and Cibrian, Franceli L. and Epstein, Daniel A. and Bhattacharya, Arpita and Ankrah, Elizabeth A. and Monteiro, Elissa and Beltran, Jesus A. and Schuck, Sabrina E. and Lakes, Kimberley D. and Hayes, Gillian R.},
doi = {10.1109/MPRV.2021.3104262},
issn = {1536-1268},
journal = {IEEE Pervasive Computing},
month = {jan},
number = {1},
pages = {48--56},
title = {{Adapting Multidevice Deployments During a Pandemic: Lessons Learned From Two Studies}},
url = {https://ieeexplore.ieee.org/document/9527153/},
volume = {21},
year = {2022}
}

@inproceedings{Chi2015,
address = {New York, NY, USA},
author = {Chi, Pei-Yu (Peggy) and Li, Yang},
booktitle = {Proceedings of the SIGCHI Conference on Human Factors in Computing Systems (CHI 2015)},
doi = {10.1145/2702123.2702451},
isbn = {9781450331456},
month = {apr},
pages = {3923--3932},
publisher = {ACM},
title = {{Weave: Scripting Cross-Device Wearable Interaction
}},
url = {http://dx.doi.org/10.1145/2702123.2702451 https://dl.acm.org/doi/10.1145/2702123.2702451},
volume = {2015-April},
year = {2015}
}

@inproceedings{Epstein2015a,
address = {New York, New York, USA},
author = {Epstein, Daniel A. and Ping, An and Fogarty, James and Munson, Sean A.},
booktitle = {Proceedings of the 2015 ACM International Joint Conference on Pervasive and Ubiquitous Computing (UbiComp 2015)},
doi = {10.1145/2750858.2804250},
isbn = {9781450335744},
month = {sep},
pages = {731--742},
publisher = {Association for Computing Machinery, Inc},
title = {{A Lived Informatics Model of Personal Informatics}},
url = {http://dl.acm.org/citation.cfm?doid=2750858.2804250},
year = {2015}
}

@article{Epstein2020,
author = {Epstein, Daniel A. and Caldeira, Clara and Figueiredo, Mayara Costa and Silva, Lucas M. and Lu, Xi and Williams, Lucretia and Lee, Jong Ho and Li, Qingyang and Ahuja, Simran and Chen, Qiuer and Hilby, Craig and Sultana, Sazeda and Yari, Payam Dowlat and Eikey, Elizabeth V. and Chen, Yunan},
doi = {https://doi.org/10.1145/3432231},
journal = {Proceedings of the ACM on Interactive, Mobile, Wearable and Ubiquitous Technologies (IMWUT 2020)},
number = {4},
title = {{Mapping and Taking Stock of the Personal Informatics Literature}},
volume = {4},
year = {2020}
}

@inproceedings{Ayobi2018,
address = {New York, NY, USA},
author = {Ayobi, Amid and Sonne, Tobias and Marshall, Paul and Cox, Anna L.},
booktitle = {Proceedings of the SIGCHI Conference on Human Factors in Computing Systems (CHI 2018)},
doi = {10.1145/3173574.3173602},
isbn = {9781450356206},
month = {apr},
pages = {1--14},
publisher = {ACM},
title = {{Flexible and Mindful Self-Tracking: Design Implications from Paper Bullet Journals}},
url = {https://dl.acm.org/doi/10.1145/3173574.3173602},
volume = {2018-April},
year = {2018}
}

@article{Jones2021a,
author = {Jones, Jasmine and Yuan, Ye and Yarosh, Svetlana},
title = {Be Consistent, Work the Program, Be Present Every Day: Exploring Technologies for Self-Tracking in Early Recovery},
year = {2022},
issue_date = {Dec 2021},
publisher = {Association for Computing Machinery},
address = {New York, NY, USA},
volume = {5},
number = {4},
url = {https://doi.org/10.1145/3494955},
doi = {10.1145/3494955},
journal = {Proc. ACM Interact. Mob. Wearable Ubiquitous Technol.},
month = dec,
articleno = {164},
numpages = {26}
}

@inproceedings{Bressa2022,
address = {New York, NY, USA},
author = {Bressa, Nathalie and Vermeulen, Jo and Willett, Wesley},
booktitle = {CHI Conference on Human Factors in Computing Systems},
doi = {10.1145/3491102.3517737},
isbn = {9781450391573},
month = {apr},
pages = {1--18},
publisher = {ACM},
title = {{Data Every Day: Designing and Living with Personal Situated Visualizations}},
url = {https://dl.acm.org/doi/10.1145/3491102.3517737},
year = {2022}
}

@inproceedings{Stefanidi2022,
author = {Stefanidi, Evropi and Sch\"{o}ning, Johannes and Feger, Sebastian S. and Marshall, Paul and Rogers, Yvonne and Niess, Jasmin},
title = {Designing for Care Ecosystems: A Literature Review of Technologies for Children with ADHD},
year = {2022},
isbn = {9781450391979},
publisher = {},
address = {New York, NY, USA},
url = {https://doi.org/10.1145/3501712.3529746},
doi = {10.1145/3501712.3529746},
booktitle = {Proceedings of the 21st Annual ACM Interaction Design and Children Conference},
pages = {13–25},
numpages = {13},
location = {Braga, Portugal},
series = {IDC '22}
}

@inproceedings{Saksono2020,
address = {New York, NY, USA},
author = {Saksono, Herman and Castaneda-Sceppa, Carmen and Hoffman, Jessica and Morris, Vivien and {Seif El-Nasr}, Magy and Parker, Andrea G},
booktitle = {Proceedings of the SIGCHI Conference on Human Factors in Computing Systems (CHI 2020)},
doi = {10.1145/3313831.3376686},
isbn = {9781450367080},
month = {apr},
pages = {1--13},
publisher = {ACM},
title = {{Storywell: Designing for Family Fitness App Motivation by Using Social Rewards and Reflection}},
url = {https://doi.org/10.1145/3313831.3376686 https://dl.acm.org/doi/10.1145/3313831.3376686},
year = {2020}
}

@inproceedings{Brush2008,
address = {New York, New York, USA},
author = {Brush, A.J. Bernheim and Inkpen, Kori M. and Tee, Kimberly},
booktitle = {Proceedings of the ACM conference on Computer supported cooperative work (CSCW 2008)},
doi = {10.1145/1460563.1460661},
isbn = {9781605580074},
pages = {629},
publisher = {ACM Press},
title = {{SPARCS: Exploring Sharing Suggestions to Enhance Family Connectedness}},
url = {https://doi.org/10.1145/1460563.1460661},
year = {2008}
}

@article{richards2021a,
author = {Richards, Olivia K. and Choi, Adrian and Marcu, Gabriela},
title = {Shared Understanding in Care Coordination for Children's Behavioral Health},
year = {2021},
issue_date = {April 2021},
publisher = {Association for Computing Machinery},
address = {New York, NY, USA},
volume = {5},
number = {CSCW1},
url = {https://doi.org/10.1145/3449095},
doi = {10.1145/3449095},
journal = {Proc. ACM Hum.-Comput. Interact.},
month = {apr},
articleno = {21},
numpages = {25}
}

@inproceedings{Gouveia2023,
address = {New York, NY, USA},
author = {Gouveia, Ruben and Epstein, Daniel A.},
booktitle = {Proceedings of the 2023 CHI Conference on Human Factors in Computing Systems (CHI ’23)},
doi = {10.1145/3544548.3580955},
isbn = {},
month = {April},
pages = {},
pmid = {},
publisher = {ACM},
title = {{This Watchface Fits with my Tattoos: Investigating Customisation Needs and Preferences in Personal Tracking}},
url = {https://doi.org/10.1145/3544548.3580955},
year = {2023}
}

@article{Classi2012,
author = {Classi, Peter and Milton, Den{\'{a}}i and Ward, Sarah and Sarsour, Khaled and Johnston, Joseph},
doi = {10.1186/1753-2000-6-33},
issn = {1753-2000},
journal = {Child and Adolescent Psychiatry and Mental Health},
number = {1},
pages = {33},
title = {{Social and emotional difficulties in children with ADHD and the impact on school attendance and healthcare utilization}},
url = {https://doi.org/10.1186/1753-2000-6-33},
volume = {6},
year = {2012}
}

@article{Gisladottir2017,
author = {Gisladottir, M. and Svavarsdottir, E. K.},
title = {The effectiveness of therapeutic conversation intervention for caregivers of adolescents with ADHD: a quasi-experimental design},
journal = {Journal of Psychiatric and Mental Health Nursing},
volume = {24},
number = {1},
pages = {15-27},
doi = {https://doi.org/10.1111/jpm.12335},
url = {https://onlinelibrary.wiley.com/doi/abs/10.1111/jpm.12335},
eprint = {https://onlinelibrary.wiley.com/doi/pdf/10.1111/jpm.12335},
year = {2017}
}

@article{jang2023,
author = {Sangsu Jang and Kyung-Ryong Lee and Geonil Goh and Dohee Kim and Gahui Yun and Nanum Kim and Byeol Kim Lux and Choong-Wan Woo and Hyungsook Kim and Young-Woo Park},
title = {Design and field trial of EmotionFrame: exploring self-journaling experiences in homes for archiving personal feelings about daily events},
journal = {Human–Computer Interaction},
volume = {0},
number = {0},
pages = {1-26},
year  = {2023},
publisher = {Taylor & Francis},
doi = {10.1080/07370024.2023.2219259},
URL = {https://doi.org/10.1080/07370024.2023.2219259}
}

@article{snapi,
author = {Wang, Dennis and Chheang, Marawin and Ji, Siyun and Mohta, Ryan and Epstein, Daniel A.},
title = {SnapPI: Understanding Everyday Use of Personal Informatics Data Stickers on Ephemeral Social Media},
year = {2022},
issue_date = {November 2022},
publisher = {Association for Computing Machinery},
address = {New York, NY, USA},
volume = {6},
number = {CSCW2},
url = {https://doi.org/10.1145/3555652},
doi = {10.1145/3555652},
journal = {Proc. ACM Hum.-Comput. Interact.},
month = {nov},
articleno = {539},
numpages = {27}
}

@article{stefanidi2025supporting,
author = {Stefanidi, Evropi and Wagener, Nadine and Chatzakis, Ioannis and Wo\'{z}niak, Pawe\l{} W. and Ntoa, Stavroula and Margetis, George and Rogers, Yvonne and Niess, Jasmin},
title = {Supporting Communication and Well-being with a Multi-Stakeholder Mobile App: Lessons Learned from a Field Study with ADHD Children and their Caregivers},
year = {2025},
issue_date = {May 2025},
publisher = {Association for Computing Machinery},
address = {New York, NY, USA},
volume = {9},
number = {2},
url = {https://doi.org/10.1145/3711075},
doi = {10.1145/3711075},
journal = {Proc. ACM Hum.-Comput. Interact.},
month = may,
articleno = {CSCW177},
numpages = {37}
}

@inproceedings{pina2017personal,
author = {Pina, Laura R. and Sien, Sang-Wha and Ward, Teresa and Yip, Jason C. and Munson, Sean A. and Fogarty, James and Kientz, Julie A.},
title = {From Personal Informatics to Family Informatics: Understanding Family Practices around Health Monitoring},
year = {2017},
isbn = {9781450343350},
publisher = {Association for Computing Machinery},
address = {New York, NY, USA},
url = {https://doi.org/10.1145/2998181.2998362},
doi = {10.1145/2998181.2998362},
booktitle = {Proceedings of the 2017 ACM Conference on Computer Supported Cooperative Work and Social Computing},
pages = {2300–2315},
numpages = {16},
location = {Portland, Oregon, USA},
series = {CSCW '17}
}

@inproceedings{murnane2018social,
author = {Murnane, Elizabeth L. and Snyder, Jaime and Voida, Stephen and Bietz, Matthew J. and Matthews, Mark and Munson, Sean and Pina, Laura R.},
title = {Social Issues in Personal Informatics: Design, Data, and Infrastructure},
year = {2018},
isbn = {9781450360180},
publisher = {Association for Computing Machinery},
address = {New York, NY, USA},
url = {https://doi.org/10.1145/3272973.3273016},
doi = {10.1145/3272973.3273016},
booktitle = {Companion of the 2018 ACM Conference on Computer Supported Cooperative Work and Social Computing},
pages = {471–478},
numpages = {8},
location = {Jersey City, NJ, USA},
series = {CSCW '18 Companion}
}

@inproceedings{saksono2019social,
author = {Saksono, Herman and Castaneda-Sceppa, Carmen and Hoffman, Jessica and Seif El-Nasr, Magy and Morris, Vivien and Parker, Andrea G.},
title = {Social Reflections on Fitness Tracking Data: A Study with Families in Low-SES Neighborhoods},
year = {2019},
isbn = {9781450359702},
publisher = {Association for Computing Machinery},
address = {New York, NY, USA},
url = {https://doi.org/10.1145/3290605.3300543},
doi = {10.1145/3290605.3300543},
booktitle = {Proceedings of the 2019 CHI Conference on Human Factors in Computing Systems},
pages = {1–14},
numpages = {14},
location = {Glasgow, Scotland Uk},
series = {CHI '19}
}

@article{pina2020dreamcatcher,
author = {Pina, Laura and Sien, Sang-Wha and Song, Clarissa and Ward, Teresa M. and Fogarty, James and Munson, Sean A. and Kientz, Julie A.},
title = {DreamCatcher: Exploring How Parents and School-Age Children can Track and Review Sleep Information Together},
year = {2020},
issue_date = {May 2020},
publisher = {Association for Computing Machinery},
address = {New York, NY, USA},
volume = {4},
number = {CSCW1},
url = {https://doi.org/10.1145/3392882},
doi = {10.1145/3392882},
journal = {Proc. ACM Hum.-Comput. Interact.},
month = {may},
articleno = {70},
numpages = {25}
}

@inproceedings{saksono2015spaceship,
author = {Saksono, Herman and Ranade, Ashwini and Kamarthi, Geeta and Castaneda-Sceppa, Carmen and Hoffman, Jessica A. and Wirth, Cathy and Parker, Andrea G.},
title = {Spaceship Launch: Designing a Collaborative Exergame for Families},
year = {2015},
isbn = {9781450329224},
publisher = {Association for Computing Machinery},
address = {New York, NY, USA},
url = {https://doi.org/10.1145/2675133.2675159},
doi = {10.1145/2675133.2675159},
booktitle = {Proceedings of the 18th ACM Conference on Computer Supported Cooperative Work \& Social Computing},
pages = {1776–1787},
numpages = {12},
location = {Vancouver, BC, Canada},
series = {CSCW '15}
}

@article{oygur2020raising,
author = {Oyg\"{u}r, I\c{s}il and Epstein, Daniel A. and Chen, Yunan},
title = {Raising the Responsible Child: Collaborative Work in the Use of Activity Trackers for Children},
year = {2020},
issue_date = {October 2020},
publisher = {Association for Computing Machinery},
address = {New York, NY, USA},
volume = {4},
number = {CSCW2},
url = {https://doi.org/10.1145/3415228},
doi = {10.1145/3415228},
journal = {Proc. ACM Hum.-Comput. Interact.},
month = {oct},
articleno = {157},
numpages = {23}
}

@inproceedings{silva2023unpacking,
author = {Silva, Lucas M. and Cibrian, Franceli L. and Monteiro, Elissa and Bhattacharya, Arpita and Beltran, Jesus A. and Bonang, Clarisse and Epstein, Daniel A. and Schuck, Sabrina E. B. and Lakes, Kimberley D. and Hayes, Gillian R.},
title = {Unpacking the Lived Experiences of Smartwatch Mediated Self and Co-Regulation with ADHD Children},
year = {2023},
isbn = {9781450394215},
publisher = {Association for Computing Machinery},
address = {New York, NY, USA},
url = {https://doi.org/10.1145/3544548.3581316},
doi = {10.1145/3544548.3581316},
booktitle = {Proceedings of the 2023 CHI Conference on Human Factors in Computing Systems},
articleno = {90},
numpages = {19},
location = {Hamburg, Germany},
series = {CHI '23}
}

@inproceedings{saksono2020storywell,
author = {Saksono, Herman and Castaneda-Sceppa, Carmen and Hoffman, Jessica and Morris, Vivien and Seif El-Nasr, Magy and Parker, Andrea G.},
title = {Storywell: Designing for Family Fitness App Motivation by Using Social Rewards and Reflection},
year = {2020},
isbn = {9781450367080},
publisher = {Association for Computing Machinery},
address = {New York, NY, USA},
url = {https://doi.org/10.1145/3313831.3376686},
doi = {10.1145/3313831.3376686},
booktitle = {Proceedings of the 2020 CHI Conference on Human Factors in Computing Systems},
pages = {1–13},
numpages = {13},
location = {Honolulu, HI, USA},
series = {CHI '20}
}

@inproceedings{schaefbauer2015snack,
author = {Schaefbauer, Christopher L. and Khan, Danish U. and Le, Amy and Sczechowski, Garrett and Siek, Katie A.},
title = {Snack Buddy: Supporting Healthy Snacking in Low Socioeconomic Status Families},
year = {2015},
isbn = {9781450329224},
publisher = {Association for Computing Machinery},
address = {New York, NY, USA},
url = {https://doi.org/10.1145/2675133.2675180},
doi = {10.1145/2675133.2675180},
booktitle = {Proceedings of the 18th ACM Conference on Computer Supported Cooperative Work \& Social Computing},
pages = {1045–1057},
numpages = {13},
location = {Vancouver, BC, Canada},
series = {CSCW '15}
}

@article{faraone2019genetics,
  title={Genetics of attention deficit hyperactivity disorder},
  author={Faraone, Stephen V and Larsson, Henrik},
  journal={Molecular psychiatry},
  volume={24},
  number={4},
  pages={562--575},
  year={2019},
  publisher={Nature Publishing Group}
}

@article{Dennis2024,
author = {Wang, Dennis and Eng, Jocelyn and Turpitka, Mykyta and Epstein, Daniel A.},
title = {Exploring Activity-Sharing Response Differences Between Broad-Purpose and Dedicated Online Social Platforms},
year = {2024},
issue_date = {November 2024},
publisher = {Association for Computing Machinery},
address = {New York, NY, USA},
volume = {8},
number = {CSCW2},
url = {https://doi.org/10.1145/3686898},
doi = {10.1145/3686898},
journal = {Proc. ACM Hum.-Comput. Interact.},
month = nov,
articleno = {359},
numpages = {37}
}

@article{Wang2017quantbaby,
author = {Wang, Junqing and O'Kane, Aisling Ann and Newhouse, Nikki and Sethu-Jones, Geraint Rhys and de Barbaro, Kaya},
title = {Quantified Baby: Parenting and the Use of a Baby Wearable in the Wild},
year = {2017},
issue_date = {November 2017},
publisher = {Association for Computing Machinery},
address = {New York, NY, USA},
volume = {1},
number = {CSCW},
url = {https://doi.org/10.1145/3134743},
doi = {10.1145/3134743},
journal = {Proc. ACM Hum.-Comput. Interact.},
month = dec,
articleno = {108},
numpages = {19}
}

@article{grevenstein2019better,
  title={Better family relationships----higher well-being: The connection between relationship quality and health related resources},
  author={Grevenstein, Dennis and Bluemke, Matthias and Schweitzer, Jochen and Aguilar-Raab, Corina},
  journal={Mental health \& prevention},
  volume={14},
  pages={200160},
  year={2019},
  publisher={Elsevier}
}

@article{Nikkhah2022,
author = {Nikkhah, Sarah and John, Swaroop and Yalamarti, Krishna Supradeep and Mueller, Emily L. and Miller, Andrew D.},
title = {Family Care Coordination in the Children's Hospital: Phases and Cycles in the Pediatric Cancer Caregiving Journey},
year = {2022},
issue_date = {November 2022},
publisher = {Association for Computing Machinery},
address = {New York, NY, USA},
volume = {6},
number = {CSCW2},
url = {https://doi.org/10.1145/3555187},
doi = {10.1145/3555187},
journal = {Proc. ACM Hum.-Comput. Interact.},
month = nov,
articleno = {296},
numpages = {30}
}

@article{berge2015all,
  title={All in the family: Correlations between parents' and adolescent siblings' weight and weight-related behaviors},
  author={Berge, Jerica M and Meyer, Craig and MacLehose, Richard F and Crichlow, Renee and Neumark-Sztainer, Dianne},
  journal={Obesity},
  volume={23},
  number={4},
  pages={833--839},
  year={2015},
  publisher={Wiley Online Library}
}

@article{bugelmayer2018family,
  title={Is it the family or the neighborhood? Evidence from sibling and neighbor correlations in youth education and health},
  author={B{\"u}gelmayer, Elisabeth and Schnitzlein, Daniel D},
  journal={The Journal of Economic Inequality},
  volume={16},
  number={3},
  pages={369--388},
  year={2018},
  publisher={Springer}
}

@inproceedings{Merel2025wearables,
author = {van den Berg, Merel K. N. and Karahano\u{g}lu, Arma\u{g}an and Noordzij, Matthijs L. and Maeckelberghe, Els L. M. and Ludden, Geke D. S.},
title = {Facilitators and Barriers of Wearable Stress Management Technology: A Narrative Review of User Perspectives},
year = {2025},
isbn = {9798400713941},
publisher = {Association for Computing Machinery},
address = {New York, NY, USA},
url = {https://doi.org/10.1145/3706598.3713802},
doi = {10.1145/3706598.3713802},
booktitle = {Proceedings of the 2025 CHI Conference on Human Factors in Computing Systems},
articleno = {92},
numpages = {24},
location = {
},
series = {CHI '25}
}

@inbook{charitos2025watch,
author = {Charitos, Sydney and Thompson, Lauren and Brigden, Amberly and Russell, Abby and Slovak, Petr and Bird, Jon},
title = {“I Like My Own Watch Independence”: Exploring Low-Burden Customisation of a Technology Probe by Children with ADHD and their Parents},
year = {2025},
isbn = {9798400714733},
publisher = {Association for Computing Machinery},
address = {New York, NY, USA},
url = {https://doi.org/10.1145/3713043.3733254},
booktitle = {Proceedings of the 24th Interaction Design and Children},
pages = {728–742},
numpages = {15}
}

@inproceedings{silva2024Codesign,
author = {Silva, Lucas M. and Cibrian, Franceli L. and Bonang, Clarisse and Bhattacharya, Arpita and Min, Aehong and Monteiro, Elissa M and Beltran, Jesus Armando and Schuck, Sabrina and Lakes, Kimberley D and Hayes, Gillian R. and Epstein, Daniel A.},
title = {Co-Designing Situated Displays for Family Co-Regulation with ADHD Children},
year = {2024},
isbn = {9798400703300},
publisher = {Association for Computing Machinery},
address = {New York, NY, USA},
url = {https://doi.org/10.1145/3613904.3642745},
doi = {10.1145/3613904.3642745},
booktitle = {Proceedings of the 2024 CHI Conference on Human Factors in Computing Systems},
articleno = {124},
numpages = {19},
location = {Honolulu, HI, USA},
series = {CHI '24}
}

@article{Lee2024FamilyScope,
author = {Lee, Hyunsoo and Jung, Yugyeong and Shin, Youwon and Park, Hyesoo and Choi, Woohyeok and Lee, Uichin},
title = {FamilyScope: Visualizing Affective Aspects of Family Social Interactions using Passive Sensor Data},
year = {2024},
issue_date = {April 2024},
publisher = {Association for Computing Machinery},
address = {New York, NY, USA},
volume = {8},
number = {CSCW1},
url = {https://doi.org/10.1145/3637334},
doi = {10.1145/3637334},
journal = {Proc. ACM Hum.-Comput. Interact.},
month = apr,
articleno = {57},
numpages = {27}
}

@article{saksono2024SocioCog,
author = {Saksono, Herman and Parker, Andrea G.},
title = {Socio-Cognitive Framework for Personal Informatics: A Preliminary Framework for Socially-Enabled Health Technologies},
year = {2024},
issue_date = {June 2024},
publisher = {Association for Computing Machinery},
address = {New York, NY, USA},
volume = {31},
number = {3},
issn = {1073-0516},
url = {https://doi.org/10.1145/3674504},
doi = {10.1145/3674504},
journal = {ACM Trans. Comput.-Hum. Interact.},
month = aug,
articleno = {42},
numpages = {41}
}

@inproceedings{Carman2006FamilyCalendar,
author = {Neustaedter, Carman and Bernheim Brush, A. J.},
title = {"LINC-ing" the family: the participatory design of an inkable family calendar},
year = {2006},
isbn = {1595933727},
publisher = {Association for Computing Machinery},
address = {New York, NY, USA},
url = {https://doi.org/10.1145/1124772.1124796},
doi = {10.1145/1124772.1124796},
booktitle = {Proceedings of the SIGCHI Conference on Human Factors in Computing Systems},
pages = {141–150},
numpages = {10},
location = {Montr\'{e}al, Qu\'{e}bec, Canada},
series = {CHI '06}
}

@inproceedings{Armagan2011,
author = {Karahano\u{g}lu, Arma\u{g}an and Erbu\u{g}, \c{C}i\u{g}dem},
title = {Perceived qualities of smart wearables: determinants of user acceptance},
year = {2011},
isbn = {9781450312806},
publisher = {Association for Computing Machinery},
address = {New York, NY, USA},
url = {https://doi.org/10.1145/2347504.2347533},
doi = {10.1145/2347504.2347533},
booktitle = {Proceedings of the 2011 Conference on Designing Pleasurable Products and Interfaces},
articleno = {26},
numpages = {8},
location = {Milano, Italy},
series = {DPPI '11}
}

@inproceedings{Le2025uEMA,
author = {Le, Ha and Potter, Veronika and Lakshminarayanan, Rithika and Mishra, Varun and Intille, Stephen},
title = {Feasibility and Utility of Multimodal Micro Ecological Momentary Assessment on a Smartwatch},
year = {2025},
isbn = {9798400713941},
publisher = {Association for Computing Machinery},
address = {New York, NY, USA},
url = {https://doi.org/10.1145/3706598.3714086},
doi = {10.1145/3706598.3714086},
booktitle = {Proceedings of the 2025 CHI Conference on Human Factors in Computing Systems},
articleno = {1182},
numpages = {22},
location = {
},
series = {CHI '25}
}

@inproceedings{epsteinBabyTemptrack,
author = {Louie, Julianne and Mukund, Tara and Vu, Chau and Epstein, Daniel A. and Papoutsaki, Alexandra},
title = {Understanding Temporality of Reflection in Personal Informatics through Baby Tracking},
year = {2025},
isbn = {9798400713941},
publisher = {Association for Computing Machinery},
address = {New York, NY, USA},
url = {https://doi.org/10.1145/3706598.3713197},
doi = {10.1145/3706598.3713197},
booktitle = {Proceedings of the 2025 CHI Conference on Human Factors in Computing Systems},
articleno = {1191},
numpages = {18},
location = {
},
series = {CHI '25}
}

@inproceedings{Froehlich2012Display,
author = {Froehlich, Jon and Findlater, Leah and Ostergren, Marilyn and Ramanathan, Solai and Peterson, Josh and Wragg, Inness and Larson, Eric and Fu, Fabia and Bai, Mazhengmin and Patel, Shwetak and Landay, James A.},
title = {The design and evaluation of prototype eco-feedback displays for fixture-level water usage data},
year = {2012},
isbn = {9781450310154},
publisher = {Association for Computing Machinery},
address = {New York, NY, USA},
url = {https://doi.org/10.1145/2207676.2208397},
doi = {10.1145/2207676.2208397},
booktitle = {Proceedings of the SIGCHI Conference on Human Factors in Computing Systems},
pages = {2367–2376},
numpages = {10},
location = {Austin, Texas, USA},
series = {CHI '12}
}

@article{Richards2025FamilyCareRoutines,
author = {Richards, Olivia K. and Veinot, Tiffany C.},
title = {Reconceptualizing Technology for Chronic Disease Management Activities in the Family: Supporting Collective Routines},
year = {2025},
issue_date = {June 2025},
publisher = {Association for Computing Machinery},
address = {New York, NY, USA},
volume = {32},
number = {3},
issn = {1073-0516},
url = {https://doi.org/10.1145/3719347},
doi = {10.1145/3719347},
journal = {ACM Trans. Comput.-Hum. Interact.},
month = jun,
articleno = {23},
numpages = {82}
}

@inproceedings{Moon2025Fluidtrack,
author = {Moon, Junhyung and Lee, Sukhyun and Kim, Youngchan and Go, Juhee and Ku, Han Mo and Jung, Yeohyun and Hwang, Seonyeong and Lee, Bongshin and Lee, Yong Seung and Lee, Hyun-Kyung and Lee, Kyoungwoo and Choe, Eun Kyoung},
title = {FluidTrack: Investigating Child-Parent Collaborative Tracking for Pediatric Voiding Dysfunction Management},
year = {2025},
isbn = {9798400713941},
publisher = {Association for Computing Machinery},
address = {New York, NY, USA},
url = {https://doi.org/10.1145/3706598.3713878},
doi = {10.1145/3706598.3713878},
booktitle = {Proceedings of the 2025 CHI Conference on Human Factors in Computing Systems},
articleno = {559},
numpages = {18},
location = {
},
series = {CHI '25}
}

@inproceedings{shin2023BedtimePals,
author = {Shin, Ji Youn and Li, Tongxin and Peng, Wei and Lee, Hee Rin},
title = {Bedtime Pals: A Deployment Study of Sleep Management Technology for Families with Young Children},
year = {2023},
isbn = {9781450398930},
publisher = {Association for Computing Machinery},
address = {New York, NY, USA},
url = {https://doi.org/10.1145/3563657.3596068},
doi = {10.1145/3563657.3596068},
booktitle = {Proceedings of the 2023 ACM Designing Interactive Systems Conference},
pages = {1610–1629},
numpages = {20},
location = {Pittsburgh, PA, USA},
series = {DIS '23}
}

@article{hennessy2020self,
  title={Self-regulation mechanisms in health behavior change: A systematic meta-review of meta-analyses, 2006--2017},
  author={Hennessy, Emily A and Johnson, Blair T and Acabchuk, Rebecca L and McCloskey, Kiran and Stewart-James, Jania},
  journal={Health psychology review},
  volume={14},
  number={1},
  pages={6--42},
  year={2020},
  publisher={Taylor \& Francis}
}

@inproceedings{parkCollabTrackingReview,
author = {Cha, Yoon Jeong and Chen, Jiongyu and Gunal, Yasemin and Zhu, Qiying and Newman, Mark W and Park, Sun Young},
title = {Collaborative Health-Tracking Technologies for Children and Parents: A Review of Current Studies and Directions for Future Research},
year = {2025},
isbn = {9798400713941},
publisher = {Association for Computing Machinery},
address = {New York, NY, USA},
url = {https://doi.org/10.1145/3706598.3713596},
doi = {10.1145/3706598.3713596},
booktitle = {Proceedings of the 2025 CHI Conference on Human Factors in Computing Systems},
articleno = {1038},
numpages = {13},
location = {
},
series = {CHI '25}
}

@inproceedings{sellen2006HomeNote,
author = {Sellen, Abigail and Harper, Richard and Eardley, Rachel and Izadi, Shahram and Regan, Tim and Taylor, Alex S. and Wood, Ken R.},
title = {HomeNote: supporting situated messaging in the home},
year = {2006},
isbn = {1595932496},
publisher = {Association for Computing Machinery},
address = {New York, NY, USA},
url = {https://doi.org/10.1145/1180875.1180933},
doi = {10.1145/1180875.1180933},
booktitle = {Proceedings of the 2006 20th Anniversary Conference on Computer Supported Cooperative Work},
pages = {383–392},
numpages = {10},
location = {Banff, Alberta, Canada},
series = {CSCW '06}
}

@inproceedings{kawsar2013homecomputingunplugged,
author = {Kawsar, Fahim and Brush, A.J. Bernheim},
title = {Home computing unplugged: why, where and when people use different connected devices at home},
year = {2013},
isbn = {9781450317702},
publisher = {Association for Computing Machinery},
address = {New York, NY, USA},
url = {https://doi.org/10.1145/2493432.2493494},
doi = {10.1145/2493432.2493494},
booktitle = {Proceedings of the 2013 ACM International Joint Conference on Pervasive and Ubiquitous Computing},
pages = {627–636},
numpages = {10},
location = {Zurich, Switzerland},
series = {UbiComp '13}
}

@inproceedings{egelman2008familyaccounts,
author = {Egelman, Serge and Brush, A.J. Bernheim and Inkpen, Kori M.},
title = {Family accounts: a new paradigm for user accounts within the home environment},
year = {2008},
isbn = {9781605580074},
publisher = {Association for Computing Machinery},
address = {New York, NY, USA},
url = {https://doi.org/10.1145/1460563.1460666},
doi = {10.1145/1460563.1460666},
booktitle = {Proceedings of the 2008 ACM Conference on Computer Supported Cooperative Work},
pages = {669–678},
numpages = {10},
location = {San Diego, CA, USA},
series = {CSCW '08}
}

@InProceedings{brush2007yoursmineours,
author="Brush, A. J. Bernheim
and Inkpen, Kori M.",
editor="Krumm, John
and Abowd, Gregory D.
and Seneviratne, Aruna
and Strang, Thomas",
title="Yours, Mine and Ours? Sharing and Use of Technology in Domestic Environments",
booktitle="UbiComp 2007: Ubiquitous Computing",
year="2007",
publisher="Springer Berlin Heidelberg",
address="Berlin, Heidelberg",
pages="109--126",
isbn="978-3-540-74853-3"
}

@InProceedings{brown2007whereaboutsclock,
author="Brown, Barry
and Taylor, Alex S.
and Izadi, Shahram
and Sellen, Abigail
and Kaye, Joseph Jofish'
and Eardley, Rachel",
editor="Krumm, John
and Abowd, Gregory D.
and Seneviratne, Aruna
and Strang, Thomas",
title="Locating Family Values: A Field Trial of the Whereabouts Clock",
booktitle="UbiComp 2007: Ubiquitous Computing",
year="2007",
publisher="Springer Berlin Heidelberg",
address="Berlin, Heidelberg",
pages="354--371",
isbn="978-3-540-74853-3"
}

@inproceedings{sellen2006whereaboutsClock,
author = {Sellen, Abigail and Eardley, Rachel and Izadi, Shahram and Harper, Richard},
title = {The whereabouts clock: early testing of a situated awareness device},
year = {2006},
isbn = {1595932984},
publisher = {Association for Computing Machinery},
address = {New York, NY, USA},
url = {https://doi.org/10.1145/1125451.1125694},
doi = {10.1145/1125451.1125694},
booktitle = {CHI '06 Extended Abstracts on Human Factors in Computing Systems},
pages = {1307–1312},
numpages = {6},
location = {Montr\'{e}al, Qu\'{e}bec, Canada},
series = {CHI EA '06}
}

@article{lindley2009BubbleBoard,
  title={Resilience in the face of innovation: Household trials with BubbleBoard},
  author={Lindley, Si{\^a}n E and Banks, Richard and Harper, Richard and Jain, Anab and Regan, Tim and Sellen, Abigail and Taylor, Alex S},
  journal={International Journal of Human-Computer Studies},
  volume={67},
  number={2},
  pages={154--164},
  year={2009},
  publisher={Elsevier}
}

@article{ModEatMobileHCI,
author = {Silva, Lucas M. and Ankrah, Elizabeth A. and Huai, Yuqi and Epstein, Daniel A.},
title = {Exploring Opportunities for Multimodality and Multiple Devices in Food Journaling},
year = {2023},
issue_date = {September 2023},


volume = {7},
number = {MHCI},
url = {https://doi.org/10.1145/3604256},
doi = {10.1145/3604256},
journal = {Proc. ACM Hum.-Comput. Interact.},
month = {sep},
articleno = {209},
numpages = {27}
}

@inproceedings{silva2025Foodytalk,
author = {Silva, Lucas M. and Lu, Xi and Liang, Emily X. and Epstein, Daniel A.},
title = {Foody Talk: Exploring Opportunities for Conversational Food Journaling},
year = {2025},
isbn = {9798400713941},
publisher = {Association for Computing Machinery},
address = {New York, NY, USA},
url = {https://doi.org/10.1145/3706598.3713875},
doi = {10.1145/3706598.3713875},
booktitle = {Proceedings of the 2025 CHI Conference on Human Factors in Computing Systems},
articleno = {1183},
numpages = {19},
location = {
},
series = {CHI '25}
}

@inproceedings{kim222MyMove,
author = {Kim, Young-Ho and Chou, Diana and Lee, Bongshin and Danilovich, Margaret and Lazar, Amanda and Conroy, David E. and Kacorri, Hernisa and Choe, Eun Kyoung},
title = {MyMove: Facilitating Older Adults to Collect In-Situ Activity Labels on a Smartwatch with Speech},
year = {2022},
isbn = {9781450391573},
publisher = {Association for Computing Machinery},
address = {New York, NY, USA},
url = {https://doi.org/10.1145/3491102.3517457},
doi = {10.1145/3491102.3517457},
booktitle = {Proceedings of the 2022 CHI Conference on Human Factors in Computing Systems},
articleno = {416},
numpages = {21},
location = {New Orleans, LA, USA},
series = {CHI '22}
}

@inproceedings{gouveaGlanceable2016,
author = {Gouveia, R\'{u}ben and Pereira, F\'{a}bio and Karapanos, Evangelos and Munson, Sean A. and Hassenzahl, Marc},
title = {Exploring the design space of glanceable feedback for physical activity trackers},
year = {2016},
isbn = {9781450344616},
publisher = {ACM},
address = {New York, NY, USA},
url = {https://doi.org/10.1145/2971648.2971754},
doi = {10.1145/2971648.2971754},
booktitle = {Proceedings of the 2016 ACM International Joint Conference on Pervasive and Ubiquitous Computing},
pages = {144–155},
numpages = {12},
location = {Heidelberg, Germany},
series = {UbiComp '16}
}

@inproceedings{chi2026familybloom,
  author = {Silva, Lucas M. and Min, Aehong and Stefanidi, Evropi and Cibrian, Franceli L. and Beltran, Jesus A. and Zeiler, Cassie and Schuck, Sabrina and Lakes, Kimberley D and Hayes, Gillian R. and Epstein, Daniel A.},
title = {FamilyBloom: Examining Ecologies of Collaboration in Family-Centered Health Tracking},
year = {2026},
isbn = {9798400722783},
publisher = {Association for Computing Machinery},
address = {New York, NY, USA},
url = {https://doi.org/10.1145/3772318.3791277},
doi = {10.1145/3772318.3791277},
booktitle = {Proceedings of the 2026 CHI Conference on Human Factors in Computing Systems},
articleno = {652},
numpages = {24},
location = {
},
series = {CHI '26}
}

@article{everydayResistance,
author = {Lu, Alex Jiahong},
title = {Toward everyday negotiation and resistance under data-driven surveillance},
year = {2022},
issue_date = {March - April 2022},
publisher = {Association for Computing Machinery},
address = {New York, NY, USA},
volume = {29},
number = {2},
issn = {1072-5520},
url = {https://doi.org/10.1145/3516427},
doi = {10.1145/3516427},
journal = {Interactions},
month = feb,
pages = {34–38},
numpages = {5}
}

@inproceedings{minorResistance2025,
author = {Hsueh, Stacy and Van Dusen, Danielle and Caspi, Anat and Mankoff, Jennifer},
title = {Minor Resistance: The Everyday Politics and Power Dynamics of Assistive Technology Adoption},
year = {2025},
isbn = {9798400706769},
publisher = {Association for Computing Machinery},
address = {New York, NY, USA},
url = {https://doi.org/10.1145/3663547.3746465},
doi = {10.1145/3663547.3746465},
booktitle = {Proceedings of the 27th International ACM SIGACCESS Conference on Computers and Accessibility},
articleno = {57},
numpages = {16},
location = {
},
series = {ASSETS '25}
}

@inproceedings{kaiwenSmartHome2026,
author = {Sun, Kaiwen and Xiaoyi Li, Jade and Chung, Irene and Radesky, Jenny and Yip, Jason and Brooks, Christopher and Schaub, Florian},
title = {“Families are messy”: From Parent-Child Tensions to Family-Centered Design of Smart Home Technologies},
year = {2026},
isbn = {9798400722783},
publisher = {Association for Computing Machinery},
address = {New York, NY, USA},
url = {https://doi.org/10.1145/3772318.3790865},
doi = {10.1145/3772318.3790865},
booktitle = {Proceedings of the 2026 CHI Conference on Human Factors in Computing Systems},
articleno = {656},
numpages = {21},
location = {
},
series = {CHI '26}
}

@inproceedings{RikuCalmReminder,
author = {Arakawa, Riku and Bali, Shreya and Sitaraman, Anupama and Seo, Woosuk and Shaaban, Sam and Lindhiem, Oliver and Kennedy, Traci M. and Goel, Mayank},
title = {CalmReminder: A Design Probe for Parental Engagement with Children with Hyperactivity, Augmented by Real-Time Motion Sensing with a Watch},
year = {2026},
isbn = {9798400722783},
publisher = {Association for Computing Machinery},
address = {New York, NY, USA},
url = {https://doi.org/10.1145/3772318.3791871},
doi = {10.1145/3772318.3791871},
booktitle = {Proceedings of the 2026 CHI Conference on Human Factors in Computing Systems},
articleno = {1022},
numpages = {17},
location = {
},
series = {CHI '26}
}

@inproceedings{liangChi2026CollabCareWork,
author = {Liang, Shichen and Salim, Alvina and Yarosh, Svetlana and Johnson, Amanda and Shin, Ji Youn},
title = {“I’ll Just Do It”: Designing for the Hidden Work of Collaborative Family Caregiving},
year = {2026},
isbn = {9798400722783},
publisher = {Association for Computing Machinery},
address = {New York, NY, USA},
url = {https://doi.org/10.1145/3772318.3791609},
doi = {10.1145/3772318.3791609},
booktitle = {Proceedings of the 2026 CHI Conference on Human Factors in Computing Systems},
articleno = {663},
numpages = {19},
location = {
},
series = {CHI '26}
}

\appendix



\end{document}